\documentclass[11pt]{article}
\usepackage{latexsym}
\usepackage{amsfonts}
\usepackage{amssymb}

\newcommand{\mysection}[1]{\section{#1}\setcounter{equation}{0}}

\newcommand{\bea}{\begin{eqnarray}} 
\newcommand{\eea}{\end{eqnarray}}
\newcommand{\beann}{\begin{eqnarray*}} 
\newcommand{\eeann}{\end{eqnarray*}}
\newcommand{\beq}{\begin{equation}} 
\newcommand{\eeq}{\end{equation}}
\newcommand{\ba}{\begin{array}} 
\newcommand{\ea}{\end{array}}
\newcommand{\ben}{\begin{enumerate}} 
\newcommand{\een}{\end{enumerate}}
\newcommand{\bit}{\begin{itemize}} 
\newcommand{\eit}{\end{itemize}}

\newcommand{\cA}{{\cal A}}

\newcommand{\cD}{{\cal D}}

\newcommand{\cF}{{\cal F}}

\newcommand{\cR}{{\cal R}}

\newcommand{\lra}{\leftrightarrow}
\newcommand{\LRA}{\Leftrightarrow}
\newcommand{\then}{\Rightarrow}
\newcommand{\umat}{\mbox{{\bf 1}}}

\newcommand{\5}{\bar }  
\newcommand{\6}{\partial } 
\newcommand{\7}{\hat } 
\newcommand{\4}{\tilde }

\newcommand{\sfrac}[2]{\mbox{$\frac{{#1}}{{#2}}$}}
\newcommand{\f}[3]{{f_{#1#2}}^{#3}}

\newcommand{\Ii}{{\mathrm{i}}}

\newcommand{\ep}{\epsilon}

\newcommand{\da}{{\dot\alpha}} 
\newcommand{\dbeta}{{\dot\beta}} 
\newcommand{\dgamma}{{\dot\gamma}}

\newcommand{\Mdex}{M}
\newcommand{\Ndex}{N}
\newcommand{\Pdex}{P}
\newcommand{\Qdex}{Q}
\newcommand{\Rdex}{R}
\newcommand{\Adex}{A}

\newcommand{\FRAME}[1]{\fbox{\mbox{$#1$}}}

\def\da{{\dot \alpha}} \def\dbe{{\dot \beta}}
\def\dg{{\dot \gamma}} \def\dde{{\dot \delta}}
\def\ua{{\underline{\alpha}}} \def\ube{{\underline{\beta}}}

\def\dR{{(R)}}
\def\T#1#2#3{{T_{#1#2}}^{#3}} 
\def\F#1#2#3{{F_{#1#2}}^{#3}}
\def\Gg#1#2#3{{g_{#1#2}}^{#3}}
\def\G#1#2#3{{G_{#1#2}}^{#3}}
\def\f#1#2#3{{f_{#1#2}}^{#3}}
\def\A#1#2{A_{#2}^{#1}}
\def\CF#1#2#3{{{\cal F}_{#1#2}}^{#3}}

\def\csum#1#2{\sum_{#1}\hspace{-1.#2em}\circ\ \ \ }

\begin{document}
\begin{flushright}
hep-th/0204035
\end{flushright}

\begin{center}
{\bf\LARGE Lectures on Supergravity
}
\end{center}

\begin{center}
{\large Friedemann Brandt}\\
Max-Planck-Institute for Mathematics in the Sciences,\\ 
Inselstra\ss e 22--26, D--04103 Leipzig, Germany
\end{center}

\begin{center}
Published in Fortschr.\ Phys.\ 50 (2002) 1126-1172
\end{center}

\begin{abstract}
The text is an essentially self-contained introduction to 
four-dimensional N=1 supergravity, including its
couplings to super Yang-Mills and chiral matter multiplets,
for readers with basic knowledge of standard
gauge theories and general relativity.
Emphasis is put on showing how supergravity
fits in the general framework of gauge theories and
how it can be derived from a tensor calculus for gauge theories 
of a standard form.\\[8pt]

\begin{tabular}{cl}
&{\bf Contents}\\[4pt]
\ref{intro} & 
Introduction \dotfill \pageref{`i'}\\
\ref{jet}   & 
Gauge symmetries in the jet space approach 
              \dotfill\dots \pageref{`ii'}\\
\ref{pure}  & 
D=4, N=1 pure supergravity \dotfill \pageref{`iii'}\\
\ref{tensor}& 
Tensor calculus for standard gauge theories 
              \dotfill \pageref{`iv'}\\
\ref{matter}& 
Off-shell formulations of D=4, N=1 supergravity with matter 
              \dots\dots \pageref{`v'}\\
\ref{spin}  & 
Lorentz algebra, spinors, Grassmann parity 
              \dotfill \pageref{`A'}\\
\ref{ver}   & 
Explicit verification of local supersymmetry
              \dotfill \pageref{`B'}
\end{tabular}
\end{abstract}

\mysection{Introduction}\label{`i'}\label{intro}

Supergravity (SUGRA) is for several reasons an interesting concept
in modern high energy physics. It raises supersymmetry (SUSY) to
a gauge symmetry and thus
combines two principles of major interest, namely
gauge invariance which underlies our present models of
fundamental interactions, and SUSY, one of
the most promising theoretical concepts for extending
these models. In addition SUGRA includes
and extends general relativity (GR) which makes it an interesting
framework for describing gravitational interactions 
in high energy physics.
In particular SUGRA theories arise in string theory,
one of the presently most favoured approaches in the field
of ``quantum gravity''. 

SUGRA had an important impact
on the development of general concepts in the field of gauge theories,
such as the reformulation and refinement of the BRST-approach,
because it exhibited properties which are not encountered in
more familiar gauge theories, such as Yang-Mills (YM) theories or
standard GR. Such properties are gauge transformations
whose commutator algebra does not close off-shell 
or involves field dependent structure functions.
Therefore SUGRA can serve as an instructive example
to illustrate the general structure of gauge theories.

As SUGRA is a generalization of GR, it may be worthwhile to
compile further differences from standard GR.
The most fundamental
difference is that SUGRA theories have more gauge symmetries
than standard GR. In particular they have of course local SUSY
and contain corresponding gauge fields, so-called
gravitinos (Rarita-Schwinger fields) which are spinor fields.
Owing to the presence of spinor fields, SUGRA theories
are formulated in the vielbein formulation (Cartan formulation)
of GR rather than in the metric formulation, and therefore
they are always invariant under local Lorentz transformations.
Many SUGRA theories, especially in higher dimensions,
contain in addition $p$-form gauge fields which generalize the 
electromagnetic gauge potential and are invariant under corresponding gauge
transformations of these fields. 

Fields which occur typically in
SUGRA theories are thus the vielbein which we denote by
$e_\mu^a$ ($\mu$ is a ``world index'' of the same as type as the
indices of the metric in GR, $a$ is a Lorentz vector index),
the gravitino(s) $\psi_\mu$ (whose spinor
index has been suppressed), $p$-form gauge fields $A_{\mu_1\dots\mu_p}$
which are totally antisymmetric in their world indices 
(the electromagnetic gauge field and YM gauge fields are 1-form
gauge fields in this terminology), and standard matter fields such as
spacetime scalar fields $\phi$ or ordinary spinor fields $\lambda$
(again the spinor index has been suppressed). A standard SUGRA
theory contains always the vielbein and at least one gravitino.
Whether and which other fields are present depends on the
particular SUGRA theory. 

An important restriction on the possible
field content is that the number of bosonic degrees of freedom and
the number of fermionic degrees of freedom must coincide on-shell (for SUGRA 
theories of standard type). This is required
by SUSY because SUSY relates bosonic fields
and fermionic fields in a particular way.
The number of degrees of freedom (DOF) relevant here is the number
of linearly independent plane wave solutions
of the free field equations with given Fourier-momentum, 
up to linearized gauge transformations. 
The free field equations are
the linearized equations of motion (EOM) in a flat gravitational
background. For the standard
fields, with standard free field equations, 
these DOF are compiled
in (\ref{DOF}) and (\ref{Fdim})
where $D$ is the spacetime dimension and $f$
is the real dimension of the smallest nontrivial irreducible spinor
representation (for Minkowski signature of the metric).
Details of the derivation of the numbers in (\ref{DOF}) and (\ref{Fdim})
can be found, for example, in \cite{LouisdeWit} and \cite{toine},
some elementary facts about spinors in various dimensions are
provided in appendix \ref{spin}.
\bea
&\begin{array}{c|c|c}
\mbox{field} & \mbox{DOF off-shell} & \mbox{DOF on-shell 
($D\geq 3$)}\\
\hline\rule{0em}{3ex}
\mbox{vielbein}\ e_\mu^a & D(D-1)/2 & D(D-3)/2 \\
\rule{0em}{3ex}
\mbox{gravitino}\ \psi_\mu & f(D-1) & f(D-3)/2 \\
\rule{0em}{3ex}
\mbox{$p$-form gauge field}\ A_{\mu_1\dots\mu_p} & 
{D-1 \choose p} & {D-2 \choose p}\\
\rule{0em}{3ex}
\mbox{real scalar field}\ \phi & 1 & 1\\
\rule{0em}{3ex}
\mbox{spinor field}\ \lambda & f & f/2
\end{array}&
\label{DOF}\\[14pt]
& f=\left\{
\ba{ll}
2^{D/2-1}\ \mbox{for $D$ mod 8 $=2$} & \mbox{Majorana-Weyl spinors}\\
2^{\lfloor D/2\rfloor}\ \mbox{for $D$ mod 8 $=0,1,3,4,6$} & 
\mbox{Weyl or Majorana spinors$^*$}\quad\\
2^{(D+1)/2}\ \mbox{for $D$ mod 8 $=5,7$} & \mbox{Dirac spinors}
\ea
\right.&
\label{Fdim}
\\
&\mbox{$^*$ Weyl spinors only if $D=2k$, no Majorana spinors 
if $D$ mod 8 $=6$}.&
\nonumber\eea

(\ref{DOF}) contains also the number
of DOF off-shell
given by the number of independent components of the respective field
up to gauge transformations (taking reducibility relations into
account, if any).  
These numbers are relevant for
so-called off-shell formulations of SUGRA theories, i.e., formulations
where the commutator algebra of the gauge transformations
closes off-shell. Namely in such formulations 
the number of bosonic DOF and
the number of fermionic DOF must coincide both on-shell and off-shell (again,
for SUGRA  theories of standard type).

An additional restriction on possible SUGRA theories of standard type
is the upper bound on the number of real SUSYs. This upper bound is
32 when one requires that dimensional reduction to $D=4$
must not yield fields with spin $\geq 5/2$ (this requirement
reflects that theories with spin $\geq 5/2$ are believed to be
inconsistent or physically unacceptable). The number of SUSYs is often given
in terms of the number $N$ of
{\em sets} of SUSYs where each set has $f$ elements
with $f$ as in (\ref{Fdim}) [i.e., the corresponding
gauge parameters sit in an irreducible spinor representation].
Hence, if in this terminology one says that a theory has $N$ SUSYs, it has
thus actually $N f$ real SUSYs. The bound of at most 32 SUSYs
limits standard SUGRA theories, which can be characterized
in this way by a value of $N$, to spacetime dimensions $D\leq 11$
because for $D\geq 12$ one has $f\geq 64$.%
\footnote{In ``non-standard theories'' the number of SUSYs need not be an
integer multiple of $f$.}
Therefore, SUGRA theories of standard type exist only up 
to eleven dimensions.

In (\ref{explicit}) the values of $f$ and the on-shell DOF
for some fields are spelled out explicitly for $4\leq D\leq 11$.
In addition the maximal value $N_\mathrm{max}$ of $N$
is given. One sees from these numbers, for example,
that in $D=4$ it might be possible to construct
an $N=1$ SUGRA theory whose only fields are
the vielbein $e_\mu^a$ (called ``vierbein'' in $D=4$) and one gravitino
$\psi_\mu$ ($N$ is also the number of gravitinos
because these sit in spinor representations with dimension $f$). 
This theory does indeed exist and will be presented in some detail later.
Other possibilities would be, for example,
a $D=4$, $N=2$ SUGRA theory with vierbein $e_\mu^a$, two gravitinos
$\psi_\mu$, $\psi'_\mu$, and one ``photon'' $A_\mu$,
or a $D=11$, $N=1$ SUGRA theory with ``elfbein'' $e_\mu^a$,
one gravitino $\psi_\mu$ and one 3-form gauge field $A_{\mu\nu\rho}$.
Both SUGRA theories do also exist.
\bea
\begin{array}{c|c|c|c|c|c|c|c|c}
D & 4 & 5 & 6 &7 &8 &9 & 10 & 11
\\
\hline\hline\rule{0em}{3ex}
f & 4 & 8 & 8 & 16 & 16 & 16 & 16 & 32
\\
\hline\rule{0em}{3ex}
N_\mathrm{max} &8&4&4&2&2&2&2&1
\\
\hline\rule{0em}{3ex}
\mbox{DOF of\ } \psi_\mu \mbox{\ on-shell}
& 2 & 8 & 12 & 32 & 40 & 48 & 56 & 128
\\
\hline\rule{0em}{3ex}
\mbox{DOF of\ }e_\mu^a \mbox{\ on-shell}
& 2 & 5 & 9 & 14 & 20 & 27 & 35 & 44
\\
\hline\rule{0em}{3ex}
\mbox{DOF of\ }A_{\mu} \mbox{\ on-shell}
& 2 & 3 & 4 & 5 & 6 & 7 & 8 & 9
\\
\hline\rule{0em}{3ex}
\mbox{DOF of\ }A_{\mu\nu} \mbox{\ on-shell}
& 1 & 3 & 6 & 10 & 15 & 21 & 28 & 36
\\
\hline\rule{0em}{3ex}
\mbox{DOF of\ }A_{\mu\nu\rho} \mbox{\ on-shell}
& 0 & 1 & 4 & 10 & 20 & 35 & 56 & 84
\end{array}
\label{explicit}
\eea

The remainder of this text is devoted to
$D=4$, $N=1$ SUGRA. It aims at giving an essentially self-contained
introduction to the structure of this theory at a non-expert level, 
for readers
with basic knowledge of GR and standard
gauge theories (in particular, some knowledge of YM theory
might be helpful). The text is limited to basic material.
In particular it does not cover more technical stuff, such as the use of
superspace techniques, or a discussion of phenomenological aspects,
which may be found in textbooks or reviews on SUGRA, such as 
\cite{PvN,WessBagger,1001,Nilles,West,DragonES,LouisdeWit,Sergei}.
Rather I have tried to emphasize that and how
SUGRA fits in with general principles of gauge theories. These
principles are briefly reviewed
in section \ref{jet}. Section \ref{pure}
presents in some detail the simplest $D=4$, $N=1$ SUGRA theory 
mentioned above, whose only
fields are the vierbein and the gravitino.
Section \ref{tensor} introduces the concept of
a tensor calculus for a class of standard gauge theories. This
calculus is used in section \ref{matter} as a framework
to present the ``old minimal'' and the ``new minimal'' 
off-shell formulations of $D=4$, $N=1$ SUGRA
including the coupling to matter multiplets (super YM multiplets,
chiral multiplets). Conventions, especially concerning
spinors, and the explicit verification that the SUGRA actions
given in the text are indeed supersymmetric are
relegated to the appendix. 

\mysection{Gauge symmetries in the jet space approach}
\label{`ii'}\label{jet}

This section assembles the 
general definition and basic properties
of Lagrangean gauge theories, using
the jet space approach.

\subsection{Jet spaces}

The concept of jet-spaces originates
from the theory of partial differential equations (see, for example,
\cite{goldschmidt,pommaret,olver,saunders,anderson}). It
provides
a mathematically rigorous, simple and general framework for the
discussion of many aspects of symmetries.
For our purposes it suffices to know
that jet spaces are spaces whose coordinates are 
the ordinary coordinates $x^\mu$ of a base space $M$
(in our case: spacetime), and additional
variables $\6_{\mu_1\dots\mu_k}\phi^i$ representing fields
$\phi^i$ ($k=0$) and their 
first and higher order derivatives ($k=1,2,\dots$)\footnote{Henceforth
we shall work in the infinite jet space, containing all derivatives.}.
The fields and their derivatives
are thus regarded as algebraic objects.
The conception of fields as functions of the
coordinates $x^\mu$, or as mappings from $M$
to some space $F$ arise
only as sections $s$ of 
the corresponding jet bundle over $M$
where the jet variables $\6_{\mu_1\dots\mu_k}\phi^i$
turn into functions $M\rightarrow F$ according to
\bea
\6_{\mu_1\dots\mu_k}\phi^i|_s
=\frac{\6^k\phi^i(x)}{\6x^{\mu_1}\dots \6x^{\mu_k}}\ .
\label{section}\eea
As the partial derivatives commute on smooth functions, 
we identify jet coordinates
$\6_{\mu_1\dots\mu_k}\phi^i$ which
differ only by permutations of the derivative indices:
\bea
\forall r,s:\quad
\6_{\mu_1\dots\mu_r\dots\mu_s\dots\mu_k}\phi^i
=\6_{\mu_1\dots\mu_s\dots\mu_r\dots\mu_k}\phi^i.
\label{jets}
\eea
One may either work with a set of independent jet coordinates,
such as $\{x^\mu,\6_{\mu_1\dots\mu_k}\phi^i:\mu_i\leq\mu_{i+1},k=0,1,\dots\}$,
or with the redundant set of all jet variables.
We prefer the second option because it allows one to avoid
inconvenient combinatorical factors.
The partial derivatives are then represented in the infinite jet space
by algebraic operations $\6_\mu$ defined by
\bea
\6_\mu=\frac{\6}{\6x^\mu}+
\sum_{k\geq 0}\6_{\mu\mu_1\dots\mu_k}\phi^i\,
\frac{\6^S}{\6 \6_{\mu_1\dots\mu_k}\phi^i}\ ,
\label{6}
\eea
where
the derivatives $\6^S/\6\6_{\mu_1\dots\mu_k}\phi^i$
act on the jet variables according to
\bea
\frac{\6^S\6_{\nu_1\dots\nu_k}\phi^j}
{\6\6_{\mu_1\dots\mu_k}\phi^i}=
\delta^j_i\,\delta_{(\nu_1}^{\mu_1}\dots\delta_{\nu_k)}^{\mu_k}\ ,\quad
k\neq r:\ \frac{\6^S\6_{\nu_1\dots\nu_r}\phi^j}
{\6\6_{\mu_1\dots\mu_k}\phi^i}=0.
\label{6S}
\eea
Here the round parantheses denote symmetrization with weight one, such as
$\delta_{(\mu}^\rho\delta_{\nu)}^\sigma=1/2(
\delta_{\mu}^\rho\delta_{\nu}^\sigma
+\delta_{\nu}^\rho\delta_{\mu}^\sigma)$.
With these definitions the derivatives $\6_\mu$
have indeed the same algebraic properties as the
partial derivatives of smooth functions -- 
as they should, in order that
(\ref{section}) makes sense. In particular they satisfy
\bea
\6_\mu \6_{\mu_1\dots\mu_k}\phi^i=\6_{\mu\mu_1\dots\mu_k}\phi^i
\eea
and they commute,
\bea
[\6_\mu,\6_\nu]=0.
\label{6,6}
\eea
A basic and important
fact is that a function $f$ on the jet space is a total divergence
if and only if it has vanishing Euler-Lagrange derivatives with respect to
all fields on which it depends,
\bea
f(x,[\phi])=\6_\mu K^\mu(x,[\phi])\quad\LRA\quad
\frac{\7\6 f(x,[\phi])}{\7\6\phi^i}=0\quad \forall \phi^i,
\label{APL}
\eea
where $\7\6 f/\7\6\phi^i$ denotes the Euler-Lagrange derivative
of $f$ with respect to $\phi^i$, 
\bea
\frac{\7\6 f(x,[\phi])}{\7\6\phi^i}
=\sum_{k\geq 0}(-)^k\6_{\mu_1}\dots\6_{\mu_k}\,
\frac{\6^Sf(x,[\phi])}{\6\6_{\mu_1\dots\mu_k}\phi^i}\ .
\eea
Here
$[\phi]$ indicates local dependence on the fields (which usually means
dependence on derivatives
up to some arbitrary but finite order).

\subsection{Gauge symmetries of a Lagrangian}

An (infinitesimal) gauge transformation is a
transformation of the fields involving linearly
``gauge parameters'' or their derivatives.
The gauge parameters are arbitrary fields and therefore
these parameters and their derivatives are also treated
as jet variables. Hence, when dealing with
gauge symmetries, we work in an enlarged jet space
involving also these extra variables
in addition to the coordinates $x^\mu$ and the
``fields'' $\phi^i$ and their derivatives.
The basic difference between the gauge parameters and
the fields is that the former do not occur in the Lagrangian and
the field equations (Euler-Lagrange equations of motion) derived from it. 
Hence the Lagrangian is a function on the ``original'' jet space
with coordinates $x^\mu$ and $\6_{\mu_1\dots\mu_k}\phi^i$.
Each jet variable (including the gauge parameters and their derivatives) 
has a Grassmann parity
which is 0 for ``bosonic'' (commuting) fields or 1 for
``fermionic'' (anticommuting) fields, cf.\ appendix
\ref{spin} for our conventions in the SUGRA context.
By assumption, the Lagrangian is a Grassmann even function
on the jet space and the gauge transformations are Grassmann
even operations.

Gauge transformations $\delta_\xi\phi^i$
of the fields are
given by operators $R^i_\Mdex$ which act on
gauge parameters $\xi^\Mdex$ and may depend on the fields
and their derivatives:
\beq
\delta_\xi\phi^i=R^i_\Mdex\xi^\Mdex,\quad
R^i_\Mdex=\sum_{k=0}^m
r^{i\mu_1\dots\mu_k}_\Mdex(x,[\phi])\6_{\mu_1}\dots\6_{\mu_k}.
\label{gdef1}
\eeq
These transformations are extended to derivatives of
the fields and to local functions on jet spaces according to
\bea
\delta_\xi =\sum_{k\geq 0}
(\6_{\mu_1}\dots\6_{\mu_k}\delta_\xi\phi^i)\,
\frac{\6^S }{\6\6_{\mu_1\dots\mu_k}\phi^i}\ .
\label{gdef2'}\eea
In particular this gives
\bea
&[\delta_\xi,\6_\mu]=0,&
\label{gdef2a}\\
&\delta_\xi x^\mu=0,&
\label{gdef2b}\\
&\delta_\xi(ab)=(\delta_\xi a)b+a(\delta_\xi b).&
\label{gdef2}
\eea
(\ref{gdef2a}) means simply that
the gauge transformations of the derivatives of the
fields are so-called `prolongations' of the gauge transformations
of the undifferentiated fields ($\delta_\xi\6_\mu\phi^i=
\6_\mu (\delta_\xi\phi^i)$ etc). (\ref{gdef2b}) means that
explicit coordinates $x^\mu$ are never transformed, i.e., 
when evaluated on a section of the jet bundle,
$(\delta_\xi\phi^i)(x)$ represents the
transformation of $\phi^i(x)$ as a function
of its arguments but not of the arguments themselves (it represents
thus the Lie derivative of $\phi^i$).
(\ref{gdef2}) is the Leibniz rule and means that
the gauge transformations are derivations on the jet space.
We can now define gauge symmetries of a Lagrangian:
\medskip

{\bf Definition.} A gauge transformation $\delta_\xi$ 
is called a gauge symmetry of a Lagrangian $L(x,[\phi])$
if it leaves the Lagrangian invariant up to a total divergence:
\beq
\delta_\xi L(x,[\phi])=\6_\mu K^\mu (x,[\phi,\xi]).
\label{gdef3}
\eeq

\subsection{Noether identities and gauge symmetry of the EOM}

Owing to (\ref{APL}), the gauge invariance condition (\ref{gdef3}) imposes
\bea
\frac{\7\6 \delta_\xi L(x,[\phi])}{\7\6\xi^\Mdex}=0\quad
\forall \xi^\Mdex
\label{NI1}
\eea
and
\bea
\frac{\7\6 \delta_\xi L(x,[\phi])}{\7\6\phi^i}=0\quad
\forall \phi^i.
\label{EOM1}
\eea
(\ref{NI1}) are the Noether identities
corresponding to the gauge symmetry.
Explicitly they read
\bea
(-)^{|\phi^i|}R^{i+}_\Mdex\,
\frac{\7\6 L(x,[\phi])}{\7\6\phi^i}=0
\label{NI}
\eea
where $|\phi^i|$ is the Grassmann parity of $\phi^i$ and
$R^{i+}_\Mdex$ is the operator adjoint to
the operator $R^i_\Mdex$ which defines the gauge symmetry
according to (\ref{gdef1}). This adjoint operator is given, on all
functions $f$ on the jet space, by
\bea
R^{i+}_\Mdex f=\sum_{k\geq 0}
(-)^k\6_{\mu_1}\dots\6_{\mu_k}[f\, r^{i\mu_1\dots\mu_k}_\Mdex(x,[\phi])].
\eea
(\ref{EOM1}) yields the gauge transformations of the
EOM. Explicitly one obtains (see, for example, formula (6.43) of
\cite{report}):
\bea
\delta_\xi\,\frac{\7\6 L(x,[\phi])}{\7\6\phi^i}=
-\sum_{k\geq 0}(-)^k\6_{\mu_1}\dots\6_{\mu_k}
\Big[
\frac{\6^S\delta_\xi\phi^j}{\6\6_{\mu_1\dots\mu_k}\phi^i}\
\frac{\7\6 L(x,[\phi])}{\7\6\phi^j}
\Big].
\label{superformel}
\eea

{\bf Remark.} Actually (\ref{NI}) is equivalent to (\ref{gdef3})
(Noether's second theorem \cite{noether}). The reason is
that every term in $\delta_\xi L$ is linear in the
$\xi$'s (or their derivatives) which implies
\[
\delta_\xi L(x,[\phi])=\xi^\Mdex\,
\frac{\7\6 \delta_\xi L(x,[\phi])}{\7\6\xi^\Mdex}
+\6_\mu K^\mu(x,[\phi,\xi]).
\]
Hence (\ref{NI}) implies indeed (\ref{gdef3}), 
and thus it also implies (\ref{EOM1}).

\subsection{Trivial gauge symmetries}\label{trivg}

Consider the transformations
\beq
\delta^\mathrm{triv}\phi^i=
\sum_{k,m\geq 0}(-)^k
\partial_{\mu_1}\dots\partial_{\mu_k}\Big[
M^{j(\nu_1\dots\nu_m)i(\mu_1\dots\mu_k)}(x,[\phi,\xi])
\partial_{\nu_1}\dots\partial_{\nu_m}\,\frac{\7\6  L}{\7\6 \phi^j}\Big],
\label{trivg1}
\eeq
where 
\beq
M^{j(\nu_1\dots\nu_m)i(\mu_1\dots\mu_k)}(x,[\phi,\xi])
= -(-)^{|\phi^i|\,|\phi^j|} 
M^{i(\mu_1\dots\mu_k)j(\nu_1\dots\nu_m)}(x,[\phi,\xi]).
\label{trivg2}
\eeq
The transformation (\ref{trivg1}),
extended to the whole jet space as in (\ref{gdef2'}),
is a gauge symmetry of $L$ according to our definition because of
\[
\delta^\mathrm{triv} L \sim
M^{j(\nu_1\dots\nu_m)i(\mu_1\dots\mu_k)}(x,[\phi,\xi])
\Big[\partial_{\mu_1}\dots\partial_{\mu_k}\,
\frac{\hat\partial  L}{\hat\partial \phi^i}\Big]
\Big[\partial_{\nu_1}\dots\partial_{\nu_m}\,
\frac{\hat\partial  L}{\hat\partial \phi^j}\Big]
=0
\]
where $\sim$ denotes equality up to a total divergence and
the last equality ($=0$) holds because of the graded antisymmetry
of the $M$'s as in (\ref{trivg2}).
For obvious reasons, such transformations 
are called trivial gauge symmetries. They exist for every Lagrangian
and vanish on-shell, i.e.,
they vanish on all solutions of the EOM.
Conversely one can prove
under fairly general assumptions (regularity conditions) 
that a
gauge symmetry which vanishes on-shell takes necessarily the
form (\ref{trivg1}) \cite{henteit,report}.

\subsection{Generating set of gauge symmetries}\label{basis}

When trying to characterize the gauge symmetries of a model
satisfactorily one faces two
complications. On the one hand, one has to deal with the
trivial gauge symmetries which one wants to ``mod out''. On the other
hand one has to take the following fact into account: whenever
$\delta_\xi \phi^i=R^i_\Mdex\xi^\Mdex$ is a gauge symmetry, then
$\7\delta_{\7\xi} \phi^i=\7R^{i}_{\Adex}\7\xi^{\Adex}$ 
with $\7R^{i}_{\Adex}=R^i_\Mdex K^\Mdex_{\Adex}$ 
is also a gauge symmetry, for any (possibly
field dependent) local operators $K^\Mdex_{\Adex}$: 
indeed, when $\delta_\xi$ is a gauge symmetry, it satisfies
(\ref{gdef3}) for all $\xi$'s and thus
in particular for $\xi^\Mdex=K^\Mdex_{\Adex}\7\xi^{\Adex}$,
whatever operators $K^\Mdex_{\Adex}$ we choose and
for arbitrary $\7\xi^{\Adex}$. Notice
that even the range of the index $\Adex$ may differ
from the range of the index $\Mdex$. 
But clearly $\7\delta_{\7\xi}$ is not a new gauge symmetry
as it arises from $\delta_\xi$ just by substituting
$f^\Mdex(x,[\phi,\7\xi])=K^\Mdex_{\Adex}\7\xi^{\Adex}$ 
for $\xi^\Mdex$.
This motivates the following definition:
we say a set of operators $\{R^i_\Mdex\}$ provides a
generating set of the gauge symmetries
of a Lagrangian if any gauge symmetry of the Lagrangian 
can be generated through them according to
\beq
\delta_\xi L=\6_\mu K^\mu\quad \then\quad
\delta_\xi\phi^i=R^i_\Mdex f^\Mdex(x,[\phi,\xi])
+\delta^\mathrm{triv}\phi^i,
\label{basis1}
\eeq
for some local functions $f^\Mdex(x,[\phi,\xi])$.

The concept of a generating set of gauge symmetries
is of fundamental importance for the theory of gauge symmetries.
It is somewhat analogous to the concept of a basis of a vector space
although the analogy must be used with great care because
a generating set evidently is {\em not}
a basis of gauge symmetries in the vector space sense.
Within the analogy,  
(\ref{basis1}) corresponds to the completeness
of a basis of a vector space. The independence of the elements of a basis
also has a counterpart: it is the irreducibility of
a generating set. The latter requires that the
operators $R^i_\Mdex$ have no nontrivial `zero mode', i.e.,
\bea
R^i_\Mdex f^\Mdex(x,[\phi,\xi])=\delta^\mathrm{triv}\phi^i
\quad \then\quad f^\Mdex(x,[\phi,\xi])\approx 0\quad
(\mbox{irreducibility})
\label{basis2}
\eea
where $\approx$ is equality on-shell, 
\bea
f\approx g\quad:\LRA\quad f-g=M^i\,\frac{\7\6 L(x,[\phi])}{\7\6\phi^i}
\label{approx}
\eea
for some local operators $M^i$.
However, unlike the situation in the case of (finite
dimensional) vector spaces, it is not always possible to
choose an irreducible set 
because locality may obstruct this. So, one sometimes
has to deal with reducible generating sets of gauge transformations.

The choice of a generating set of gauge 
transformations is by no means unique; switching from one generating set to
another one corresponds in the above analogy to changing the
basis of a vecor space, albeit the freedom in the choice of
a generating set evidently exceeds by far the freedom
in the choice of a basis of a vector space.
The relation between two generating
sets $\{R^i_\Mdex\}$ and $\{\7R^i_{\Adex}\}$ is of the
type discussed above,
\bea
\7R^i_{\Adex}\approx R^i_\Mdex K^\Mdex_{\Adex}\ ,\quad
R^i_{\Mdex}\approx \7R^i_{\Adex} \7K^{\Adex}_{\Mdex}\ ,
\label{newbasis1}
\eea
for some local, generally field dependent operators
$K^\Mdex_{\Adex}$ and $\7K^{\Adex}_{\Mdex}$.
Again, the ranges of the indices $\Mdex$ and $\Adex$ may
differ; in particular one may switch from an irreducible
to a reducible set.
Notice that switching between different generating
sets is accompanied by redefinitions
of the corresponding sets of gauge parameters because
(\ref{newbasis1}) yields
\bea
&& \7\delta_{\7\xi}=\delta_{K\7\xi}+\delta^{\mathrm{triv}},
\quad (K\7\xi)^\Mdex=K^\Mdex_{\Adex}\7\xi^{\Adex},\nonumber\\
&&\delta_{\xi}=\7\delta_{\7K\xi}+\delta^{\mathrm{triv}},
\quad (\7K\xi)^{\Adex}=\7K^{\Adex}_{\Mdex}\xi^{\Mdex}.
\label{newbasis2}
\eea

\paragraph{Example.} Let us
consider 3-dimensional abelian Chern-Simons theory with
Lagrangian
\[
L=\ep^{\mu\nu\rho}A_\mu\6_\nu A_\rho.
\]
The set of fields $\phi^i$ is in this case given
by the components of the gauge field,
\[
\{\phi^i\}\equiv\{A_\mu\}.
\]
It can be proved that a generating set of gauge symmetries
of the abelian Chern-Simons Lagrangian is 
given by the abelian gauge transformations
\[
\delta_\xi A_\mu=\6_\mu\xi.
\]
The corresponding set of operators $R^i_N$ is thus given
just by the derivatives $\6_\mu$:
\[
\{R^i_N\}\equiv\{\6_\mu\}.
\]
Now, if this provides really a generating set, it must be possible to
express every gauge symmetry of the abelian Chern-Simons Lagrangian
in terms of these operators up
to trivial gauge symmetries, as in (\ref{basis1}). 
Let us verify that this holds for the
spacetime diffeomorphisms [the latter are indeed gauge symmetries
because the Chern-Simons Lagrangian is a scalar density with weight one
under spacetime diffeomorphisms]:
\beann
\delta_\mathrm{diffeo} A_\mu&=&\xi^\nu\6_\nu A_\mu+\6_\mu\xi^\nu A_\nu
\\
&=&\xi^\nu(\6_\nu A_\mu-\6_\mu A_\nu)
+\xi^\nu\6_\mu A_\nu+\6_\mu\xi^\nu A_\nu
\\
&=&
\frac 12\,\xi^\nu\ep_{\nu\mu\rho}\,\frac{\7\6 L}{\7\6 A_\rho}
+\6_\mu(\xi^\nu A_\nu).
\eeann
Note that the first term in the last line is a trivial
symmetry as in (\ref{trivg1}) (with
$M^{ji}\equiv \frac 12\,\xi^\nu\ep_{\nu\mu\rho}$), while the second term
is of the form $R^i_\Mdex f^\Mdex(\xi,\phi)$ (with
$f^M\equiv \xi^\nu A_\nu$).

\subsection{Algebra of gauge symmetries}

The concept of a generating set of gauge symmetries
allows one to derive the general form of
the commutator algebra of gauge symmetries.
The commutator of two gauge symmetries $\delta_{\xi_1}$ and
$\delta_{\xi_2}$ is again a gauge symmetry because it
leaves the Lagrangian invariant (simply because
$\delta_\xi L=\6_\mu K^\mu (x,[\phi,\xi])$ and $[\delta_\xi,\6_\mu]=0$
imply $\delta_{\xi_1}\delta_{\xi_2}L=\delta_{\xi_1}
\6_\mu K^\mu (x,[\phi,\xi_2])=\6_\mu \delta_{\xi_1}K^\mu (x,[\phi,\xi_2])$)
and is a derivation (because the commutator of two derivations is
again a derivation). 
Owing to (\ref{basis1}) it can thus
be expressed through the operators $R^i_\Mdex$ of the
generating set and a trivial gauge symmetry. In particular
one has
\bea
\delta_{\xi_1}\phi^i=R^i_\Mdex\xi^\Mdex_1,\ 
\delta_{\xi_2}\phi^i=R^i_\Mdex\xi^{\Mdex}_2\ \then\ 
[\delta_{\xi_1},\delta_{\xi_2}]\phi^i=
R^i_\Mdex f^\Mdex(x,[\phi,\xi_1,\xi_2])+
\delta^\mathrm{triv}\phi^i.
\label{basis3}
\eea
Notice that $R^i_\Mdex f^\Mdex$ is a gauge transformation
$\delta_f$ as $\delta_{\xi_1}\phi^i$ and $\delta_{\xi_2}\phi^i$
but with ``composite'' (possibly
field dependent) parameter $f^\Mdex(x,[\phi,\xi_1,\xi_2])$.
Owing to (\ref{superformel}), 
the commutator of a trivial gauge symmetry
and any other gauge symmetry (trivial or non-trivial)
vanishes on-shell,
\bea
[\delta^\mathrm{triv},\delta_{\xi}]\phi^i\approx 0\quad\forall \phi^i.
\label{basis4}
\eea
As already remarked at the end of section \ref{trivg}
this implies under fairly
general assumptions that
this commutator is again a trivial gauge symmetry,
\bea
[\delta^\mathrm{triv},\delta_{\xi}]=\4\delta^\mathrm{triv}.
\label{basis5}
\eea
Hence the only possibly nontrivial part of the commutator algebra
of gauge symmetries is made up of the terms 
$R^i_\Mdex f^\Mdex(x,[\phi,\xi_1,\xi_2])$ 
in the commutators of two nontrivial gauge symmetries as in
(\ref{basis3}).
If these commutators involve a 
nonvanishing $\delta^\mathrm{triv}$
on the right hand side, the commutator
algebra is called an ``open gauge algebra''.
Notice that it may depend on the choice of the
generating set whether or not the algebra is open.

\mysection{D=4, N=1 pure SUGRA}\label{`iii'}\label{pure}

This section presents the Lagrangian and gauge transformations
of the simplest four-dimensional SUGRA theory \cite{FNF,DZ} 
($N=1$ SUGRA without
matter multiplets) in the basic formulation with open gauge algebra,
using the Weyl-spinor notation as
in appendix \ref{spin}.

\subsection{Lagrangian}\label{lagrangian}

\paragraph{Vielbein formulation.}
Owing to the presence of spinor fields, SUGRA theories
are constructed in the vielbein formulation (Cartan formulation)
of general relativity. In $D$ dimensions the vielbein
is a real $D\times D$-matrix field denoted by $e^a_\mu$
and related to the
spacetime metric $g_{\mu\nu}$ according to
\bea
g_{\mu\nu}=e^a_\mu e^b_\nu \eta_{ab}
\label{metric}\eea
where $\eta_{ab}$ is the Minkowski metric.
In order that the metric be invertible, the vielbein
must be invertible. We denote its inverse by
$E^\mu_a$,
\bea
e_\mu^a E^\mu_b=\delta^a_b\ ,\quad
e_\mu^a E^\nu_a=\delta^\nu_\mu\ .
\eea
In contrast to the standard (metric) formulation
of general relativity, the metric is thus not treated
as an elementary field but constructed from the vielbein
according to (\ref{metric}).
Conversely,
given a metric (with the same signature as the Minkowski metric), 
one can always construct a vielbein satisfying
(\ref{metric}): as the metric is symmetric, it can
at each point be diagonalized by some orthogonal matrix $O$ and
one may choose the vielbein as $DO$ where $D=
\mathrm{diag}(|r_1|^{1/2},\dots,|r_D|^{1/2})$
is a diagonal matrix and the $r$'s are the 
eigenvalues of the metric. Of course,
both $O$ and the $r$'s in general depend on the
point, i.e., they are fields, and so is the vielbein.
Actually this choice of the vielbein is not unique because
(\ref{metric}) determines the vielbein
only modulo arbitrary local Lorentz transformations as these
leave the Minkowski metric invariant. Hence GR
has in the vielbein formulation
more gauge symmetries than in the metric formulation
because it is also invariant under local Lorentz transformations
in addition to the diffeomorphism invariance.

\paragraph{SUGRA Lagrangian in first order formulation.}
In four dimensions the vielbein is called vierbein.
The gravitino is denoted by $\psi_\mu^\alpha$ where
$\alpha$ are Weyl spinor indices, see appendix \ref{spin}. 
Hence, for each value
of $\mu$, $\psi_\mu^\alpha$ is a complex 2-component
Weyl spinor field. Its complex conjugate is denoted
by $\5\psi_\mu^\da$. Our index notation is thus:
Greek indices from the beginning of the alphabet
denote Weyl spinor indices, Greek indices from the
middle of the alphabet denote world indices and
lower case Latin indices from the beginning of the alphabet denote
Lorentz vector indices. The spinor indices
and the Lorentz vector indices indicate the transformation
properties under local Lorentz transformations, the 
world indices the transformation
properties under spacetime diffeomorphisms.

In addition to the vielbein and the gravitino
one may introduce the so-called spin
connection $\omega_\mu{}^{ab}=-\omega_\mu{}^{ba}$ 
as an independent field. It serves as
the gauge field for the local Lorentz transformations.
However, it is only an auxiliary field, i.e.,  it can be
eliminated by solving algebraically its EOM.
The formulation with the spin connection as
an auxiliary field is called
first order formulation, the one which uses from the
very beginning only the vielbein and the gravitino is
called second order formulation. We shall first
introduce the first order formulation and then
focus on the second order formulation.

In the first order formulation, the Lagrangian is a
function of the vielbein, gravitino,
spin connection and their derivatives, given by
\bea
\FRAME{\ 
L=\sfrac 12 e E_b^\mu E_a^\nu R_{\mu\nu}{}^{ab}+
2(\nabla_\mu\psi_{\nu}\sigma_\rho\5\psi_\sigma+
\psi_\sigma\sigma_\rho\nabla_\mu\5\psi_{\nu})
\ep^{\mu\nu\rho\sigma}
\ }
\label{action1}
\eea
where:
\beann
e&=&\det(e_\mu^a),
\\
R_{\mu\nu}{}^{ab}&=&\6_\mu\omega_\nu{}^{ab}-
   \6_\nu\omega_\mu{}^{ab}+
   \omega_\mu{}^{ac}\omega_{\nu c}{}^{b}-
   \omega_\nu{}^{ac}\omega_{\mu c}{}^{b}
   \quad
   (\mbox{field strength of $\omega_\mu{}^{ab}$}),
\\
\nabla_\mu\psi_\nu^\alpha
   &=&\6_\mu\psi_\nu^\alpha-\sfrac 12\omega_\mu{}^{ab}
   (\psi_\nu\sigma_{ab})^\alpha
   \quad(\mbox{Lorentz-covariant derivative of $\psi_\mu$}),
\\
\nabla_\mu\5\psi_\nu^\da
   &=&\6_\mu\5\psi_\nu^\da+\sfrac 12\omega_\mu{}^{ab}
   (\5\sigma_{ab}\5\psi_\nu)^\da
   \quad(\mbox{Lorentz-covariant derivative of $\5\psi_\mu$}),
\\
\sigma_{\rho\alpha\da}&=&e_\rho^a\,\sigma_{a\alpha\da}
   \quad(\mbox{field dependent!}),
\\
\ep^{\mu\nu\rho\sigma}&=& e\underbrace{E^\mu_a E^\nu_b E^\rho_c E^\sigma_d
   \ep^{abcd}}_{\propto\det(E^\mu_a)=1/e}\in\{0,1,-1\}
   \quad(\mbox{field independent!}).
\eeann
\newpage

\noindent
{\bf Remarks:}
\bit
\item
The Lorentz-covariant derivative $\nabla_\mu$
is built in the standard manner (cf. electrodynamics, YM theory,
GR, \dots) by means of the 
gauge field $\omega_\mu{}^{ab}$.
It is defined not only on
spinor fields, but also on any other Lorentz-covariant fields by
\[
\nabla_\mu=\6_\mu-\sfrac 12\omega_\mu{}^{ab}l_{ab}\quad
(\mbox{conventional factor $1/2$ because of $l_{ab}=-l_{ba}$}).
\]
\item $E_b^\mu E_a^\nu R_{\mu\nu}{}^{ab}$
is a spacetime curvature scalar built from
$R_{\mu\nu}{}^{ab}$,
\[
R=E_b^\mu E_a^\nu R_{\mu\nu}{}^{ab}
    \quad(\mbox{``curvature scalar''}).
\]
\item Because of the antisymmetry of $\ep^{\mu\nu\rho\sigma}$, the
derivatives of $\psi_\mu$ and $\5\psi_\mu$ occur
only through the combinations
\beann
   \nabla_\mu\psi_\nu-\nabla_\nu\psi_\mu\quad
   (\mbox{``field strength of $\psi_\mu$''}),
\\
   \nabla_\mu\5\psi_\nu-\nabla_\nu\5\psi_\mu
   \quad(\mbox{``field strength of $\5\psi_\mu$''}).
\eeann
\item 
In terms of Majorana-spinors $\Psi_\mu$ (see appendix \ref{spin}), 
one has
\beann
(\nabla_\mu\psi_{\nu}\sigma_\rho\5\psi_\sigma+
\psi_\sigma\sigma_\rho\nabla_\mu\5\psi_{\nu})
\ep^{\mu\nu\rho\sigma}&=&
-\overline{\Psi}_{\mu}\7\gamma\gamma_\nu\nabla_\rho\Psi_{\sigma}
\ep^{\mu\nu\rho\sigma}\\
&=&\Ii\,e\,\overline{\Psi}_{\mu}\gamma^{[\mu}\gamma^\nu\gamma^{\rho]}
\nabla_\nu\Psi_{\rho},\quad
\Psi_{\mu}={\psi_{\mu\alpha}\choose \5\psi_\mu^\da}.
\eeann
\item Notice that $e=\pm\sqrt{-\det(g_{\mu\nu})}$ because of 
(\ref{metric}).
\item The definition of $\sigma_\mu$ illustrates the
general rule how one converts Lorentz-indices into
world-indices  and vice versa by means of the vierbein
and its inverse:
\[
X_\mu=e_\mu^a X_a,\quad X_a=E_a^\mu X_\mu, 
\quad X_{\mu\nu}=e_\mu^a e_\nu^b X_{ab}
\quad etc.
\]
\eit

\paragraph{Determination of 
{\boldmath ${\omega_\mu}^{ab}$} from its
EOM and second order formulation.} Varying $\omega_\mu{}^{ab}$
in the Lagrangian (\ref{action1}) yields
\beann
&&L([e,\psi,\omega+\delta\omega])-L([e,\psi,\omega])=
\frac e2 E_b^\mu E_a^\nu (\nabla_\mu\delta\omega_\nu{}^{ab}
-\nabla_\nu\delta\omega_\mu{}^{ab})
\\
&&\phantom{L([e,\psi,\omega+\delta\omega])-L([e,\psi,\omega])=}
-\delta\omega_\mu{}^{ab}\ep^{\mu\nu\rho\sigma}
\psi_\nu
\underbrace{(\sigma_{ab}\sigma_\rho+\sigma_\rho\5\sigma_{ab})}_
{=\Ii\ep_{ab\rho c}\sigma^c\atop
=\Ii e E_a^\lambda E_b^\kappa E_c^\tau
\ep_{\lambda\kappa\rho\tau}\sigma^c}
\5\psi_\sigma\\
&\then&\frac{\7\6 L([e,\psi,\omega])}{\7\6\omega_\mu{}^{ab}}
=\sfrac12\nabla_\nu(eE_b^\mu E_a^\nu -eE_a^\mu E_b^\nu )
+6e\Ii E_{[a}^\mu E_b^\nu E_{c]}^\sigma \psi_\nu\sigma^c\5\psi_\sigma
\\
&&\phantom{\frac{\7\6 L(e,\psi,\omega)}{\7\6\omega_\mu{}^{ab}}}
=\sfrac12\6_\nu(eE_b^\mu E_a^\nu -eE_a^\mu E_b^\nu )
+e\omega_{[ab]}{}^\mu-e\omega_c{}^c{}_{[a}E_{b]}^\mu
+6e\Ii E_{[a}^\mu E_b^\nu E_{c]}^\sigma \psi_\nu\sigma^c\5\psi_\sigma
\eeann
where $[\dots]$ denotes complete antisymmetrization with ``weight one'',
and the above rules for conversion of world and Lorentz indices were used,
e.g.: 
\[
\omega_{[ab]}{}^\mu=\sfrac12 (\omega_{ab}{}^\mu-\omega_{ba}{}^\mu),
\quad 
\omega_{ab}{}^\mu=E_a^\nu\omega_\nu{}^{cd} \eta_{bc}E^\mu_d.
\]
The EOM for the $\omega_\mu{}^{ab}$
are $\7\6 L/\7\6\omega_\mu{}^{ab}=0$.
They can be solved algebraically for the $\omega_\mu{}^{ab}$
(the $\omega_\mu{}^{ab}$ appear only linearly and undifferentiated in 
$\7\6 L/\7\6\omega_\mu{}^{ab}$). To do so, one may first
determine $\omega_{ab}{}^a$ by contracting
the equation $\7\6 L/\7\6\omega_\mu{}^{ab}=0$ with
$e_\mu^a$, then insert the result into $\7\6 L/\7\6\omega_\mu{}^{ab}=0$
and solve the latter for $\omega_{[ab]}{}^\mu$
(hint: use the identity
$\6_\mu e=eE^\nu_a\6_\mu e_\nu^a$). 
The result is,
written in convenient form:
\bea
\FRAME{\ 
\omega_{[\mu\nu]}{}^a=\6^{}_{[\mu} e_{\nu]}^a
-2\Ii\psi_{[\mu}\sigma^a\5\psi_{\nu]}.
\ }\label{omega1}
\eea
This yields $\omega_\mu{}^{ab}$ because 
$\omega_{\mu\nu\rho}=-\omega_{\mu\rho\nu}$ implies
\beann
\omega_{\mu\nu\rho}=\omega_{[\mu\nu]\rho}-\omega_{[\nu\rho]\mu}
+\omega_{[\rho\mu]\nu}
\ \then\ 
\omega_\mu{}^{ab}=\omega_{[\nu\rho]}{}^c(
2\delta^\nu_\mu E^{\rho[a}_{}\delta^{b]}_c-e_{\mu c}E^{\nu a}E^{\rho b}).
\eeann
Using (\ref{omega1}), we obtain:
\begin{equation}
\omega_\mu{}^{ab}=E^{\nu a}\6^{}_{[\mu} e_{\nu]}^{b}
-E^{\nu b}\6^{}_{[\mu} e_{\nu]}^{a}
-e_{\mu c}E^{\nu a}E^{\rho b}\6^{}_{[\nu} e_{\rho]}^c
+2\Ii(\psi_\mu\sigma^{[a}\5\psi^{b]}+\psi^{[a}\sigma^{b]}\5\psi_\mu
+\psi^{[a}\sigma_\mu\5\psi^{b]}).
\label{omega2}\end{equation}
The Lagrangian in the second order formulation is given by
(\ref{action1}) with $\omega_\mu{}^{ab}$ as in (\ref{omega2}).

\subsection{EOM}

From now on we shall always work in the second order formulation,
i.e., with $\omega_\mu{}^{ab}$ as in (\ref{omega2}).
The Euler-Lagrange derivatives of the second order Lagrangian
(\ref{action1}) 
with respect to the vierbein and gravitino
are (one may apply the ``1.5 order formalism'' here, see appendix \ref{ver1}):
\bea
&&\frac{\7\6 L}{\7\6 e_\mu^a}=
e(\sfrac 12 E_a^\mu R-R_{\rho\nu}{}^{bc}E^\rho_a E^\nu_b E^\mu_c)
+2\ep^{\mu\nu\rho\sigma}(\nabla_\nu\psi_{\rho}\sigma_a\5\psi_\sigma+
\psi_\sigma\sigma_a\nabla_\nu\5\psi_{\rho}),
\label{EOMe}\\
&&\frac{\7\6 L}{\7\6 \psi_\mu^\alpha}=
-4\ep^{\mu\nu\rho\sigma}(\sigma_\nu\nabla_\rho\5\psi_\sigma)_\alpha,
\quad\frac{\7\6 L}{\7\6 \5\psi_\mu^\da}=
4\ep^{\mu\nu\rho\sigma}(\nabla_\rho\psi_\sigma\sigma_\nu)_\da.
\label{EOMpsi}
\eea
The EOM are thus obtained by
setting the Euler-Lagrange derivatives in
(\ref{EOMe}) and (\ref{EOMpsi}) to zero.
In particular (\ref{EOMe}) yields
Einstein's field equations with a stress-energy tensor containing
the gravitino and its derivatives.
Notice that $R_{\mu\nu}{}^{ab}$ contains gravitino dependent terms
via the gravitino dependence of $\omega_\mu{}^{ab}$.
Hence, in order to cast Einstein's field equations in the familiar form,
one not only has to divide by $e$ and convert the Lorentz index $a$
into a world index by means of the vierbein, but in addition
one has to separate the
gravitino dependent terms contained in $R_{\mu\nu}{}^{ab}$
from those terms which depend only on the vierbein (the latter give
rise to the standard Einstein tensor on the
``left hand side'' of Einstein's field equations).

\subsection{Gauge symmetries}\label{gaugetrafos}

The nontrivial gauge symmetries of the SUGRA action (\ref{action1}) 
may be grouped into three types:
\ben
\item 
Invariance under
spacetime diffeomorphisms with four real gauge parameters $\xi^\mu$:
\bea
\delta_\mathrm{diffeo}e_\mu^a&=&\xi^\nu\6_\nu e_\mu^a+\6_\mu\xi^\nu e_\nu^a,
\label{diffe}\\
\delta_\mathrm{diffeo}\psi_\mu&=&
\xi^\nu\6_\nu \psi_\mu+\6_\mu\xi^\nu \psi_\nu,
\label{diffpsi}\\
\delta_\mathrm{diffeo}\5\psi_\mu&=&
\xi^\nu\6_\nu \5\psi_\mu+\6_\mu\xi^\nu \5\psi_\nu.
\label{diff5psi}
\eea
The invariance under these transformations can be deduced from the
fact that the Lagrangian is by construction a scalar density
with weight one under spacetime diffeomorphisms, as is familiar
from standard GR (the induced transformation of the
spin connection (\ref{omega2}) is $\delta_\mathrm{diffeo}\omega_\mu{}^{ab}=
\xi^\nu\6_\nu \omega_\mu{}^{ab}+\6_\mu\xi^\nu \omega_\nu{}^{ab}$).

\item Invariance under local
Lorentz transformations with six real gauge 
parameters $\xi^{ab}=-\xi^{ba}$:
\bea
\delta_\mathrm{Lorentz}e_\mu^a&=&\xi_b{}^a e_\mu^b,
\label{Lore}\\
\delta_\mathrm{Lorentz}\psi^\alpha_\mu&=&
\sfrac 12 \xi^{ab}(\psi_\mu\sigma_{ab})^\alpha,
\label{Lorpsi}\\
\delta_\mathrm{Lorentz}\5\psi^\da_\mu&=&
-\sfrac 12 \xi^{ab}(\5\sigma_{ab}\5\psi_\mu)^\da.
\label{Lor5psi}\eea
The Lagrangian is invariant under local Lorentz transformations because
it is composed of Lorentz-covariant objects
whose Lorentz-vector and
spinor indices are ``correctly contracted''
(the induced transformation of the spin connection 
is $\delta_\mathrm{Lorentz}\omega_\mu{}^{ab}=
\nabla_\mu\xi^{ab}=\6_\mu\xi^{ab}-\omega_{\mu c}{}^a\xi^{cb}
-\omega_{\mu c}{}^b\xi^{ac}$, i.e., $\omega_\mu{}^{ab}$ transforms
indeed as a gauge field for Lorentz transformations;
$R_{\mu\nu}{}^{ab}$ is Lorentz-covariant
because it is the field strength of $\omega_\mu{}^{ab}$,
and $\nabla_\mu$ is the Lorentz-covariant derivative).

\item Local SUSY with gauge parameters $\xi^\alpha$ that are
complex Weyl spinors (and thus make up four real gauge parameters):
\bea
\delta_\mathrm{susy}e_\mu^a&=&
2\Ii\xi\sigma^a\5\psi_\mu-2\Ii\psi_\mu\sigma^a\5\xi,
\label{susye}\\
\delta_\mathrm{susy}\psi^\alpha_\mu&=&
\nabla_\mu \xi^\alpha=\6_\mu \xi^\alpha
-\sfrac 12\omega_\mu{}^{ab}(\xi\sigma_{ab})^\alpha,
\label{susypsi}\\
\delta_\mathrm{susy}\5\psi^\da_\mu&=&
\nabla_\mu \5\xi^\da=\6_\mu \5\xi^\da
+\sfrac 12\omega_\mu{}^{ab}(\5\sigma_{ab}\5\xi)^\da.
\label{susy5psi}\eea
The invariance under these transformations is explicitly
demonstrated in appendix \ref{ver1} using the ``1.5 order formalism''.
\een

\subsection{Algebra of gauge transformations}\label{Algebra}

Let us first compute the
commutator of two SUSY transformations on the vierbein.
We shall use the notation $\delta_\mathrm{susy}(\xi)$
meaning a SUSY transformation with parameters $\xi^\alpha$,
and analogous notation for diffeomorphism and local Lorentz transformations.
\bea
[\delta_\mathrm{susy}(\xi_1),\delta_\mathrm{susy}(\xi_2)]e_\mu^a
&=&
\delta_\mathrm{susy}(\xi_1)
(2\Ii\xi_2\sigma^a\5\psi_\mu-2\Ii\psi_\mu\sigma^a\5\xi_2)
-(1\lra 2)
\nonumber\\
&=&
2\Ii\xi_2\sigma^a\nabla_\mu\5\xi_1-2\Ii\nabla_\mu\xi_1\sigma^a\5\xi_2
-2\Ii\xi_1\sigma^a\nabla_\mu\5\xi_2+2\Ii\nabla_\mu\xi_2\sigma^a\5\xi_1
\nonumber\\
&=&
\nabla_\mu(2\Ii\xi_2\sigma^a\5\xi_1-2\Ii\xi_1\sigma^a\5\xi_2).
\label{alg1}\eea
Notice that this expression does not contain derivatives of the gravitino
and at most first order derivatives of the vierbein. Hence, in the
second order formulation it cannot
contain a trivial gauge transformation
discussed in section \ref{trivg} because the Euler-Lagrange
derivatives (\ref{EOMe}) and (\ref{EOMpsi}) contain second order
derivatives of the vierbein and first order derivatives of the
gravitino, respectively. Therefore the commutator
(\ref{alg1}) should be a combination of the
gauge transformations (\ref{diffe}), (\ref{Lore}) and (\ref{susye})
with composite parameters depending on the fields and
the gauge parameters $\xi_1^\alpha$, $\xi_2^\alpha$ (and their derivatives).
To verify that this is indeed the case, we examine a 
general gauge transformation
of the vierbein (diffeomorphism $+$ local Lorentz $+$ SUSY transformation):
\beann
\lefteqn{
\delta_\mathrm{gauge}\, e_\mu^a=
\xi^\nu\6_\nu e_\mu^a+\6_\mu\xi^\nu e_\nu^a
+\xi_b{}^a e_\mu^b
+2\Ii\xi\sigma^a\5\psi_\mu-2\Ii\psi_\mu\sigma^a\5\xi
}
\\
&=&
\xi^\nu\hspace{-.3cm}\underbrace{(\6_\nu e_\mu^a-\6_\mu e_\nu^a)}_{\stackrel{
(\ref{omega1})}{=}\omega_{\nu\mu}{}^a-\omega_{\mu\nu}{}^a\atop
+2\Ii(\psi_{\nu}\sigma^a\5\psi_{\mu}-\psi_{\mu}\sigma^a\5\psi_{\nu})} 
\hspace{-.15cm}+\hspace{.1cm}\underbrace{\xi^\nu\6_\mu e_\nu^a+\6_\mu\xi^\nu e_\nu^a}_{\6_\mu
(\xi^\nu e_\nu^a)\atop =\nabla_\mu(\xi^\nu e_\nu^a)
+\omega_{\mu b}{}^ae_\nu^b\xi^\nu}
+\ \xi_b{}^a e_\mu^b
+2\Ii(\xi\sigma^a\5\psi_\mu-\psi_\mu\sigma^a\5\xi)
\\
&=&
\nabla_\mu(\xi^\nu e_\nu^a)+(\xi_b{}^a+\xi^\nu\omega_{\nu b}{}^a) e_\mu^b
+2\Ii(\xi+\xi^\nu\psi_\nu)\sigma^a\5\psi_\mu
-2\Ii\psi_\mu\sigma^a(\5\xi+\xi^\nu\5\psi_\nu).
\eeann
Hence we have
\bea
\FRAME{\ 
\delta_\mathrm{gauge}\, e_\mu^a=
\nabla_\mu\7\xi^a+\7\xi_b{}^a e_\mu^b
+2\Ii\7\xi\sigma^a\5\psi_\mu-2\Ii\psi_\mu\sigma^a\,\5{\!\7\xi}
\ }
\label{alg2}
\eea
where
\bea
\FRAME{\ 
\7\xi^a=\xi^\nu e_\nu^a,\quad \7\xi^{ab}=\xi^{ab}+\xi^\nu\omega_\nu{}^{ab},
\quad \7\xi^\alpha=\xi^\alpha+\xi^\nu\psi_\nu^\alpha,\quad
\5{\!\!\7\xi}\,{}^\da=\5\xi^\da+\xi^\nu\5\psi_\nu^\da.
\ }
\label{newparas}
\eea
Consider now gauge transformations with $\7\xi^{ab}=0$
and $\7\xi^\alpha=0$. These are combinations of
diffeomorphism transformations
of $e_\mu^a$
with parameters $\xi^\nu$, Lorentz transformations of $e_\mu^a$ with
composite parameters $\xi^{ab}=-\xi^\nu\omega_\nu{}^{ab}$
(as this is equivalent to $\7\xi^{ab}=0$), and SUSY transformations
of $e_\mu^a$ with composite parameters $\xi^\alpha=-\xi^\nu\psi_\nu^\alpha$
($\LRA \7\xi^\alpha=0$). Since 
the right hand side of 
(\ref{alg2}) reduces for $\7\xi^{ab}=\7\xi^\alpha=0$
to $\nabla_\mu\7\xi^a$, one has thus:
\bea
\delta_\mathrm{diffeo}(\xi^\nu)\,e_\mu^a
+\delta_\mathrm{Lorentz}(-\xi^\nu\omega_\nu{}^{ab})\,e_\mu^a
+\delta_\mathrm{susy}(-\xi^\nu\psi_\nu^\alpha)\, e_\mu^a=\nabla_\mu\7\xi^a.
\label{alg3}\eea
Using this in (\ref{alg1}) we obtain that, on the vielbein,
$[\delta_\mathrm{susy}(\xi_1),\delta_\mathrm{susy}(\xi_2)]$
is the sum of a diffeomorphism transformation with parameters
$\xi^\nu_{1,2}=2\Ii(\xi_2\sigma^\nu\5\xi_1-\xi_1\sigma^\nu\5\xi_2)$,
a local Lorentz transformation with parameters
$-\xi^\nu_{1,2}\omega_\nu{}^{ab}$ and a SUSY transformation with parameters
$-\xi^\nu_{1,2}\psi_\nu^\alpha$. On the gravitino
this holds only on-shell as can be explicitly verified but
the computation is cumbersome because one must
compute $\delta_\mathrm{susy}\omega_\mu{}^{ab}$
with $\omega_\mu{}^{ab}$ given by (\ref{omega2}), and use
the EOM of the gravitino. 
We shall not perform this computation here because
its result can be obtained more elegantly from the
supercovariant tensor calculus to be discussed later. One obtains thus
\bea
\ba{c}
\FRAME{\ 
[\delta_\mathrm{susy}(\xi_1),\delta_\mathrm{susy}(\xi_2)]
=\delta_\mathrm{diffeo}(\xi_{1,2}^\nu)
+\delta_\mathrm{Lorentz}(-\xi_{1,2}^\nu\omega_\nu{}^{ab})
+\delta_\mathrm{susy}(-\xi_{1,2}^\nu\psi_\nu^\alpha)+\delta^\mathrm{triv}
\ }
\\[8pt]
\mbox{with}\quad
\FRAME{\ 
\xi^\nu_{1,2}=2\Ii(\xi_2\sigma^\nu\5\xi_1-\xi_1\sigma^\nu\5\xi_2),\ }
\ea
\label{alg5}
\eea
where $\delta^\mathrm{triv}$ is a trivial gauge transformation
as in section \ref{trivg} involving the Euler-Lagrange derivatives
(\ref{EOMpsi}).
The remaining part of the algebra is quite standard and can be easily derived:
\bea
&[\delta_\mathrm{diffeo}(\xi_1),\delta_\mathrm{diffeo}(\xi_2)]
=\delta_\mathrm{diffeo}(\xi_{1,2})&\mbox{with}\quad
\xi_{1,2}^\mu=\xi^\nu_2\6_\nu\xi^\mu_1-\xi^\nu_1\6_\nu\xi^\mu_2,
\label{alg6}
\\
&[\delta_\mathrm{Lorentz}(\xi_1),\delta_\mathrm{Lorentz}(\xi_2)]
=\delta_\mathrm{Lorentz}(\xi_{1,2})&\mbox{with}\quad
\xi_{1,2}^{ab}=\xi_1^{ac}\xi_{2c}{}^b-\xi_2^{ac}\xi_{1c}{}^b,
\label{alg7}
\\
&[\delta_\mathrm{diffeo}(\xi_1),\delta_\mathrm{Lorentz}(\xi_2)]
=\delta_\mathrm{Lorentz}(\xi_{1,2})&\mbox{with}\quad
\xi_{1,2}^{ab}=-\xi^\mu_1\6_\mu\xi^{ab}_2,
\label{alg8}
\\
&[\delta_\mathrm{diffeo}(\xi_1),\delta_\mathrm{susy}(\xi_2)]
=\delta_\mathrm{susy}(\xi_{1,2})&\mbox{with}\quad
\xi_{1,2}^\alpha=-\xi^\mu_1\6_\mu\xi^\alpha_2,
\label{alg9}
\\
&[\delta_\mathrm{Lorentz}(\xi_1),\delta_\mathrm{susy}(\xi_2)]
=\delta_\mathrm{susy}(\xi_{1,2})&\mbox{with}\quad
\xi_{1,2}^\alpha=-\sfrac 12\xi^{ab}_1(\xi_2\sigma_{ab})^\alpha.
\label{alg10}
\eea

Owing to the trivial gauge transformations in (\ref{alg5})
the algebra is open.
This is one difference as compared to simpler gauge theories
such as YM theory or standard GR. Another difference is that
the composite parameters of the gauge transformations which occur on the
right hand side of (\ref{alg5}) are field dependent, whereas
in YM theory or standard GR one has
$[\delta_{\xi_1},\delta_{\xi_2}]=\delta_{\xi_{1,2}}$ with
$\xi_{1,2}$ depending only on $\xi_1$, $\xi_2$ and their derivatives, 
as in (\ref{alg6})--(\ref{alg10}).

Remark: Note that the $\7\xi$ in
(\ref{newparas}) are related to the $\xi$
by gauge parameter redefinitions of the type
discussed already in section \ref{basis}, namely $\7\xi^N=K^N_M\xi^M$
with field dependent $K^N_M$. We are free to use the $\7\xi$
as gauge parameters instead of the $\xi$. 
As explained in section \ref{basis}, this  is equivalent
to changing the generating set of gauge transformations.
This alternative form of the gauge transformations arises naturally
within an approach to SUGRA based on a supercovariant
tensor calculus to be discussed in the following sections.
The gauge transformations
of the vierbein in terms of these parameters are given by
(\ref{alg2}), the corresponding transformations of the gravitino read:
\bea
&\delta_\mathrm{gauge}\,\psi_\mu^\alpha=
\xi^\nu\6_\nu \psi_\mu^\alpha+\6_\mu\xi^\nu \psi_\nu^\alpha
+\sfrac 12 (\7\xi^{ab}-\xi^\nu\omega_\nu{}^{ab})(\psi_\mu\sigma_{ab})^\alpha
+\nabla_\mu(\7\xi^\alpha-\xi^\nu\psi_\nu^\alpha)&
\nonumber\\
&\LRA\quad\FRAME{\ 
\delta_\mathrm{gauge} \psi_\mu^\alpha=
\7\xi^a E_a^\nu(\nabla_\nu \psi_\mu-\nabla_\mu \psi_\nu)^\alpha
+\sfrac 12 \7\xi^{ab}(\psi_\mu\sigma_{ab})^\alpha
+\nabla_\mu\7\xi^\alpha
\ }&
\label{alg11}
\eea

\mysection{Tensor calculus for standard gauge theories}
\label{`iv'}\label{tensor}

So far we discussed pure $D=4$, $N=1$ SUGRA with
field content made up only of the vierbein and gravitino fields.
In that basic formulation the gauge transformations 
form an open algebra in the terminology of section \ref{basis}.
There is an alternative formulation
\cite{offshell1,offshell2}, often called ``off-shell formulation'' 
because in that formulation the
commutator algebra of the gauge transformations closes off-shell.
This is made possible by the inclusion of additional fields which
do not carry physical degrees of freedom and
can be eliminated algebraically using their equations of motion
(analogously to the spin connection $\omega_\mu{}^{ab}$ in the
first order fomulation, see section \ref{lagrangian}).
Therefore they are called auxiliary fields.
Elimination of the auxiliary fields reproduces 
the ``on-shell formulation'' of pure $D=4$, $N=1$ SUGRA discussed
in section \ref{pure}. An off-shell formulation does not only
exist for pure $D=4$, $N=1$ SUGRA but also for
its coupling to standard ``matter multiplets'' which is of great
help for the construction of matter couplings to $D=4$, $N=1$ SUGRA.

These off-shell formulations
can be derived within a scheme that is not restricted to
$D=4$, $N=1$ SUGRA but extends to a more general class of
gauge theories. I refer to this class of gauge theories
as standard gauge theories because it is characterized
by properties familiar from YM theories or GR. The scheme itself
may be called
``tensor calculus for standard gauge theories'' and is
presented in this section\footnote{Actually the scheme can be
extended to rather general gauge theories and thus to
a general tensor calculus
\cite{ten,jetletter} but the explanation
of this extension is beyond the scope of this work.}. 
In section \ref{matter} we shall specify
how it can be used to derive the off-shell 
formulation of $D=4$, $N=1$ SUGRA.

\subsection{Basic input}\label{Input}

The tensor calculus centers round the notion of gauge covariance,
in particular gauge covariant quantities and operations, such as
tensor fields and covariant derivatives. 
Its structure resembles properties familiar from
YM theories and GR. However we shall introduce it in a somewhat unfamiliar
manner which starts off from formulae for
the gauge transformations
and the ``partial derivatives'' ($\6_\mu$) of tensor fields.
The formula for the gauge transformations characterizes
tensor fields through
a certain transformation law and is thus analogous
to the definition of tensor fields in GR through
transformation properties under general coordinate transformation, 
for instance.
The formula for the derivatives
of tensor fields is an
unusual but quite useful way to introduce gauge covariant derivatives.

We denote the gauge parameters by $\7\xi^\Mdex$. 
The hat on $\xi$ indicates that these parameters might correspond to
an unusal formulation
of the gauge transformations.
For instance, in pure SUGRA this formulation corresponds to the parameters
in equation (\ref{newparas})
rather than to those used in section \ref{gaugetrafos}.
At the end of section \ref{CONREQ} we shall cast
the gauge transformations in more standard form 
with ``unhatted'' parameters.
Tensor fields are now characterized as follows: a
tensor field $T$ is a local function
of the fields whose gauge transformations do not contain
derivatives of gauge parameters $\7\xi^\Mdex$ and thus
take the form $\delta_{\7\xi}T=\7\xi^\Mdex X_\Mdex$, for some
local functions $X_\Mdex$. Moreover we require that
these functions are themselves tensor fields and
that they can be written in terms of operators $\Delta_\Mdex$ 
(graded derivations, see below) according to $\Delta_\Mdex T=X_\Mdex$.
Basically, the latter just means that we can define $\Delta_\Mdex$ on $T$
through $\Delta_\Mdex T:=X_\Mdex$. Hence, tensor fields transform
in this setting according to
\bea
\FRAME{\
\delta_{\7\xi}T=\7\xi^\Mdex \Delta_\Mdex T.\ }
\label{T2}
\eea
This is the formula for the gauge transformations
of tensor fields announced above.
The formula for the derivatives of tensor fields takes
a similar form. In terms of the exterior derivative 
on the jet space, $d=dx^\mu\6_\mu$, it reads
\bea
\FRAME{\
d\,T=A^\Mdex\Delta_\Mdex T.
\ }
\label{T3a}\eea
This expresses the
exterior derivative of a tensor field as a
linear combination of the operations $\Delta_M$ with
coefficients that are 1-forms $A^\Mdex$ (because $d$ has form-degree 1).
In general these 1-forms will
not be tensor fields because $dT=dx^\mu\6_\mu T$ is a combination
of the derivatives of $T$ which are usually not tensor fields
(cf.\ GR or YM theories).
Rather we shall see that the $A^\Mdex$ should be interpreted as
``connections'' built of gauge fields $\A \Mdex\mu$ according to
\bea
A^\Mdex=dx^\mu \A \Mdex\mu.
\label{T3b}
\eea
(\ref{T3a}) will now be used to introduce gauge covariant derivatives.
To that end we assume that a subset of the gauge fields
$\A \Mdex\mu$ forms a field dependent invertible matrix
(in the SUGRA case this will be the vierbein). We denote that
subset by $\{V_\mu^a\}$, and the remaining gauge fields by
$\A {\7\Mdex}\mu$ where we have split the index set $\{\Mdex\}$
into subsets $\{a\}$ and $\{\7\Mdex\}$:
\bea
\{\Mdex\}=\{a,\7\Mdex\},\quad 
\{\A \Mdex\mu\}=\{V_\mu^a,\A {\7\Mdex}\mu\},\quad
a\in\{0,\dots,D-1\}.
\label{split}
\eea
Equation (\ref{T3a}) can now be interpreted as a definition of the
operators $\Delta_a$:
\bea
\Delta_a T= (V^{-1})_a^\mu\,(\6_\mu-\A {\7\Mdex}\mu\Delta_{\7\Mdex})\,T.
\label{covderiv}
\eea 
Notice that $\Delta_a$ has a form analogous to covariant derivatives
in YM theory or GR. Therefore we interpret it as a gauge covariant
derivative. It is indeed gauge covariant
if $\Delta_\Mdex T$ is a tensor field for any $\Mdex$ and every 
tensor field $T$, as
we have assumed. Let us elaborate in some more detail on this assumption.
It demands that
the $\Delta$'s are graded derivations in the space
of tensor fields, i.e., they
map tensor fields to tensor fields and
satisfy the Leibniz rule
\bea
\Delta_\Mdex (T_1T_2)=(\Delta_\Mdex T_1)T_2
+(-)^{|\Mdex|\,|T_1|}T_1(\Delta_\Mdex T_2),
\label{T3}
\eea
where $|\Mdex|$ denotes the Grassmann parity of the 
gauge parameter $\7\xi^\Mdex$.
(\ref{T3}) must hold because the gauge transformations are to be
Grassmann even derivations, cf.\ (\ref{gdef2}), and shows that
$\Delta_\Mdex$ has the same Grassmann parity as
the corresponding gauge parameter; moreover $\Delta_a$
should have even Grassmann parity (the same as $\6_\mu$),
\bea
|\Delta_\Mdex|=|\7\xi^\Mdex|=|\Mdex|,\quad
|\Delta_a|=|\7\xi^a|=|a|=0.
\label{T4}
\eea
Owing to (\ref{T3a}) (and because $d$ is Grassmann odd, as it
contains the differentials $dx^\mu$),
this also fixes the Grassmann parities of the
gauge fields:
\bea
|A^\Mdex|=|\Mdex|+1\quad(\mbox{mod 2}),\quad |\A \Mdex\mu|=|\Mdex|.
\label{T3bb}
\eea
{\bf Remark:}
(\ref{T2}) and (\ref{T3a})
establish a formal similarity
of the gauge transformations and the derivatives of
tensor fields which
might be surprising at first glance. However, at a second glance
it makes quite some sense:
from a purely algebraic point of view (in particular in the jet space approach)
the gauge transformations and the derivatives
are actually quite similar and differ basically only in their 
commutation relations (the derivatives are required to commute among
themselves and with
the gauge transformations, whereas the latter in general do not 
necessarily commute
among themselves, see equations (\ref{comm1}) through (\ref{comm3}) below). 
Furthermore,
it may be worthwhile to compare with the fiber bundle formulation
of YM theories: there the gauge transformations and the partial
derivatives are also similar operations in the sense that 
the former correspond to 
displacements in the fiber, the latter to displacements
in the base manifold.

\subsection{Consistency requirements}\label{CONREQ}

We proceed by working out the consistency conditions which must be
satisfied in order that (\ref{T2}) and (\ref{T3a}) 
can provide an off-shell formulation of a gauge theory.
These consistency conditions
arise from the algebra of gauge transformations
and partial derivatives which is to read
\bea
&&[\delta_{\7\xi_1},\delta_{\7\xi_2}]=\delta_{f},\quad
f^{\Mdex}=f^{\Mdex}(x,[\7\xi_1,\7\xi_2,\phi]),
\label{comm1}\\[4pt]
&&[d,\delta_{\7\xi}]=0,
\label{comm2}\\[4pt]
&&d^2=0.
\label{comm3}
\eea
In (\ref{comm1}), $\delta_{f}$ is to be a gauge transformation
of the same form as $\delta_{\7\xi_1}$ and $\delta_{\7\xi_2}$,
but with ``composite parameters'' $f^\Mdex$, in order that the
commutator algebra of the gauge transformations closes
off-shell.
(\ref{comm2}) is equivalent to $[\6_\mu,\delta_{\7\xi}]=0$
and thus expresses (\ref{gdef2a}).
(\ref{comm3}) is equivalent to $[\6_\mu,\6_\nu]=0$ and
is included because (\ref{T3a}) is to be consistent
with these basic commutation relations of the derivatives.

We start with the commutator of
two gauge transformations on tensor fields.
Using (\ref{T2}) and that the $\Delta_\Mdex T$
are tensor fields, we obtain
\bea
[\delta_{\7\xi_1},\delta_{\7\xi_2}]\,T
&=&\delta_{\7\xi_1}(\7\xi_2^\Ndex\Delta_\Ndex T)-(1\lra 2)=
\7\xi_2^\Ndex\7\xi_1^\Mdex\Delta_\Mdex\Delta_\Ndex T-
\7\xi_1^\Ndex\7\xi_2^\Mdex\Delta_\Mdex\Delta_\Ndex T
\nonumber\\
&=&\7\xi_2^\Ndex\7\xi_1^\Mdex[\Delta_\Mdex,\Delta_\Ndex]\,T,
\label{T5}\eea
where $[\ ,\ ]$ is the graded commutator
\bea
{}[X,Y]=XY-(-)^{|X|\,|Y|}YX.
\label{T5a}\eea
On a tensor field,
the right hand side of (\ref{comm1}) must again be a gauge transformation
of the form (\ref{T2}) when we impose
off-shell closure of the gauge algebra,
i.e., it must be
a combination of the
$\Delta_\Mdex T$ with certain coefficient functions $f^{\Mdex}$.
Since the
gauge transformations of a tensor field do not involve
derivatives of the gauge parameters, these coefficient functions
do not involve derivatives of the $\7\xi$'s, cf.\ (\ref{T5}). 
Hence we require
\bea
{}[\delta_{\7\xi_1},\delta_{\7\xi_2}]\,T=
\7\xi_1^\Mdex\7\xi_2^\Ndex\CF \Ndex\Mdex\Pdex\Delta_\Pdex T,
\label{T6}
\eea
for some tensor fields $\CF \Mdex\Ndex\Pdex$ [that these must
be tensor fields is also seen by comparing with (\ref{T5}), since
$\Delta_\Mdex T$ is to be a tensor field whenever $T$ is].
As $[\delta_{\7\xi_1},\delta_{\7\xi_2}]$
is skew-symmetric under exchange of
$\7\xi_1$ and $\7\xi_2$, these tensor fields are subject to the
symmetry property
\bea
\CF \Mdex\Ndex\Pdex=-(-)^{|\Mdex|\,|\Ndex|}\CF \Ndex\Mdex\Pdex.
\label{T7}
\eea
Since (\ref{T5}) and (\ref{T6}) must coincide for all
gauge parameters and all tensor fields, we 
require that
the $\Delta$'s satisfy the graded commutator algebra%
\footnote{Note that (\ref{T8}) is a sufficient
condition for the compatibility of (\ref{T5}) and (\ref{T6}). In special
cases it might not be
a necessary condition. Analogously, equations (\ref{T11}) and (\ref{T12})
are only sufficient for consistency, but in general not necessary.}
\bea
\FRAME{\ 
[\Delta_\Mdex,\Delta_\Ndex]=
-\CF \Mdex\Ndex\Pdex\Delta_\Pdex\ .
\ }
\label{T8}
\eea
[The minus sign is due to 
$\7\xi_1^\Mdex\7\xi_2^\Ndex\CF \Ndex\Mdex\Pdex
=-\7\xi_2^\Ndex\7\xi_1^\Mdex\CF \Mdex\Ndex\Pdex$.]
This algebra implies consistency conditions for
the tensor fields $\CF \Mdex\Ndex\Pdex$ and their $\Delta$-transformations.
These follow from the following identity for graded commutators:
\bea
\csum {\Mdex\Ndex\Pdex}{5}
[\Delta_\Mdex,[\Delta_\Ndex,\Delta_\Pdex]]= 0
\label{T9}\eea
where the graded cyclic sum was used defined by
\bea
\csum {\Mdex\Ndex\Pdex}{5}X_{\Mdex\Ndex\Pdex}
=(-)^{|\Mdex|\,|\Pdex|}X_{\Mdex\Ndex\Pdex}+
(-)^{|\Ndex|\,|\Mdex|}X_{\Ndex\Pdex\Mdex}
+(-)^{|\Pdex|\,|\Ndex|}X_{\Pdex\Mdex\Ndex}\ .
\label{cyclicsum}
\eea
(\ref{T8}) and (\ref{T9}) yield
\bea
\FRAME{\ \displaystyle{
\csum {\Mdex\Ndex\Pdex}{5}
(\Delta_\Mdex\cF_{\Ndex\Pdex}{}^\Qdex
+\cF_{\Mdex\Ndex}{}^\Rdex\cF_{\Rdex\Pdex}{}^\Qdex)= 0.
}\ }
\label{T10}\eea
As we shall see, these equations are the crucial
consistency requirements.
 
Next we consider the commutators of the
exterior derivative and gauge transformations on tensor fields.
Using (\ref{T2}), (\ref{T3a}), (\ref{T8}) and that
the $\Delta_M T$ are tensor fields, we obtain
\bea
[d,\delta_{\7\xi}]\,T&=&
d(\7\xi^\Mdex\Delta_\Mdex T)-\delta_{\7\xi}(A^\Mdex\Delta_\Mdex T)
\nonumber\\
&=&(d\7\xi^\Mdex)\Delta_\Mdex T+(-)^{|\Mdex|}\7\xi^\Mdex d(\Delta_\Mdex T)
-(\delta_{\7\xi}A^\Mdex)\Delta_\Mdex T-A^\Mdex\delta_{\7\xi}(\Delta_\Mdex T)
\nonumber\\
&=&(d\7\xi^\Mdex)\Delta_\Mdex T
+(-)^{|\Mdex|}\7\xi^\Mdex A^\Ndex\Delta_\Ndex\Delta_\Mdex T
-(\delta_{\7\xi}A^\Mdex)\Delta_\Mdex T
-A^\Mdex\7\xi^\Ndex\Delta_\Ndex\Delta_\Mdex T
\nonumber\\
&=&(d\7\xi^\Mdex-\delta_{\7\xi}A^\Mdex)\Delta_\Mdex T
-A^\Mdex\7\xi^\Ndex[\Delta_\Ndex ,\Delta_\Mdex ]T
\nonumber\\
&=&
(d\7\xi^\Mdex-\delta_{\7\xi}A^\Mdex
+A^\Pdex\7\xi^\Ndex\CF \Ndex\Pdex\Mdex)\Delta_\Mdex T.
\label{T11'}
\eea
According to (\ref{comm2}), these
commutators must vanish for all $T$. Therefore we require that
the sum of the
terms in parantheses in the last line of (\ref{T11'}) vanishes
for each $\Mdex$. This fixes the
gauge transformations of the connections:
\bea
\FRAME{\ 
\delta_{\7\xi}A^\Mdex=d\7\xi^\Mdex+A^\Pdex\7\xi^\Ndex\CF \Ndex\Pdex\Mdex
\ }
\label{T11}
\eea
i.e., for the gauge fields:
\bea
\FRAME{\ 
\delta_{\7\xi}\A \Mdex\mu=\6_\mu\,\7\xi^\Mdex
+\A \Pdex\mu\7\xi^\Ndex\CF \Ndex\Pdex\Mdex.
\ }
\label{T11a}\eea

Last but not least we compute $d^2$ on tensor fields using
(\ref{T3a}). We obtain
\beann
d^2T&=&d(A^\Mdex\Delta_\Mdex T)
\\
&=&(dA^\Mdex)\Delta_\Mdex T+(-)^{|M|+1}A^\Mdex d(\Delta_\Mdex T)
\\
&=&(dA^\Mdex)\Delta_\Mdex T
+(-)^{|\Mdex|+1}A^\Mdex A^\Ndex\Delta_\Ndex\Delta_\Mdex T
\\
&=&(dA^\Mdex)\Delta_\Mdex T+\sfrac 12(-)^{|\Mdex|+1}(A^\Mdex A^\Ndex
+(-)^{(|\Mdex|+1)(|\Ndex|+1)}A^\Ndex A^\Mdex)\Delta_\Ndex\Delta_\Mdex T
\\
&=&(dA^\Mdex)\Delta_\Mdex T
+\sfrac 12(-)^{|\Mdex|+1}A^\Mdex A^\Ndex\Delta_\Ndex\Delta_\Mdex T
+\sfrac 12(-)^{(|\Mdex|+1)|\Ndex|}A^\Ndex A^\Mdex\Delta_\Ndex\Delta_\Mdex T
\\
&=&(dA^\Mdex)\Delta_\Mdex T
+\sfrac 12(-)^{|\Mdex|+1}A^\Mdex A^\Ndex\Delta_\Ndex\Delta_\Mdex T
+\sfrac 12(-)^{(|\Ndex|+1)|\Mdex|}A^\Mdex A^\Ndex\Delta_\Mdex\Delta_\Ndex T
\\
&=&(dA^\Mdex)\Delta_\Mdex T
+\sfrac 12(-)^{|\Mdex|+1}A^\Mdex A^\Ndex
(\Delta_\Ndex\Delta_\Mdex + (-)^{(|\Ndex|+1)|\Mdex|+|\Mdex|+1}
\Delta_\Mdex\Delta_\Ndex) T
\\
&=&(dA^\Mdex)\Delta_\Mdex T
+\sfrac 12(-)^{|\Mdex|+1}A^\Mdex A^\Ndex
(\Delta_\Ndex\Delta_\Mdex - (-)^{|\Ndex|\,|\Mdex|}
\Delta_\Mdex\Delta_\Ndex) T
\\
&=&(dA^\Mdex)\Delta_\Mdex T+\sfrac 12(-)^{|\Mdex|+1}A^\Mdex A^\Ndex
[\Delta_\Ndex,\Delta_\Mdex] T
\\
&=&(dA^\Mdex-\sfrac 12(-)^{|\Pdex|+1}A^\Pdex A^\Ndex
\CF \Ndex\Pdex\Mdex)\Delta_\Mdex T
\eeann
where we used
(\ref{T3a}),  (\ref{T3bb}), (\ref{T8}) and, again, that
the $\Delta_M T$ are tensor fields [note that
(\ref{T3bb}) implies $A^\Mdex A^\Ndex=
(-)^{(|\Mdex|+1)(|\Ndex|+1)}A^\Ndex A^\Mdex$]. As
$d^2T$ must vanish for all $T$ we require
\bea
\FRAME{\ 
dA^\Mdex+\sfrac 12(-)^{|\Pdex|}A^\Pdex A^\Ndex
\CF \Ndex\Pdex\Mdex=0.
\ }
\label{T12}
\eea
This equation looks at first glance like a differential equation
for $A^\Mdex$.%
\footnote{Notice also that it looks
formally like a Maurer-Cartan equation,
or a 
``zero-curvature condition''. Actually
it is indeed a 
zero-curvature condition, but just for the derivatives
as it expresses $[\6_\mu,\6_\nu]T=0$.}
However, actually it 
determines the curvatures of the covariant derivatives
[this is similar -- and related -- to the fact that
(\ref{T3a}) is no differential equation for tensor fields
but the definition of the covariant derivatives].
To see this we spell it out in components.
Using 
\[
dA^\Mdex=dx^\mu\6_\mu A^\Mdex=dx^\mu dx^\nu \6_\mu \A \Mdex\nu
=\sfrac 12 dx^\mu dx^\nu (\6_\mu \A \Mdex\nu-\6_\nu \A \Mdex\mu)
\]
and 
\[
A^\Pdex A^\Ndex=dx^\mu \A \Pdex\mu dx^\nu \A \Ndex\nu
=(-)^{|\Pdex|} dx^\mu dx^\nu\A \Pdex\mu \A \Ndex\nu
\]
we obtain from (\ref{T12})
\bea
\FRAME{\ 
\6_\mu \A \Mdex\nu-\6_\nu \A \Mdex\mu
+\A\Pdex\mu \A\Ndex\nu\CF \Ndex\Pdex\Mdex=0.
\ }
\label{T12a}
\eea
[One has $\A \Pdex\mu \A \Ndex\nu\CF \Ndex\Pdex\Mdex=
-\A \Pdex\nu \A \Ndex\mu\CF \Ndex\Pdex\Mdex$ owing to
(\ref{T3bb}) and (\ref{T7}).]
Now, $\A \Pdex\mu \A \Ndex\nu\CF \Ndex\Pdex\Mdex$
contains $V_\mu^a V_\nu^b\CF ba\Mdex$, cf.\ (\ref{split}).
We can thus write (\ref{T12a}) as
\bea
\FRAME{\ 
V_\mu^a V_\nu^b\,\CF ab\Mdex=
\6^{}_\mu \A \Mdex\nu-\6^{}_\nu \A \Mdex\mu
+\A {\hat\Pdex}\mu \A {\hat\Ndex}\nu\CF {\hat\Ndex}{\hat\Pdex}\Mdex
+V_\mu^a\A {\hat\Ndex}\nu\CF {\hat\Ndex}{a}\Mdex
-V_\nu^a\A {\hat\Ndex}\mu\CF {\hat\Ndex}{a}\Mdex
\ }
\label{T12b}
\eea
where we used $\CF {a}\Ndex\Mdex=-\CF \Ndex{a}\Mdex$
which follows from (\ref{T7}) owing to $|a|=0$, see (\ref{T4}).
As $V$ is assumed to be invertible, (\ref{T12b})
can be solved for $\CF cd\Mdex$ by contracting it
with $(V^{-1})^\mu_c$ and $(V^{-1})^\nu_d$.
Hence (\ref{T12})
can be viewed as an equation for the $\CF ab\Mdex$ which can
indeed be interpreted as curvatures or torsions for the 
covariant derivatives, as (\ref{T8}) reads
for $\Mdex=a$ and $\Ndex=b$:
\[
{}[\Delta_a,\Delta_b]\, T=-\CF ab\Mdex\Delta_\Mdex T .
\]

This ends the discussion of (\ref{comm1}) through (\ref{comm3})
on tensor fields. 
What about the gauge fields? It turns out that
(\ref{comm1}) through (\ref{comm3}) do automatically
hold also on the gauge fields as a consequence
of (\ref{T10}), with the same $f^\Mdex$ 
as in (\ref{T6}) (note that the latter is required
because the commutator algebra of the gauge transformations
must of course coincide on tensor field and gauge fields in an off-shell
formulation).
Indeed one obtains, using the formulae
derived so far:
\bea
[\delta_{\7\xi_1},\delta_{\7\xi_2}]\, A^\Mdex&=&
\delta_{\7\xi_1}(d\7\xi_2^\Mdex+A^\Pdex\7\xi_2^\Ndex\CF \Ndex\Pdex\Mdex)
-(1\lra 2)
\nonumber\\
&=&(\delta_{\7\xi_1}A^\Pdex)\7\xi_2^\Ndex\CF \Ndex\Pdex\Mdex
+A^\Pdex\7\xi_2^\Ndex(\delta_{\7\xi_1}\CF \Ndex\Pdex\Mdex)-(1\lra 2)
\nonumber\\
&=&(d\7\xi_1^\Pdex+A^\Qdex\7\xi_1^\Rdex\CF \Rdex\Qdex\Pdex)
\7\xi_2^\Ndex\CF \Ndex\Pdex\Mdex
+A^\Pdex\7\xi_2^\Ndex\7\xi_1^\Qdex\Delta_\Qdex\CF \Ndex\Pdex\Mdex-(1\lra 2)
\nonumber\\
&=&d(\7\xi_1^\Pdex\7\xi_2^\Ndex\CF \Ndex\Pdex\Mdex)
+A^\Pdex\7\xi_1^\Qdex\7\xi_2^\Rdex\CF \Rdex\Qdex\Pdex\CF \Ndex\Pdex\Mdex
\nonumber\\
&&+(-)^{|\Qdex|\,|\Pdex|}A^\Pdex\7\xi_2^\Ndex\7\xi_1^\Qdex
\csum {\Mdex\Ndex\Pdex}{5}
(\Delta_\Qdex\cF_{\Ndex\Pdex}{}^\Mdex
+\cF_{\Qdex\Ndex}{}^\Rdex\cF_{\Rdex\Pdex}{}^\Mdex),
\label{T13a}\\
{}[d,\delta_{\7\xi}]\,A^\Mdex&=&
d(d\7\xi^\Mdex+A^\Pdex\7\xi^\Ndex\CF \Ndex\Pdex\Mdex)
-\delta_{\7\xi}(-\sfrac 12(-)^{|\Pdex|}A^\Pdex A^\Ndex
\CF \Ndex\Pdex\Mdex)=\dots
\nonumber\\
&=&\sfrac 12(-)^{|\Pdex|(1+|\Qdex|)}A^\Pdex A^\Ndex\7\xi^\Qdex
\csum {\Mdex\Ndex\Pdex}{5}
(\Delta_\Qdex\cF_{\Ndex\Pdex}{}^\Mdex
+\cF_{\Qdex\Ndex}{}^\Rdex\cF_{\Rdex\Pdex}{}^\Mdex),
\label{T13b}\\
d^2A^\Mdex&=&d(-\sfrac 12(-)^{|\Pdex|}A^\Pdex A^\Ndex
\CF \Ndex\Pdex\Mdex)=\dots
\nonumber\\
&=&-\sfrac 16(-)^{|\Ndex|+|\Qdex|\,|\Pdex|}A^\Pdex A^\Ndex A^\Qdex
\csum {\Mdex\Ndex\Pdex}{5}
(\Delta_\Qdex\cF_{\Ndex\Pdex}{}^\Mdex
+\cF_{\Qdex\Ndex}{}^\Rdex\cF_{\Rdex\Pdex}{}^\Mdex).
\label{T13c}\eea
Hence (\ref{comm1}) through (\ref{comm3}) are indeed
satisfied on $A^M$ when (\ref{T10}) holds.
This
emphasizes the central importance of (\ref{T10}).
Furthermore, we can now specify (\ref{comm1}):
\bea
\FRAME{\ 
[\delta_{\7\xi_1},\delta_{\7\xi_2}]=\delta_{f}\ ,\quad
f^\Pdex=\7\xi_1^\Mdex\7\xi_2^\Ndex\CF \Ndex\Mdex\Pdex.
\ }
\label{T14}\eea

Let us finally rewrite the gauge transformations
in terms of parameters $\xi^\mu,\xi^{\7\Mdex}$ related
to the $\7\xi^\Mdex$ analogously to (\ref{newparas}):
\bea
\FRAME{\ 
\7\xi^a=\xi^\mu V_\mu^a,\quad 
\7\xi^{\7\Mdex}=\xi^{\7\Mdex}+\xi^\mu \A {\7\Mdex}\mu.
\ }
\label{new1}
\eea
For the gauge transformations of tensor fields we have
\beann
\delta_{\7\xi}T=\7\xi^\Mdex\Delta_\Mdex T=
\xi^\mu V_\mu^a\Delta_a T+\7\xi^{\7\Mdex}\Delta_{\7\Mdex}T
=\xi^\mu(\6_\mu-\A {\7\Mdex}\mu\Delta_{\7\Mdex})T+
\7\xi^{\7\Mdex}\Delta_{\7\Mdex}T
\eeann
where we used $V_\mu^a\Delta_a T=
(\6_\mu-\A {\7\Mdex}\mu\Delta_{\7\Mdex})T$ which is nothing
but a rewriting of (\ref{T3a}).
Hence the gauge transformations of tensor fields read
in terms of the $\xi$'s:
\bea
\FRAME{\ 
\delta_{\xi}T=\xi^\mu\6_\mu T+\xi^{\7\Mdex}\Delta_{\7\Mdex}T
\ }
\label{new2}
\eea
For the gauge transformations of the gauge fields $\A {\7\Mdex}\mu$
we obtain from (\ref{T11a}):
\beann
\delta_{\7\xi} \A {\7\Mdex}\mu&=&\6_\mu\,\7\xi^{\7\Mdex}
+\A \Pdex\mu\7\xi^{\Ndex}\CF {\Ndex}\Pdex{\7\Mdex}
\\
&=&\6_\mu(\xi^{\7\Mdex}+\xi^\nu \A {\7\Mdex}\nu)
+\A \Pdex\mu(\xi^{\7\Ndex}+\xi^\nu \A {\7\Ndex}\nu)\CF {\7\Ndex}\Pdex{\7\Mdex}
+\A \Pdex\mu\xi^\nu V_\nu^a\CF a\Pdex{\7\Mdex}
\\
&=&\6_\mu\xi^{\7\Mdex}+\6_\mu\xi^\nu \A {\7\Mdex}\nu
+\xi^\nu(\6_\mu\A {\7\Mdex}\nu-\6_\nu\A {\7\Mdex}\mu)
+\xi^\nu\6_\nu\A {\7\Mdex}\mu
\\
&&+\A \Pdex\mu(\xi^{\7\Ndex}
+\xi^\nu \A {\7\Ndex}\nu)\CF {\7\Ndex}\Pdex{\7\Mdex}
+\A \Pdex\mu\xi^\nu V_\nu^a\CF a\Pdex{\7\Mdex}.
\eeann
Using now equation (\ref{T12a}), i.e.,
$\6_\mu\A {\7\Mdex}\nu-\6_\nu\A {\7\Mdex}\mu=
-\A {\Pdex}\mu\A {\Ndex}\nu\CF {\Ndex}\Pdex{\7\Mdex}$,
we obtain
\bea
\FRAME{\ 
\delta_{\xi} \A {\7\Mdex}\mu=
\xi^\nu\6_\nu \A {\7\Mdex}\mu+\6_\mu\xi^\nu \A {\7\Mdex}\nu
+\6_\mu\,\xi^{\7\Mdex}
+\A \Pdex\mu\xi^{\7\Ndex}\CF {\7\Ndex}\Pdex{\7\Mdex}.
\ }
\label{new3}
\eea
An analogous computation for $V^a_\mu$ yields
\bea
\FRAME{\ 
\delta_{\xi} V^a_\mu=
\xi^\nu\6_\nu V^a_\mu+\6_\mu\xi^\nu V^a_\nu
+\A \Pdex\mu\xi^{\7\Ndex}\CF {\7\Ndex}\Pdex{a}.
\ }
\label{new4}
\eea
Notice that the right hand sides of equations
(\ref{new2}), (\ref{new3}) and (\ref{new4}) involve
$\xi^\mu$ only via the ``Lie derivative terms''
$\xi^\mu\6_\mu T$ and 
$\xi^\nu\6_\nu \A {\Mdex}\mu+\6_\mu\xi^\nu \A {\Mdex}\nu$,
respectively.
\medskip

{\bf Remark:}
Formally the formulae above look quite
familiar. For instance, (\ref{T11}) looks
formally like the gauge transformations of a gauge field in YM theory
if the $\CF \Ndex\Pdex\Mdex$ were the structure constants
of a Lie algebra. However, in general (and in particular in SUGRA)
the $\CF \Ndex\Pdex\Mdex$ are {\em not} constant but rather
they are tensor fields, and therefore the algebra (\ref{T8}) is
{\em not} a (graded) Lie algebra but a more general structure.
In fact, Lie algebras are just the simplest examples of this structure,
because in these examples the $\CF \Ndex\Pdex\Mdex$ are constants
and (\ref{T10}) turns into the Jacobi identity for the
structure constants of a Lie algebra. Hence (\ref{T10}) generalizes the
Jacobi identity for Lie algebras to the
more general algebras (\ref{T8}).

\mysection{Off-shell formulations of D=4,\,N=1\,SUGRA with matter}
\label{`v'}\label{matter}

\subsection{Supercovariant tensor calculus}
\label{tensorcalc}

We shall now outline how an off-shell formulation of
D=4, N=1 SUGRA and its coupling to matter is obtained 
within the scheme described in
section \ref{tensor}.
The gauge symmetries to be implemented
are in this case the spacetime diffeomorphisms,
local Lorentz symmetry, SUSY and YM gauge symmetry.
The corresponding ``hatted'' gauge parameters $\7\xi$ are
\bea
\{\7\xi^\Mdex\}=\{\7\xi^a,\7\xi^\alpha,\ \5{\!\!\7\xi}\,{}^\da,
\7\xi^{ab},\7\xi^i\}
\label{T1}
\eea
where the $\7\xi^i$ are the hatted Yang-Mills gauge parameters, i.e.,
the index $i$ refers to some basis
of the Lie algebra 
of a YM gauge group (for pure SUGRA, $\{\7\xi^i\}$
is simply the empty set). The other gauge parameters and indices 
have already been introduced in section \ref{pure}.
The gauge fields $\A \Mdex\mu$ are the vierbein $e_\mu^a$, the gravitino
$\psi_\mu^\alpha$ and its complex conjugate $\5\psi_\mu^\da$, 
the spin connection $\omega_\mu{}^{ab}$ 
and Yang-Mills gauge fields $\A i\mu$,
\bea
\{\A \Mdex\mu\}=\{e_\mu^a,\psi_\mu^\alpha,\5\psi_\mu^\da,
\omega_\mu{}^{ab},A_\mu^i\}.
\label{T3bbb}
\eea
The vierbein is in this case identified with the gauge fields
$V_\mu^a$ in (\ref{split}),
\bea
V_\mu^a\equiv e_\mu^a.
\eea
The $\Delta$-operations are denoted by
\bea
\{\Delta_\Mdex\}=\{\cD_a,\cD_\alpha,\5\cD_\da,l_{ab},\delta_i\}.
\label{T3c}
\eea
Concerning summations over the indices $\Mdex$, we employ
the following convention:
\bea
X^\Mdex Y_\Mdex\equiv X^a Y_a+X^\ua Y_\ua
+\sfrac 12 X^{ab} Y_{ab}+X^i Y_i,\quad
X^\ua Y_\ua=X^\alpha Y_\alpha+X_\da Y^\da.
\label{T3d}
\eea
For instance, (\ref{T3a}) reads thus explicitly in this case:
\beann
\6_\mu T=(e_\mu^a\cD_a+\psi_\mu^\alpha\cD_\alpha+\5\psi_{\mu\da}\5\cD^\da+
\sfrac 12\omega_\mu{}^{ab}l_{ab}+A_\mu^i\delta_i)\,T .
\eeann
The covariant derivatives (\ref{covderiv}) are thus given by
\bea
\cD_a\, T=E^\mu_a(\6_\mu-\psi_\mu^\alpha\cD_\alpha-\5\psi_{\mu\da}\5\cD^\da
-\sfrac 12\omega_\mu{}^{ab}l_{ab}-A_\mu^i\delta_i)\,T.
\label{T3f}
\eea
Notice that these covariant derivatives involve not only
the spin connection and Yang-Mills gauge fields, but in addition also 
the gravitino. They are thus covariant also with respect 
to local SUSY transformations.
To distinguish them from the 
more familiar covariant derivatives in standard GR,
we shall refer to them as supercovariant derivatives
and to the corresponding tensor fields as supercovariant tensor fields.
Notice also that $\cD_a$ does not
contain a connection $\Gamma_{\mu\nu}{}^\rho$
for world indices.
The reason is that all supercovariant tensor fields must be
scalar fields with regard to spacetime diffeomorphisms because
otherwise their gauge transformations would contain derivatives
of the diffeomorphism parameters,
in contradiction to the definition of
tensor fields according to (\ref{T2}). 
Hence, according to this definition, supercovariant tensor fields do
not carry world indices, and
therefore a term with $\Gamma_{\mu\nu}{}^\rho$
is not needed in $\cD_a$. For the same reason
the supercovariant derivatives themselves must be scalar operators with
regard to diffeomorphisms which explains why the carry a Lorentz
index instead of a world index.

\subsection{Bianchi identities}
\label{BIs}

$D=4$, $N=1$ SUGRA arises now by a suitable specification 
of the tensor fields
$\CF \Mdex\Ndex\Pdex$ occurring in (\ref{T8}). This has to
be done such that the consistency conditions
(\ref{T10}) are satisfied. To describe this specification,
we introduce 
the index sets $\{A\}=\{a,\alpha,\da\}$ and $\{I\}=\{[ab],i\}$ so that
(\ref{T3c}) becomes
\bea
\{\Delta_\Mdex\}=\{\cD_A,\delta_I\},\quad
\{\cD_A\}=\{\cD_a,\cD_\alpha,\5\cD_\da\},\quad
\{\delta_I\}=\{l_{ab},\delta_i\}.
\label{T15}
\eea
The graded commutator algebra (\ref{T8}) for 
an off-shell formulation of $D=4$, $N=1$ SUGRA reads
\bea
[\cD_A,\cD_B]&=&-\T ABC\cD_C-\F ABI\delta_I\ ,
\label{DD}\\
{}[\delta_I,\cD_A]&=&-g_{IA}{}^B\cD_B\ ,
\label{Dd}\\
{}[\delta_I,\delta_J]&=&\f IJK \delta_K\ .
\label{dd}
\eea
Note that this is not the most general form
that the algebra of the $\cD_A$ and $\delta_I$ could have
because the right hand side of (\ref{Dd}) contains no term
with a $\delta_I$ while the right hand side of (\ref{dd})
contains no term with a $\cD_A$. Furthermore we impose that
the $\f IJK$ and $g_{IA}{}^B$ are constants (whereas
the $\T ABC$ and $\F ABI$ are in general field dependent),
\[
\f IJK=\mathit{constant},\quad g_{IA}{}^B=\mathit{constant}.
\]
The conditions (\ref{T10}) read then for the
various index pictures $_{\Mdex\Ndex\Pdex}{}^\Qdex$:
\bea
_{IJK}{}^L:&& \f IJM\f MKL+\f JKM\f MIL+\f KIM\f MJL=0,
\label{C1}\\
_{IJK}{}^A:&& 0=0,
\label{C2}\\
_{IJA}{}^K:&& 0=0,
\label{C3}\\
_{IJA}{}^B:&& g_{IA}{}^C g_{JC}{}^B-g_{JA}{}^C g_{IC}{}^B=\f IJK g_{KA}{}^B,
\label{C4}\\
_{IAB}{}^C:&& \delta_I \T ABC=-g_{IA}{}^D\T DBC-g_{IB}{}^D\T ADC
              +\T ABD g_{ID}{}^C,
\label{C5}\\
_{IAB}{}^J:&& \delta_I \F ABJ=-g_{IA}{}^C\F CBJ-g_{IB}{}^C\F ACJ-\f IKJ\F ABK,
\label{C6}\\
_{ABC}{}^I:&& \csum {ABC}{35}(\cD_A\F BCI+\T ABD\F DCI)=0,
\label{C7}\\
_{ABC}{}^D:&& \csum {ABC}{35}(\cD_A\T BCD+\T ABE\T ECD+\F ABI\Gg ICD)=0.
\label{C8}
\eea
(\ref{C1}) is the Jacobi identity for structure constants of Lie 
algebra. It just reflects that, according to (\ref{dd}), 
the $\delta_I$ are to
form a Lie algebra with structure constants $\f IJK$.
This Lie algebra is denoted by $\mathfrak{g}$ and chosen
to be the direct sum of the Lorentz group and
the Lie algebra $\mathfrak{g}_\mathrm{YM}$ of a YM gauge group,
$\mathfrak{g}=\mathfrak{so}(1,3)\oplus\mathfrak{g}_\mathrm{YM}$.

(\ref{C4}) imposes that
the constants $g_{IA}{}^B$ are the entries of 
matrices $g_I$ representing $\mathfrak{g}$ on the $\cD$'s because
in matrix notation it reads just $[g_I,g_J]=\f IJK g_K$.
To fulfill it, we choose
the only nonvanishing $g_I$ to be those for the Lorentz algebra
and, possibly, for two abelian elements $\delta_\dR,\delta_{(W)}
\in\mathfrak{g}_\mathrm{YM}$ which belong to
so-called $R$-transformations (these are $U(1)$-transformations
which do not commute with SUSY transformations) and
Weyl-transformations [Weyl-transformations
are included here for the sake of generality; we shall drop
them later again]:
\bea
\ba{lll} [l_{ab},\cD_c]=\eta_{cb}\cD_{a}-\eta_{ca}\cD_{b},& 
[l_{ab},\cD_\alpha]=-{\sigma_{ab\,\alpha}}^\beta\cD_\beta,&
[l_{ab},\5\cD_\da]=\5\sigma_{ab}{}^\dbe{}_\da\5\cD_\dbe,\\
{}[\delta_\dR,\cD_a]=0,&
[\delta_\dR,\cD_\alpha]=-\Ii\,\cD_\alpha,&
[\delta_\dR,\5\cD_\da]=\Ii\, \5\cD_\da ,
\\
{}[\delta_{(W)},\cD_a]=-\cD_a,&
[\delta_{(W)},\cD_\alpha]=-\frac 12\cD_\alpha,&
[\delta_{(W)},\5\cD_\da]=-\frac 12\5\cD_\da.
\ea\label{T17}
\eea

(\ref{C5}) and (\ref{C6}) require that the ``torsions'' $\T ABC$ and
``curvatures'' $\F ABI$ transform under $\mathfrak{g}$ according to 
linear representations characterized by their index pictures.
They are thus fulfilled when the $\T ABC$ and
$\F ABI$ are ordinary tensor fields with regard to the 
Lorentz group and the YM gauge group.

(\ref{C7}) and (\ref{C8}) are conditions on the
$\T ABC$ and $\F ABI$ and their $\cD_A$-transformations.
They provide in particular in part the
SUSY-transformations of these
tensor fields (recall that the gauge transformations
of a tensor fields are
$\delta_{\7\xi}T=\7\xi^\Mdex \Delta_\Mdex T$ whose
``SUSY-part'' is thus
$\7\xi^\alpha\cD_\alpha T+\,\5{\!\!\7\xi}_\da\5\cD^\da T$).
(\ref{C7}) and (\ref{C8}) are called the 
Bianchi identities of $D=4$, $N=1$ SUGRA because they
generalize the Bianchi identities of GR and YM theory
(the latter are obtained from (\ref{C7}) for
$ABC=abc$ by setting all fields with spinors indices to zero).
A set of tensor fields $\{\T ABC,\F ABI\}$ which
satisfies these equations is called a
``solution of the Bianchi identities''.
It was shown in \cite{dragon} that the 
Bianchi identities (\ref{C7}) follow
from (\ref{C8}) [using (\ref{T8}) and (\ref{C5})].

Different solutions of the Bianchi identities
lead to different formulations of
$D=4$, $N=1$ SUGRA. However, two such `different'
formulations can actually still be equivalent because they
may only differ by redefinitions of the fields or gauge parameters.
Indeed, consider redefinitions of the gauge parameters
of the form $\7\xi^{\prime \Mdex}=\7\xi^\Ndex X_\Ndex^\Mdex$ where
$X_\Ndex^\Mdex$ is a local invertible matrix whose entries
are tensor fields.
Such redefinitions of the gauge parameters correspond
to redefinitions $\Delta_\Mdex'=(X^{-1})_\Mdex^\Ndex\Delta_\Ndex$ of the
$\Delta$'s (as these yield the same gauge transformations:
on tensor fields one has
$\7\xi^\Mdex\Delta_\Mdex T=\7\xi^{\prime \Mdex}\Delta_\Mdex'T$ 
for all tensor fields).
Hence, two solutions of the Bianchi identities differing
only by such redefinitions (which preserve
(\ref{Dd}) and (\ref{dd})) must be considered equivalent,
since such redefinitions of gauge parameters can always be
made in gauge theories (cf. section \ref{basis}).
By such redefinitions one can always achieve \cite{diss} 
that
\bea
\T \alpha\da{a}=2\Ii\sigma^a_{\alpha\da},\quad
\T \alpha\da\beta=
\T \alpha\da\dbe=
\T \alpha\beta\gamma=
\T \da\dbe\dg=
\T abc=\F \alpha\da{i}=0.
\label{C9}
\eea
Hence (\ref{C9}) can be assumed without loss of generality.
These choices are therefore called ``conventional constraints''.
The constraint $\T abc=0$ determines the spin connection because
(\ref{T12b}) yields for $\Mdex=c$:
\bea
e_\mu^a e_\nu^b\,\T abc=
\6^{}_\mu e^c_\nu-\6^{}_\nu e^c_\mu
+\psi^\ube_\mu \psi^\ua_\nu\T \ua\ube{c}
+(e_\mu^a\psi^\ua_\nu-e_\nu^a\psi^\ua_\mu)\T \ua{a}c
+(e_\mu^a\A I\nu-e_\nu^a\A I\mu) g_{Ia}{}^c
\label{C10a}
\eea
where
$\{\psi^\ua_\mu\}=\{\psi^\alpha_\mu,\5\psi^\da_\mu\}$
and summation convention as in (\ref{T3d}). Using
(\ref{T17}), the term $e_\mu^a\A I\nu g_{Ia}{}^c$
which occurs in (\ref{C10a})
reads explicitly
\[
e_\mu^a\A I\nu g_{Ia}{}^c=\omega_{\nu\mu}{}^c+e_\mu^c \A {(W)}\nu.
\]
Hence, for $\T abc=0$ we obtain from (\ref{C10a}):
\bea
\FRAME{\ 
\omega_{[\mu\nu]}{}^c=\6^{}_{[\mu} e^c_{\nu]}
+\sfrac 12\psi^\ube_\mu \psi^\ua_\nu\T \ua\ube{c}
+e_{[\mu}^a\psi^\ua_{\nu]}\T \ua{a}c
+e_{[\mu}^c\A {(W)}{\nu]}.
\ }
\label{C10b}
\eea
Note that this is analogous to (\ref{omega1}) and determines
$\omega_\mu{}^{ab}$ analogously to (\ref{omega2}),
using
$\omega_{\mu\nu\rho}=\omega_{[\mu\nu]\rho}-\omega_{[\nu\rho]\mu}
+\omega_{[\rho\mu]\nu}$.

Constraints in addition to (\ref{C9}) yield different
off-shell formulations of $D=4$, $N=1$ SUGRA.
The additional constraints cannot be arbitrarily chosen because
the Bianchi identities
(\ref{C7}) and (\ref{C8}) must be satisfied.
The simplest solutions to the Bianchi identities
are spelled out in the next subsections.

\subsection{Old minimal SUGRA}

We shall now present the so-called ``old minimal''
SUGRA theory which is certainly
the most popular off-shell formulation of $D=4$, $N=1$ SUGRA. 
We shall start from the corresponding
solution of the Bianchi identities
(\ref{C7}) and (\ref{C8}) in presence
of super-YM multiplets without
discussing how one derives this solution systematically
(for details see, e.g., \cite{grimmetal}).
Then we shall introduce chiral matter multiplets, spell
out the gauge transformations and finally the construction
of invariant actions, including the higher order invariants.

\subsubsection{Old minimal solution of the Bianchi identities}

We shall present the solution for the case that $R$-transformations are
possibly gauged (the version without gauged $R$-transformation
is obtained simply by setting all fields with an index $\dR$ to zero),
but without gauged Weyl-transformations,
\[
\delta_{(W)}\not\in\{\delta_i\}.
\]
The torsions and curvatures
are, except for those that can be obtained from the others using
the graded symmetry in $AB$, or the following relations
\[
\T \alpha{a}\beta=-(\T \da{a}\dbe)^*,\quad
\T \alpha{a}\dbe=-(\T \da{a}\beta)^*,\quad
\F \alpha{a}I=(\F \da{a}I)^*,\quad
\F \alpha\beta{ab}=-(\F \da\dbeta{ab})^*,
\]
or (\ref{T12b}) (with $\T abc=0$):
\bea
\ba{c|c|c|c}
 & AB=\da b & AB=\da\dbe & AB=\alpha\dbe \\
\hline\rule{0em}{3ex}
\T ABc & 0 & 0 & 2\Ii\sigma^c_{\alpha\dbe} \\
\rule{0em}{3ex}
\T AB\gamma & 
           \sfrac \Ii8 M\ep^{\gamma\alpha}\sigma_{b\,\alpha\da} & 0 & 0 \\
\rule{0em}{3ex}
\T AB\dg & 
              -\Ii\, (\delta_\da^\dg B_b
               +B^c\5\sigma_{cb}{}^\dg{}_\da) & 0 & 0 \\
\rule{0em}{3ex}
\F ABi & 
       \Ii\lambda^{i\,\alpha}\sigma_{b\,\alpha\da}  & 0 & 0 \\
\rule{0em}{3ex}
\F AB{cd} & 
       \Ii T^{cd\alpha}\sigma_{b\, \alpha\da}
       -2\Ii\sigma^{[c}_{\alpha\da} T^{d]}{}_b{}^\alpha &
       -M\5\sigma^{cd}{}_{\da\dbe} &
       2\Ii \ep^{abcd}\sigma_{a\alpha\dbe}B_b 
\ea
\label{table}
\eea
Here $M$ is a complex scalar field  
and $B_a$ is a real vector field. These fields are the 
auxiliary fields of the old minimal SUGRA multiplet [of course,
that these fields are indeed auxiliary ones can not really be seen
at this point but only from the action to be constructed later; however, one
may anticipate it by counting the DOF off-shell and by 
inspecting the dimensions of these fields].
The $\lambda_\alpha^i$ are the fermions (``gauginos'')
of the super-YM multiplets, i.e., the ``superpartners'' of the
YM gauge fields.
Explicitly this yields:
\bea
\ba{l}
{}[\cD_a,\cD_b]=-\sfrac 12\F ab{cd}l_{cd}-\F abi\delta_i
-\T ab\alpha\cD_\alpha-T_{ab\da}\5\cD^\da
\\[4pt]
{}[\cD_\alpha,\cD_a]=-\sfrac 12\F \alpha{a}{cd}l_{cd}
+\Ii \sigma_{a\alpha\da}\5\lambda^{\da i}\delta_i
+\Ii(B_a\delta_\alpha^\beta-B^b\sigma_{ba\alpha}{}^\beta)\cD_\beta
-\sfrac{\Ii}8\5M\sigma_{a\alpha\da}\5\cD^\da
\\[4pt]
{}[\5\cD_\da,\cD_a]=-\sfrac 12\F \da{a}{cd}l_{cd}
-\Ii \sigma_{a\alpha\da}\lambda^{\alpha i}\delta_i
-\Ii(B_a\delta_\da^\dbe+B^b\5\sigma_{ba}{}^\dbe{}_\da)\5\cD_\dbe
+\sfrac{\Ii}8 M\sigma_{a\alpha\da}\cD^\alpha
\\[4pt]
{}[\cD_\alpha,\5\cD_\da]=-2\Ii\sigma^a_{\alpha\da}\cD_a
-\Ii \ep^{abcd}\sigma_{a\alpha\da}B_bl_{cd}
=-2\Ii\cD_{\alpha\da}
+2B_{\beta\da}l_\alpha{}^\beta-2B_{\alpha\dbe}\5l_\da{}^\dbe
\\[4pt]
{}[\cD_\alpha,\cD_\beta]=\sfrac 12 \5M\sigma^{ab}{}_{\alpha\beta}\,l_{ab}
=\5Ml_{\alpha\beta}
\\[4pt]
{}[\5\cD_\da,\5\cD_\dbe]=\sfrac 12 M\5\sigma^{ab}{}_{\da\dbe}\,l_{ab}
=-M\5l_{\da\dbe}
\ea
\label{spinoralg}
\eea
where $l_{\alpha\beta}$ and $l_{\da\dbe}$ 
are the Lorentz ($\mathfrak{sl}(2,\mathbb{C})$) generators acting
on undotted and dotted spinor indices according to
\bea
l_{\alpha\beta}X_\gamma =-\ep_{\gamma(\alpha}X_{\beta)},\quad
l_{\alpha\beta}\5X_\da=0,\quad
\5l_{\da\dbe}\5X_\dg =-\ep_{\dg(\da}\5X_{\dbe)},\quad
\5l_{\da\dbe}X_\alpha=0.
\label{sl2c}
\eea
They are related to the $l_{ab}$ by
\bea
l_{ab}=\sigma_{ab}{}^{\alpha\beta}l_{\alpha\beta}-
\5\sigma_{ab}{}^{\da\dbe}\5l_{\da\dbe}.
\eea

Furthermore the Bianchi identities yield
\bea
\cD_\alpha M &=& \sfrac {16}3\,(S_\alpha-\Ii\lambda_\alpha^\dR),
\label{ca7}
\\                                           
\cD_\alpha \5M & =& 0,\label{ca8}\\
\cD_\alpha B_{\beta\dbe}
&=& \sfrac 13\ep_{\beta\alpha}(\5S_\dbe+4\Ii\5\lambda_\dbe^\dR)
                     -\5U_{\alpha\beta\dbe},  \label{ca9} \\
\cD_\alpha \lambda_\beta^i &=& \Ii\ep_{\alpha\beta}D^i
                             +\G \alpha\beta{i},\label{ca10}\\
\cD_\alpha\5\lambda_\da^i & =& 0.  \label{ca11}\\
\cD_\alpha D^i &=&
\cD_{\alpha\da}\5\lambda^{i\da}+
\sfrac {3\Ii}2 B_{\alpha\da}\5\lambda^{i\da}\ .
\label{ca12}\eea
where $D^i$ are real auxiliary fields of the
super-YM multiplets and $S_\alpha$, $U_{\da\dbe\gamma}$
and $\G \alpha\beta{i}$ are given by
\bea
S_\alpha=\T ab\beta\sigma^{ab}{}_{\alpha\beta},\
U_{\da\dbe\gamma}=T_{ab\gamma}\5\sigma^{ab}{}_{\da\dbe},\
W_{\alpha\beta\gamma}=T^{}_{ab(\alpha}\sigma^{ab}{}_{\beta\gamma)},\
\G \alpha\beta{i}=-\F abi\sigma^{ab}{}_{\alpha\beta}.
\label{ca13}\eea
Notice that the fields $D^i$ do not occur in any of the torsions or
curvatures. They arise only `indirectly' from the Bianchi identities 
because the latter determine
$\cD_\alpha\lambda^i_\beta$ only up to the piece which
is antisymmetric in $\alpha$ and $\beta$ and purely imaginary.
That piece is written as $\Ii\ep_{\alpha\beta}D^i$ which
introduces thus additional fields $D^i$. That these
fields are really needed, i.e., that they cannot be set 
to zero off-shell is then seen by imposing the algebra (\ref{spinoralg})
on the $\lambda^i$ and $\5\lambda^i$ with the result given in
(\ref{ca12}) (the right hand side of (\ref{ca12})
does not vanish off-shell and therefore the $D^i$ cannot be set 
to zero off-shell either).

The tensor fields (\ref{ca13}) arise when one decomposes
$\T ab\alpha$ and $\F abi$ into Lorentz-irreducible parts
by expressing them in terms of spinor indices 
(using $\F {\alpha\da\,}{\beta\dbe}i
=\sigma^a_{\alpha\da}\sigma^b_{\beta\dbe}\F abi$ etc)
and then decomposing the resulting expressions into
pieces which are totally symmetric in all undotted
and all undotted spinor indices, respectively (splitting off 
$\ep$'s):
\bea
&&
T_{\alpha\da\,\beta\dbe\,\gamma}=
\ep_{\alpha\beta}U_{\da\dbe\gamma}
+\ep_{\dbe\da}(W_{\alpha\beta\gamma}+\sfrac 23\ep_{\gamma(\alpha}S_{\beta)})
\nonumber\\
\LRA 
&& 
T_{ab\gamma}=\sfrac 12\5\sigma_{ab}{}^{\da\dbe}U_{\da\dbe\gamma}
+\sfrac 12\sigma_{ab}{}^{\alpha\beta}W_{\alpha\beta\gamma}
-\sfrac 13\sigma_{ab\alpha\gamma}S^\alpha,
\label{ca12a}\\
&&
\F {\alpha\da\,}{\beta\dbe}i=\ep_{\alpha\beta}\5G_{\da\dbe}{}^i
+\ep_{\da\dbe}\G \alpha\beta{i}
\nonumber\\
\LRA 
&&
\F abi=\sfrac 12\5\sigma_{ab}{}^{\da\dbe}\5G_{\da\dbe}{}^i
-\sfrac 12\sigma_{ab}{}^{\alpha\beta}\G \alpha\beta{i}.
\label{ca12b}
\eea
For the sake of completeness, and for later use, let me
also give the corresponding decomposition of the supercovariant
version of the Riemann tensor $\F ab{cd}$:
\bea
&&
F_{\alpha\da\,\beta\dbe\,\gamma\dg\,\delta\dde}=
\ep_{\da\dbe}\ep_{\dg\dde}[X_{\alpha\beta\gamma\delta}
-\sfrac 16(\ep_{\alpha\gamma}\ep_{\beta\delta}
+\ep_{\beta\gamma}\ep_{\alpha\delta})\cR]-
\ep_{\alpha\beta}\ep_{\dg\dde}Y_{\gamma\delta\da\dbe}+\mathrm{c.c.}
\nonumber\\
&&
X_{\alpha\beta\gamma\delta}=
\sigma^{ab}{}_{(\alpha\beta}\sigma^{cd}{}_{\gamma\delta)}F_{abcd},\quad
Y_{\alpha\beta\da\dbe}
=\5\sigma^{ab}{}_{\da\dbe}\sigma^{cd}{}_{\alpha\beta}F_{abcd},\quad
\cR=\F ab{ba}.
\label{ca13a}
\eea
$X_{\alpha\beta\gamma\delta}$, $Y_{\alpha\beta\da\dbe}$ and
$\cR$ are the supercovariant versions of the
Weyl tensor, traceless Ricci tensor and Riemann curvature scalar,
respectively (in spinor notation).
I also note for later use another important result:
\bea
\cD^\alpha\cD_\alpha M=\sfrac {8}3\cR+\sfrac {32}3D^\dR
+2M\5M-16B^aB_a+16\Ii\cD_a B^a.
\label{DDM}
\eea
Using the torsions in the table (\ref{table}), 
one obtains from (\ref{C10b}):
\bea
\omega_{[\mu\nu]}{}^a=\6^{}_{[\mu} e_{\nu]}^a
-2\Ii\psi_{[\mu}\sigma^a\5\psi_{\nu]}\ .
\label{omegaold1}
\eea
This is precisely the same expression as (\ref{omega1}).
Hence the spin connection of the old minimal
formulation is given again by (\ref{omega2}).

\subsubsection{Chiral matter multiplets}\label{chiral}

Next we discuss so-called chiral matter multiplets. 
These consist of tensorial matter fields
$\varphi$, $\chi_\alpha$, $F$ where $\varphi$ and $F$ are complex
scalar fields and $\chi_\alpha$ are Weyl spinor fields.
These fields may carry additional indices which refer
to the YM gauge group (more precisely, a representation thereof),
which we shall suppress. So, one should think of $\varphi$
as a column vector on which representation
matrices $T_i$ of the YM-Lie algebra $\mathfrak{g}_\mathrm{YM}$ act, 
and the same applies to $\chi_\alpha$ and $F$. These
representation matrices $T_i$
agree on $\varphi$, $\chi_\alpha$, $F$ for all $i$
except for $i=\dR$ (this exception will become clear below),
\bea
i\neq\dR:&&
\delta_i\phi=-T_i\phi,\ \delta_i\chi_\alpha=-T_i\chi_\alpha,\
\delta_i F=-T_i F;
\nonumber\\
i=\dR:&& \delta_\dR\varphi=- T_\dR\varphi,\
\delta_\dR\chi_\alpha=-(T_\dR+\Ii)\chi_\alpha,\
\delta_\dR F=-(T_\dR+2\Ii)F;
\nonumber\\
&&[T_i,T_j]=\f ijk T_k.
\label{YMmat}
\eea
The Lorentz group acts on $\varphi$, $\chi_\alpha$, $F$
in the standard way, 
\bea
l_{ab}\varphi=0,\quad 
l_{ab}\chi_\alpha=-(\sigma_{ab}\chi)_\alpha,\quad
l_{ab}F=0.
\label{lorentzmat}
\eea
Then (\ref{Dd}), (\ref{dd}) are satisfied on the
$\varphi$, $\chi_\alpha$, $F$.
(\ref{spinoralg}) is satisfied with the
following transformations:
\bea
\ba{ll}
\cD_\alpha\varphi=\chi_\alpha\,,& \5\cD_\da\varphi=0,\\[4pt]
\cD_\alpha\chi_\beta=-\ep_{\alpha\beta}F,&
\5\cD_\da\chi_\alpha=-2\Ii\sigma^a_{\alpha\da}\cD_a\varphi,\\[4pt]
\cD_\alpha F=-\sfrac 12\5M\chi_\alpha\,,&
\5\cD_\da F=-2\Ii\cD_{\alpha\da}\chi^\alpha-4\5\lambda^i_\da \delta_i\varphi
+B_{\alpha\da}\chi^\alpha.
\ea
\label{SUSYmat}
\eea
This explains in particular the relations for $\delta_\dR$ in
(\ref{YMmat}), as $\cD_\alpha$ carries $R$-weight 1,
cf.\ (\ref{T17}).

The field content of chiral matter multiplets and
the transformations (\ref{SUSYmat}) can be found as follows.
We start just with the field
$\varphi$, which is chosen to be the ``lowest'' 
component field of the multiplet to be constructed
(i.e., it has lowest dimension). We impose
$\5\cD_\da\varphi=0$ which may
be viewed as the simplest possible 
$\5\cD_\da$-transformation one may choose%
\footnote{In accordance with standard SUSY terminology,
$\5\cD_\da$-invariant fields are called ``chiral fields''.
Hence, $\varphi$ is a chiral field and that explains why the
whole multiplet is termed ``chiral multiplet''.}
(that choice is possible because (\ref{spinoralg}) requires
$[\5\cD_\da,\5\cD_\dbe]\varphi=
\sfrac 12 M\5\sigma^{ab}{}_{\da\dbe}\,l_{ab}\varphi$ which
vanishes owing to $l_{ab}\varphi=0$).
$\cD_\alpha\varphi$ is then defined to be a new field
denoted by $\chi_\alpha$ which thus becomes the second member of the
multiplet.

We have thus fixed the
$\cD_\ua$-transformations of $\varphi$ (and also of $\5\varphi$
by complex conjugation) and introduced new fields $\chi_\alpha$.
Next we have to define the transformations of these fields.
Let us first consider $\cD_\alpha\chi_\beta$. 
Using $\chi_\beta=\cD_\beta\varphi$ we obtain
\beann
\cD_\alpha\chi_\beta=\cD_\alpha\cD_\beta\varphi
=\sfrac 12(\cD_\alpha\cD_\beta+\cD_\beta\cD_\alpha)\varphi
+\sfrac 12(\cD_\alpha\cD_\beta-\cD_\beta\cD_\alpha)\varphi.
\eeann
Up to the factor $1/2$, the first term on the right hand side
is the graded commutator $[\cD_\alpha,\cD_\beta]$
(since $\cD_\alpha$ and $\cD_\beta$ are
Grassmann odd their graded commutator is the anticommutator).
(\ref{spinoralg}) imposes that this
term must vanish (owing to $l_{ab}\varphi=0$).
The second term is antisymmetric in $\alpha$ and $\beta$
and thus proportional to $\ep_{\alpha\beta}$. 
We define it to be $-\ep_{\alpha\beta}F$ where $F$ is 
a new field (an additional member of the multiplet).
This yields the transformations $\cD_\alpha\chi_\beta=-\ep_{\alpha\beta}F$
in (\ref{SUSYmat}).
To define $\5\cD_\da\chi_\alpha$ we proceed similarly:
\beann
\5\cD_\da\chi_\alpha=
\5\cD_\da\cD_\alpha\varphi
=(\5\cD_\da\cD_\alpha+\cD_\alpha\5\cD_\da)\varphi
-\cD_\alpha\5\cD_\da\varphi.
\eeann
The first term on the right hand side is the graded
commutator $[\cD_\alpha,\5\cD_\da]\varphi$.
According to (\ref{spinoralg}) it should be
equal to $-2\Ii\sigma^a_{\alpha\da}\cD_a\varphi$
(owing to $l_{ab}\varphi=0$). The second term must
vanish  because of $\5\cD_\da\varphi=0$.
This yields the transformations
$\5\cD_\da\chi_\alpha=-2\Ii\sigma^a_{\alpha\da}\cD_a\varphi$
in (\ref{SUSYmat}).
Note that this really defines $\5\cD_\da\chi_\alpha$
completely because, using
(\ref{T3f}), we obtain:
\bea
\cD_a\varphi&=&E^\mu_a(\6_\mu-\psi_\mu^\alpha\cD_\alpha
-\5\psi_{\mu\da}\5\cD^\da
-\sfrac 12\omega_\mu{}^{ab}l_{ab}-A_\mu^i\delta_i)\varphi
\nonumber\\
&=&E^\mu_a(\6_\mu\varphi-\psi_\mu^\alpha\chi_\alpha-A_\mu^i \delta_i\varphi).
\label{covphi}
\eea
As we have introduced a new field $F$, we must now determine
its transformations. $\cD_\alpha\chi_\beta=-\ep_{\alpha\beta}F$
gives $2F=\cD_\beta\chi^\beta$. Using this, we obtain
\bea
\cD_\alpha F=\sfrac 12\cD_\alpha\cD_\beta\chi^\beta
=\sfrac 12[\cD_\alpha,\cD_\beta]\chi^\beta
-\sfrac 12\cD_\beta\cD_\alpha\chi^\beta.
\label{DDchi}\eea
Using the algebra (\ref{spinoralg}), we obtain for
the first term on the right hand side of (\ref{DDchi}):
\[
\sfrac 12[\cD_\alpha,\cD_\beta]\chi^\beta=
\sfrac 12\5M l_{\alpha\beta}\chi^\beta
=-\sfrac 34\5M\chi_\alpha.\] 
Using once again $\cD_\alpha\chi_\beta=-\ep_{\alpha\beta}F$,
the second term on the right hand side
of (\ref{DDchi}) is: 
\[
-\sfrac 12\cD_\beta\cD_\alpha\chi^\beta=
-\sfrac 12\cD_\beta(\delta_\alpha^\beta F)=-\sfrac 12\cD_\alpha F.
\]
Bringing this term to the left hand side of 
(\ref{DDchi}) we obtain the transformation
$\cD_\alpha F=-\sfrac 12\5M\chi_\alpha$ in (\ref{SUSYmat}).
Finally we compute $\5\cD_\da F$ 
starting again from $2F=\cD_\beta\chi^\beta$ and then
using the results for $ \5\cD_\da\chi^\alpha$
and $\cD_\alpha\varphi$:
\beann
\5\cD_\da F&=&\sfrac 12\5\cD_\da\cD_\alpha\chi^\alpha
\\
&=&\sfrac 12[\5\cD_\da,\cD_\alpha]\chi^\alpha
-\sfrac 12\cD_\alpha\5\cD_\da\chi^\alpha
\\
&=&
\sfrac 12[\5\cD_\da,\cD_\alpha]\chi^\alpha
+\Ii\sigma^{a\alpha}{}_{\da}\cD_\alpha\cD_a\varphi
\\
&=&\sfrac 12[\5\cD_\da,\cD_\alpha]\chi^\alpha
+\Ii\sigma^{a\alpha}{}_{\da}[\cD_\alpha,\cD_a]\varphi
+\Ii\sigma^{a\alpha}{}_{\da}\cD_a\cD_\alpha\varphi
\\
&=&
\sfrac 12[\5\cD_\da,\cD_\alpha]\chi^\alpha
+\Ii\sigma^{a\alpha}{}_{\da}[\cD_\alpha,\cD_a]\varphi
+\Ii\sigma^{a\alpha}{}_{\da}\cD_a\chi_\alpha
\eeann
$[\5\cD_\da,\cD_\alpha]\chi^\alpha$ and
$\sigma^{a\alpha}{}_{\da}[\cD_\alpha,\cD_a]\varphi$ can be worked out
using the algebra (\ref{spinoralg}): the former
yields terms proportional to $\cD_{\alpha\da}\chi^\alpha$ 
and $B_{\alpha\da}\chi^\alpha$, the latter
terms proportional to $\5\lambda^i_\da\delta_i\varphi$ and
$B_{\alpha\da}\chi^\alpha$. Working out the precise coefficients
one obtains the result for $\5\cD_\da F$ given in
(\ref{SUSYmat}). This time we did not introduce any new field
and therefore this ends the derivation of the multiplet
and the transformations (\ref{SUSYmat}). The fields $F$
are the auxiliary fields of the chiral matter multiplets.

\subsubsection{Gauge transformations}\label{oldtrafos}

We can now spell out the gauge transformations
of old minimal SUGRA
coupled to super-YM multiplets and chiral matter multiplets
with field content
\bea
e_\mu^a,\psi^\alpha_\mu,M,B_a;\quad 
\A i\mu,\lambda^i_\alpha,D^i;\quad 
\varphi,\chi_\alpha,F.
\label{oldcontent}
\eea
$(e_\mu^a, \psi^\alpha_\mu, M, B_a)$ is called the old
minimal SUGRA multiplet,
$(\A i\mu, \lambda^i_\alpha, D^i)$ the super-YM multiplet(s),
$(\varphi, \chi_\alpha, F)$ the chiral matter multiplet(s).
The gauge transformations of $M$, $B_a$, $D^i$, $\varphi$, $\chi_\alpha$
and $F$ are obtained from (\ref{T2}) using 
(\ref{ca7}) through (\ref{ca12}) and (\ref{SUSYmat})
and their complex conjugates,
the gauge transformations of the gauge fields
from (\ref{T11a}) using the torsions and curvatures
of old minimal SUGRA:
\bea
\delta_{\7\xi}e^a_\mu&=&
  \6_\mu\7\xi^a
  +\sfrac 12(e^b_\mu\7\xi^{cd}-\omega_\mu{}^{cd}\7\xi^b)g_{[cd]b}{}^a
  +\psi^\ua_\mu\7\xi^\ube\T \ube\ua{a}
  \nonumber\\
  &=&\6_\mu\7\xi^a-\omega_{\mu b}{}^{a}\7\xi^b+\7\xi_b{}^a e^b_\mu
  +2\Ii\7\xi\sigma^a\5\psi_\mu-2\Ii\psi_\mu\sigma^a\,\5{\!\7\xi}
  \label{ca14}\\
\delta_{\7\xi}\psi_\mu^\alpha&=&
  \6_\mu\7\xi^\alpha
  +\sfrac 12(\psi^\beta_\mu\7\xi^{ab}-\omega_\mu{}^{ab}\7\xi^\beta)
  g_{[ab]\beta}{}^\alpha
  +(\psi^\beta_\mu\7\xi^\dR-\A \dR\mu\7\xi^\beta)g_{\dR\beta}{}^\alpha
  \nonumber\\
  &&+(e_\mu^a\7\xi^\ube-\psi^\ube_\mu\7\xi^a)\T \ube{a}\alpha
  +e_\mu^a\7\xi^b\T ba\alpha
  \nonumber\\
  &=&\6_\mu\7\xi^\alpha
  -\sfrac 12\omega_\mu{}^{ab}(\7\xi\sigma_{ab})^\alpha
  -\Ii\A \dR\mu\7\xi^\alpha
  +\sfrac 12\7\xi^{ab}(\psi_\mu\sigma_{ab})^\alpha
  +\Ii\psi_\mu^\alpha\7\xi^\dR
  \nonumber\\
  &&
  +(e_\mu^a\7\xi^\beta-\psi^\beta_\mu\7\xi^a)\T \beta{a}\alpha
  -(e_\mu^a\,\5{\!\7\xi}{}^\dbe-\5\psi_\mu^\dbe\7\xi^a)\T \dbe{a}\alpha
  +e_\mu^a\7\xi^b\T ba\alpha
  \label{ca15}\\
\delta_{\7\xi}M&=&\7\xi^a\cD_a M
  +\sfrac {16}3\,\7\xi^\alpha(S_\alpha-\Ii\lambda_\alpha^\dR)
  +2\Ii\7\xi^\dR M
  \label{ca16}\\
\delta_{\7\xi}B_{\alpha\da}&=&\7\xi^b\cD_b B_a
  -\sigma^a_{\alpha\da}\7\xi_a{}^bB_b
  +[-\sfrac 13\7\xi_\alpha(\5S_\da+4\Ii\5\lambda_\da^\dR)
  +\7\xi^\beta\5U_{\beta\alpha\da}+\mathrm{c.c.}]
  \label{ca17}\\
\delta_{\7\xi} \A i\mu&=&\6_\mu\7\xi^i
  -\A j\mu\7\xi^k\f kji
  +(\psi^\ua_\mu\7\xi^a-e^a_\mu\7\xi^\ua)\F a\ua{i}
  +e^a_\mu\7\xi^b\F bai
  \nonumber\\
  &=&\6_\mu\7\xi^i
  -\A j\mu\7\xi^k\f kji
  -\Ii\7\xi\sigma_\mu\5\lambda^i+\Ii\lambda^i\sigma_\mu\,\5{\!\7\xi}
  \nonumber\\
  &&
  +\Ii\7\xi^a(\psi_\mu\sigma_a\5\lambda^i-\lambda^i\sigma_a\5\psi_\mu)
  +e^a_\mu\7\xi^b\F bai
  \label{ca18}\\
\delta_{\7\xi} \lambda^i_\alpha&=&\7\xi^a\cD_a\lambda^i_\alpha
  -\sfrac 12\7\xi^{ab}(\sigma_{ab}\lambda^i)_\alpha
  +\7\xi^j\lambda^k\f kji
  +\Ii\7\xi^\dR\lambda^i_\alpha
  -\Ii\7\xi_\alpha D^i+\xi^\beta\G \beta\alpha{i}
  \label{ca19}\\
\delta_{\7\xi} D^i&=&\7\xi^a\cD_a D^i
  +\7\xi^jD^k\f kji+(\7\xi^\alpha\cD_{\alpha\da}\5\lambda^{i\da}+
  \sfrac {3\Ii}2 \7\xi^\alpha B_{\alpha\da}\5\lambda^{i\da}+\mathit{c.c.})
  \label{ca20}\\
\delta_{\7\xi}\varphi&=&\7\xi^a\cD_a\varphi
  +\7\xi^i\delta_i\varphi+\7\xi\chi
  \label{ca21}\\
\delta_{\7\xi}\chi_\alpha&=&\7\xi^a\cD_a\chi_\alpha
  -\sfrac 12\7\xi^{ab}(\sigma_{ab}\chi)_\alpha
  +\7\xi^i\delta_i\chi_\alpha
  +\7\xi_\alpha F+2\Ii(\sigma^a\,\5{\!\7\xi})_\alpha\cD_a\varphi
  \label{ca22}\\
\delta_{\7\xi}F&=&\7\xi^a\cD_a F
  +\7\xi^i\delta_i F-\sfrac 12\chi\7\xi\5M
  -2\Ii\cD_a\chi\sigma^a\,\5{\!\7\xi}
  -4\5\lambda^i\,\5{\!\7\xi}\delta_i\varphi
  +B_a\chi\sigma^a\,\5{\!\7\xi}.
  \label{ca23}
\eea
The gauge transformation (\ref{ca14}) of the vierbein
agrees entirely with the transformation given in equation (\ref{alg2}).
The gauge transformations (\ref{ca15}) of the gravitino involve the
torsions $\T \beta{a}\alpha$, $\T \dbe{a}\alpha$ 
given in table (\ref{table}), and $\T ba\alpha$
obtained from (\ref{T12b}). If one sets $M$, $B_a$,
$\A \dR\mu$ and $\7\xi^\dR$ to zero,
$\T ab\alpha$ reduces to
$E^\mu_aE^\nu_b(\nabla_\mu\psi_\nu^\alpha-\nabla_\nu\psi_\mu^\alpha)$
and the whole expression (\ref{ca15}) collapses to
the transformation given in equation (\ref{alg11}).
This reflects that
the auxiliary fields $M$ and $B_a$ vanish on-shell
in the off-shell formulation of pure $D=4$, $N=1$ SUGRA
when $R$-transformations are not gauged,
as we shall see below. Let us also indicate how the
transformations (\ref{ca14}) through (\ref{ca23}) read in
terms of the parameters $\xi$.
According to (\ref{new2}) the transformations of
$M$, $B_a$, $\lambda^i_\alpha$, $D^i$, $\varphi$, $\chi_\alpha$,
$F$ are obtained
from those given above simply
by the replacements $\7\xi^a\cD_a\rightarrow\xi^\mu\6_\mu$,
$\7\xi^\ua\rightarrow\xi^\ua$, $\7\xi^{ab}\rightarrow\xi^{ab}$,
$\7\xi^{i}\rightarrow\xi^{i}$. For the transformations
of the gauge fields one obtains from (\ref{new3}) and (\ref{new4}):
\bea
\delta_{\xi}e^a_\mu&=&
  \xi^\nu\6_\nu e^a_\mu+\6_\mu\xi^\nu e^a_\nu
  +\xi_b{}^a e^b_\mu
  +2\Ii\xi\sigma^a\5\psi_\mu-2\Ii\psi_\mu\sigma^a\5\xi
  \label{ca14'}\\
\delta_{\xi}\psi_\mu^\alpha&=&
  \xi^\nu\6_\nu \psi^\alpha_\mu+\6_\mu\xi^\nu \psi^\alpha_\nu
  +\6_\mu\xi^\alpha
  -\sfrac 12\omega_\mu{}^{ab}(\xi\sigma_{ab})^\alpha
  -\Ii\A \dR\mu\xi^\alpha
  \nonumber\\
  &&
  +\sfrac 12\xi^{ab}(\psi_\mu\sigma_{ab})^\alpha
  +\Ii\psi_\mu^\alpha\xi^\dR
  +e_\mu^a\xi^\beta\T \beta{a}\alpha
  -e_\mu^a\,\5\xi^\dbe\T \dbe{a}\alpha
  \nonumber\\
  &=&
  \xi^\nu\6_\nu \psi^\alpha_\mu+\6_\mu\xi^\nu \psi^\alpha_\nu
  +\sfrac 12\xi^{ab}(\psi_\mu\sigma_{ab})^\alpha
  +\Ii\psi_\mu^\alpha\xi^\dR
  \nonumber\\
  &&
  +\nabla_\mu\xi^\alpha
  -\Ii\xi^\alpha B_\mu+\Ii B^\nu(\xi\sigma_{\nu\mu})^\alpha
  +\sfrac {\Ii}8 M(\5\xi\5\sigma_\mu)^\alpha
  \label{ca15'}\\
\delta_{\xi} \A i\mu&=&
  \xi^\nu\6_\nu \A i\mu+\6_\mu\xi^\nu \A i\nu
  +\6_\mu\xi^i-\A j\mu\xi^k\f kji
  -\Ii\xi\sigma_\mu\5\lambda^i+\Ii\lambda^i\sigma_\mu\5\xi.
  \label{ca18'}
\eea

\subsubsection{Action}

It was proved in \cite{diss,sugra}
that the  most general local function
invariant up to a total divergence under the gauge 
transformations given in section \ref{oldtrafos}
is, up to a total divergence:
\bea 
L_{old}&=& e\,(\5\cD^2-4\Ii\psi_\mu\sigma^\mu\5\cD-3M
+16\psi_\mu\sigma^{\mu\nu}\psi_\nu)\, \cA+c.c.\ ,\nonumber\\
\cA&=&P(\5W,\5\lambda,\5\varphi)+(\cD^2-\5M)\, \Omega(T)
\label{act2}
\eea
where $\5W_{\da\dbe\dg}$ is the complex conjugate of
$W_{\alpha\beta\gamma}$ in (\ref{ca13}),
$\5\cD^2$ and
$\cD^2$ are shorthand notations for $\5\cD_\da\5\cD^\da$ and
$\cD^\alpha\cD_\alpha$ respectively,
\[
\cD^2=\cD^\alpha\cD_\alpha,\quad\5\cD^2=\5\cD_\da\5\cD^\da,
\]
$\Omega$ is invariant under all $\delta_I$,
$P$ is invariant under all $\delta_I$ except under $R$-transformations
and has $R$-weight $2$,
\bea
\delta_I\Omega=0\quad \forall I,\quad 
\delta_IP=0\quad \forall I\neq\dR,\quad 
\delta_\dR P=-2\Ii P.
\label{rigid}
\eea
Of course the conditions imposed by $\delta_\dR$ are present only if
we require
$R$-invariance. The invariance of (\ref{act2})
under local SUSY transformations up to a total divergence
is explicitly demonstrated in
appendix \ref{appB} (the invariance under the remaining gauge transformations
is evident). 
I emphasize that
$P$, as indicated by its arguments, depends only on the $\5W_{\da\dbe\dg}$,
$\5\lambda^i_\da$ and $\5\varphi$ but no (covariant) derivatives
thereof. In contrast, $\Omega$ is an arbitrary function of
the tensor fields only subject to (\ref{rigid}).
Let us now spell out various contributions to the
Lagrangian obtained from (\ref{act2}).

\paragraph{Pure SUGRA action.}
The off-shell version of the pure SUGRA action arises when
$\cA$ is proportional to $\5M$
(i.e., $P=0$ and $\Omega=\mathit{constant}$).
Then (\ref{ca7}) and (\ref{DDM}) (resp. their complex conjugates)
yield straightforwardly:
\bea 
&&\cA=\sfrac 3{32} \5M\quad\then
\nonumber\\[6pt]
&&L_{old}= e\, [\sfrac 12\cR
-2\Ii\psi_\mu\sigma^\mu(\5S+\Ii\5\lambda^\dR)
+2\Ii(S-\Ii\lambda^\dR)\sigma^\mu\5\psi_\mu+2D^\dR\nonumber\\
&&\phantom{L_{grav}\propto e\, [}-3B_aB^a-\sfrac 3{16} M\5M+
\sfrac 32(\5M\psi_\mu\sigma^{\mu\nu}\psi_\nu+
M\5\psi_\mu\5\sigma^{\mu\nu}\5\psi_\nu)]
\label{act3}\eea
where
$\lambda^\dR$ and $D^\dR$ contribute
of course only if $R$-transformations
are gauged -- otherwise these fields 
simply have to be set to zero.
In fact the Lagrangian (\ref{act3}) by itself
is inconsistent in presence of these fields as
one sees, for instance, from the EOM for $D^\dR$ which would
read $2e=0$. This is cured when the 
YM Lagrangian $L_{YM}$ given below is added as it contains
terms which are quadratic and of higher order in $\A \dR\mu$,
$\lambda^\dR$ and $D^\dR$. 
The locally $R$-symmetric SUGRA Lagrangian was first constructed in
\cite{Freedman}.
When $R$-transformations
are not gauged, (\ref{act3}) reduces to the old minimal
version of the pure SUGRA action (\ref{action1}) as
given first in \cite{offshell1,offshell2}:
\bea
L_{pure}=
e\,(\sfrac 12 R-3B_aB^a-\sfrac 3{16} M\5M)
+2\ep^{\mu\nu\rho\sigma}(\nabla_\mu\psi_{\nu}\sigma_\rho\5\psi_\sigma
+\psi_\sigma\sigma_\rho\nabla_\mu\5\psi_{\nu})
\label{act3a}
\eea
with $R=E_b^\mu E_a^\nu R_{\mu\nu}{}^{ab}$ as
in (\ref{action1}).
(\ref{act3a}) arises from (\ref{act3}) 
by working out the supercovariant tensor fields $\cR$ and $S_\alpha$ 
explicitly. For instance,
the supercovariant curvature scalar
$\cR$ contains gravitino dependent contributions
that combine with the term
$2\Ii S\sigma^\mu\5\psi_\mu+c.c.$ to the
familiar kinetic term for the gravitino in (\ref{act3a}). 
Furthermore, the terms linear in $B$, $M$
and $\5M$, i.e. those contained in $\cR$, $S$ and $\5S$ and
the last two terms in (\ref{act3}), cancel out exactly.
Notice that the EOM deriving from (\ref{act3a})
set indeed both $M$ and $B_a$ to zero.

\paragraph{Locally supersymmetric YM action.}
The locally supersymmetric YM Lagrangian arises from
the contribution $\sfrac 1{16}\5\lambda^i\5\lambda_i$ to $P$
(nonabelian indices $i$ are lowered with
the Cartan--Killing metric of the Yang--Mills gauge group and
Abelian ones with the unit matrix). It reads
\bea
e^{-1}L_{YM}=-\sfrac 14\, {F_{\mu\nu}}^iF^{\mu\nu}{}_i
-\sfrac {\Ii}2 \, (\lambda^i\sigma^\mu\nabla_\mu\5\lambda_i
+\5\lambda^i\5\sigma^\mu\nabla_\mu\lambda_i)+\sfrac 12\, D^i D_i
+\sfrac 32\, \lambda^i\sigma^\mu\5\lambda_i B_\mu\nonumber\\
-\sfrac 12\,e^{-1} \F \mu\nu{i}\ep^{\mu\nu\rho\sigma}
(\psi_\rho\sigma_\sigma\5\lambda_i+\lambda_i\sigma_\sigma\5\psi_\rho)
+\psi_\mu\sigma^{\mu\nu}\psi_\nu\5\lambda^i\5\lambda_i
+\5\psi_\mu\5\sigma^{\mu\nu}\5\psi_\nu \lambda^i \lambda_i
\label{act4}\eea
where $\nabla_\mu $ is
the usual covariant derivative (not the super-covariant one),
\beq \nabla_\mu=\6_\mu-\A i\mu\delta_i-\sfrac 12\,\omega_\mu{}^{ab} l_{ab}\ ,
\label{hatcov}\eeq
and ${F_{\mu\nu}}^i$ is the supercovariant Yang--Mills
field strength,
\beq {F_{\mu\nu}}^i=\6_\mu \A i\nu-\6_\nu \A i\mu
     +\f jki\A j\mu\A k\nu
     +2i\, (\lambda^i\sigma_{[\mu}\5\psi_{\nu]}
               +\psi_{[\mu}\sigma_{\nu]}\5\lambda^i).
\label{superF}\eeq

\paragraph{Contributions with chiral matter multiplets and
K\"ahler structure.}
Kinetic terms for the chiral matter multiplets
arise from a contribution to $\Omega$ 
of the form $K(\varphi,\5\varphi)$ with
$K$ invariant under all $\delta_i$.
To see this observe that
\[
\5\cD^2\cD^2K(\varphi,\5\varphi)=
(\5\cD^2\cD^2\varphi^s)\,\frac{\6K(\varphi,\5\varphi)}{\6\varphi^s}+\dots
\]
where we have introduced an index $s$ labelling the chiral
multiplets (instead of interpreting $\varphi$ as a 
``column vector'' in the representation space
of $\mathfrak{g}_\mathrm{YM}$ as before) and have
omitted a bunch of terms.
Using (\ref{SUSYmat}) it is easy to verify that 
\[
\5\cD^2\cD^2\varphi^s=-16\cD_a\cD^a\varphi^s+\dots
\]
where again we omitted many other terms.
This shows that a contribution $K(\varphi,\5\varphi)$
to $\Omega$ leads to a contribution to the Lagrangian of the form
\bea
L_{matter}&=&
-16e\,\frac{\6K(\varphi,\5\varphi)}{\6\varphi^s}\,g^{\mu\nu}
\6_\mu\6_\nu\varphi^s
-16e\,\frac{\6K(\varphi,\5\varphi)}{\6\5\varphi^{\5s}}\,
g^{\mu\nu}\6_\mu\6_\nu\5\varphi^{\5s}
+\dots
\nonumber\\
&\sim&
32e\,\frac{\6^2K(\varphi,\5\varphi)}{\6\varphi^s\6\5\varphi^{\5s}}\,
g^{\mu\nu}\6_\mu\varphi^s\6_\nu\5\varphi^{\5s}
+\dots
\label{Lmat}
\eea
I shall not spell out $L_{matter}$ in more detail.
It has quite a number of terms.
I only note that it also involves a term proportional to
\bea
e\,K(\varphi,\5\varphi)\,\cR 
\label{bransdicke}\eea
which originates from $\5\cD^2\5M K(\varphi,\5\varphi)+c.c.$
owing to (\ref{DDM}). Hence one actually obtains a
Brans-Dicke type action from (\ref{act2}) in presence of
chiral matter multiplets.
To bring this action to the standard (Einstein) form
one has to do a redefinition (``Weyl rescaling'') of the vierbein 
according to
\[
\7e^a_\mu\propto \sqrt{K}\,e^a_\mu\quad
\then\quad e\, g^{\mu\nu}\propto K^{-1}\7e\, \7g^{\mu\nu}.
\]
[In order to get a standard form of the action, one
usually also redefines  similarly the fermion fields.]
In terms of the redefined vierbein, 
(\ref{Lmat}) reads
\bea
L_{matter}\propto
\7e\, G_{s\5s}(\varphi,\5\varphi)\,
\7g^{\mu\nu}\6_\mu\varphi^s\6_\nu\5\varphi^{\5s}+\dots
\eea
where we have introduced a K\"ahler metric in the space of 
the scalar fields $\varphi^s$ and $\5\varphi^{\5s}$ 
given by
\bea
G_{s\5s}(\varphi,\5\varphi)=
\frac{\6^2\ln K(\varphi,\5\varphi)}{\6\varphi^s\6\5\varphi^{\5s}}\quad
(\mbox{K\"ahler metric}).
\label{Kaehler}
\eea
It turns out that the other terms in
$L_{matter}$ can also be expressed nicely in terms of quantities
related to the K\"ahler structure
(for instance, there are 4-fermion-terms containing
the curvature of $G_{s\5s}$). 
I refer to the textbooks for the details and only add the remark
that geometrical structures related to scalar fields
are typical of SUGRA theories, also for higher $N$ or $D$.
Of course, they are not always K\"ahler structures as above but of a
similar type. 

Notice that $L_{matter}$ can be viewed as a
generalization of the pure SUGRA action (\ref{act3a}) because
the latter arises from the special choice $K=\mathit{constant}$.
The YM part (\ref{act4})
of the Lagrangian can also be generalized in presence
of chiral matter multiplets. Namely
a contribution $(-1/2)f_{ij}(\5\varphi)\5\lambda^i\5\lambda^j$ to $P$,
with $f_{ij}(\5\varphi)$
a symmetric 2-tensor of the YM group,
results in a contribution to the Lagrangian of the form
\bea
L'_{YM}= e\,[f_{ij}(\5\varphi)+\mathrm{c.c.}]
{F_{\mu\nu}}^i F^{\mu\nu j}+\dots
\eea
This generalizes indeed (\ref{act4}) which is just the special
case of a constant $f_{ij}$.

\paragraph{Further invariants.}
Of course (\ref{act2}) can be also used to construct
other invariants. In particular, a
constant contribution $m$ to $P$ gives rise to
\beq e^{-1}L_{cosmo}= -3mM
+16m\psi_\mu\sigma^{\mu\nu}\psi_\nu+c.c.
\label{act5}\eeq
which, when included,
contributes to the cosmological constant.
Note however that $L_{cosmo}$ is neither locally nor
globally $R$-invariant and is thus forbidden when
global or local $R$-invariance is imposed.
Furthermore 
(\ref{act2}) can be used to construct higher order invariants
containing terms with more than two derivatives.
For instance, a contribution of the form
$W^2\5W^2X^{2n}\5X^{2n}$ to $\Omega$ results in an
invariant containing a contribution
$eX^{2(n+1)}\5X^{2(n+1)}$, i.e.
a term of order $4(n+1)$ in the Weyl tensor. Such invariants are
candidate counterterms in a perturbative
quantum field theoretical approach to SUGRA.

\subsection{New minimal SUGRA}

Actually new minimal SUGRA \cite{Sohnius:1981tp} 
is not fully described by the 
framework of section \ref{tensorcalc} because
it contains a 2-form
gauge potential and is thus
a reducible gauge theory. Nevertheless it can be
obtained within this framework -- it only gives rise to
additional formulas for the gauge transformations
and Bianchi identities of the 2-form
gauge potential and its field strength.
The solution to the Bianchi identities is
very similar to that of old minimal SUGRA;
the differences are that the
complex auxiliary field $M$ is zero and the
consequences thereof. These consequences arise because
$M=0$ requires that the
transformations of $M$ must also be zero by consistency.
(\ref{ca7}) and the real part of the right hand
side of (\ref{DDM}) show that this imposes the
identifications
\bea
M\equiv 0,\quad \lambda^\dR_\alpha\equiv -\Ii S_\alpha,\quad
D^\dR\equiv -\sfrac 14\cR+\sfrac 32 B_a B^a.
\label{Id1}
\eea
The imaginary part of the right hand
side of (\ref{DDM}) imposes in addition
\bea
\cD_a B^a=0.
\label{N2}
\eea
(\ref{Id1}) shows that in new minimal SUGRA
$R$-transformations must be included
among the gauge transformations and that
$\lambda^\dR_\alpha$ and $D^\dR$ disappear
from the list of independent fields.
(\ref{N2}) must hold as an identity in
elementary fields (off-shell). Hence,
$B_a$ cannot be an independent field either.
Rather we must replace it by an expression
that satisfies (\ref{N2}) identically in the
fields and their derivatives. To get an idea
how this might work, note that
(\ref{N2}) is reminiscent of
the equation $d\omega_3=0$ 
because of
\bea
d\omega_3=0,\quad
\omega_3=\sfrac 16dx^\mu dx^\nu dx^\rho f_{\mu\nu\rho}
\quad\LRA\quad \6_\mu h^\mu=0,\quad
h^\mu=\ep^{\mu\nu\rho\sigma}f_{\nu\rho\sigma}.
\label{N2a}
\eea
We know that $d\omega_3=0$ is identically solved
by
\bea
\omega_3=d\omega_2,\quad
\omega_2=\sfrac 12 dx^\mu dx^\nu f_{\mu\nu}\quad\LRA\quad
f_{\mu\nu\rho}=3\6_{[\mu}f_{\nu\rho]},
\label{N2b}\eea
where $f_{\nu\rho}$ are arbitrary functions. 
Notice that $\omega_2$ is by no means unique because,
owing to $d^2=0$,
it can be shifted by $d\omega_1$ with an arbitrary
1-form $\omega_1$.
It turns out that
(\ref{N2}) can be solved similarly 
even though it is much more complicated. In particular, it contains
gravitino dependent terms through the $\omega_\mu{}^{ab}$
occuring in the covariant derivatives $\cD_a$ {\em and}
through the terms $E_a^\mu\psi_\mu^\ua\cD_\ua B^a$
present in $\cD_a B^a$. Notice that the latter
terms involve in particular derivatives of the gravitino
because, according to (\ref{ca9}),
the transformations $\cD_\ua B^a$ contain
the torsions $\T ab\ua$ which are obtained from (\ref{T12b}).
It is therefore by no means obvious whether or not
(\ref{N2}) can be satisfied but an explicit computation shows
that this is indeed the case. The solution is surprisingly simple:
\bea
B^a\equiv e^{-1} e_\mu^a\ep^{\mu\nu\rho\sigma}
(\sfrac 12 \6_\nu A_{\rho\sigma}+\Ii\psi_\nu\sigma_\rho\5\psi_\sigma),
\label{Id2}
\eea
where $A_{\mu\nu}$ are arbitrary antisymmetric real fields analogous to
the $f_{\mu\nu}$ in (\ref{N2b}).
Obviously they are determined only up to
redefinitions of the form
\bea
A'_{\mu\nu}=A_{\mu\nu}+\6_\mu\omega_\nu-\6_\nu\omega_\mu
\label{N5}
\eea
for arbitrary $\omega_\mu$ (this is completely analogous
to the arbitrary shifts $\omega_2\rightarrow\omega_2+d\omega_1$ 
in the example above).
This indicates that $A_{\mu\nu}$ is a 2-form gauge potential.
The gauge transformations are reducible because
the gauge parameters $\omega_\mu$ can be shifted
by $\6_\mu\omega$ with arbitrary $\omega$ without
altering (\ref{N5}).

Having ``solved'' (\ref{N2}) by (\ref{Id2}), it is still not
clear whether this solution is consistent in the sense that
we can assign supersymmetry transformations to $A_{\mu\nu}$
consistently: namely
the expression on the right hand side of (\ref{Id2})
is to transform exactly as $B^a$ in old minimal SUGRA
with the identifications (\ref{Id1}) and (\ref{Id2}).
It is not obvious that this is possible
because the SUSY transformations of $B_a$ in old minimal
SUGRA are quite complicated. But, again, this turns out
to be the case and the solution is very simple.
Together with the diffeomorphism transformations
and the gauge transformations (\ref{N5}) one obtains
the following general gauge transformations of $A_{\mu\nu}$:
\bea
\delta_{\xi,\omega}A_{\mu\nu}&=&
\6_\mu\omega_\nu-\6_\nu\omega_\mu
+\xi^\rho\6_\rho A_{\mu\nu}+\6_\mu \xi^\rho A_{\rho\nu}
+\6_\nu \xi^\rho A_{\mu\rho}\nonumber\\
& &-\Ii\,(\xi\sigma_\mu\5\psi_\nu-\xi\sigma_\nu\5\psi_\mu
+\psi_\mu\sigma_\nu\5\xi-\psi_\nu\sigma_\mu\5\xi).
\label{N6}
\eea
It follows that the expression on the
right hand side of (\ref{Id2}) is a supercovariant
tensor field because in old minimal SUGRA $B^a$ is a tensor
field. The supercovariant
field strength of $A_{\mu\nu}$ can thus be
identified with the expression dual to (\ref{Id2}):
\bea
H_{abc}=E_a^\mu E_b^\nu E_c^\rho(3\6_{[\mu} A_{\nu\rho]}
+6\Ii\psi_{[\mu}\sigma_\nu\5\psi_{\rho]}).
\label{N7}
\eea
In terms of $H_{abc}$, (\ref{N2}) reads $\ep^{abcd}\cD_a H_{bcd}=0$, i.e.,
\[
\cD_{[a} H_{bcd]}=0
\] which can be
interpreted as the Bianchi identity for $H_{abc}$.

The gauge transformations of the other fields
are obtained from those 
given in section \ref{oldtrafos}
using the identifications (\ref{Id1}) and (\ref{Id2}).
Together with (\ref{N6}) they make up the gauge transformations
of new minimal SUGRA with field content
\bea
e_\mu^a,\psi^\alpha_\mu,A_{\mu\nu},\A \dR\mu;\quad 
\A i\mu,\lambda^i_\alpha,D^i\ (i\neq\dR);\quad 
\varphi,\chi_\alpha,F.
\label{newcontent}
\eea
$(e_\mu^a,\psi^\alpha_\mu,A_{\mu\nu},\A \dR\mu)$ is
the new minimal SUGRA multiplet. Notice that it consists
solely of gauge fields. Both $A_{\mu\nu}$ and $\A \dR\mu$
have three DOF off-shell, and thus the number of bosonic
and fermionic DOF match off-shell. As 
the number of DOF of $e_\mu^a$ and $\psi_\mu$ match on-shell,
neither $A_{\mu\nu}$ nor $\A \dR\mu$ must have DOF
on-shell, i.e., these fields must not propagate (in particular,
their DOF on-shell are thus not obtained from (\ref{DOF})).
This is indeed the case because the pure new minimal SUGRA
action reads
\bea
L_{pure,new}= \sfrac 12e\cR
+2\Ii e(S\sigma^\mu\5\psi_\mu-\psi_\mu\sigma^\mu\5S)
+\sfrac 12e H_{abc}H^{abc}
-2\ep^{\mu\nu\rho\sigma}\A \dR\mu \6_\nu A_{\rho\sigma} 
\nonumber\\
=\sfrac 12\, e R+
2\ep^{\mu\nu\rho\sigma}
(\nabla_\mu\psi_{\nu}\sigma_\rho\5\psi_\sigma+\mathrm{c.c.})
+\sfrac 12 e H_{abc}H^{abc}
-2\ep^{\mu\nu\rho\sigma}\A \dR\mu \6_\nu A_{\rho\sigma}
\label{newgrav}
\eea
where
$\nabla_\mu$ is covariant with respect to
Lorentz and $R$-transformations,
\[
\nabla_\mu\psi_\nu^\alpha
   =\6_\mu\psi_\nu^\alpha-\sfrac 12\omega_\mu{}^{ab}
   (\psi_\nu\sigma_{ab})^\alpha-\Ii\A \dR\mu\psi_\nu^\alpha.
\]
The EOM for $\A \dR\mu$ derived from $L_{pure,new}$
set $H_{abc}$ to zero (notice that $\A \dR\mu$ occurs
in $\nabla_\mu\psi_{\nu}$ and $\nabla_\mu\5\psi_{\nu}$).
The EOM for $A_{\mu\nu}$ set the ordinary
(non-supercovariant) field strength
of $\A \dR\mu$ proportional to $\6_\rho H^{\mu\nu\rho}$
and thus, together with the EOM for $\A \dR\mu$,
this field strength vanishes on-shell. Hence
$\A \dR\mu$ and $A_{\mu\nu}$ carry indeed no
physical DOF.

It was proved in \cite{sugra} that
the most general local function invariant up to a total divergence 
under the 
gauge transformations of new minimal SUGRA described above is,
up to a total divergence:
\bea
L_{new}&=&\mu_\dR L_{pure,new}+L_{FI}+L_2
\nonumber\\
L_{FI}&=&\sum_{i_a}
\mu_{i_a}  (eD^{i_a}
+e\lambda^{i_a}\sigma^\mu\5\psi_\mu+e\psi_\mu\sigma^\mu\5\lambda^{i_a}
+\ep^{\mu\nu\rho\sigma}\A {i_a}\mu \6_\nu A_{\rho\sigma})
\label{FI}\\
L_2&=&e(\5\cD^2-4i\psi_\mu\sigma^\mu\5\cD
+16\psi_\mu\sigma^{\mu\nu}\psi_\nu)\, \cA+\mathrm{c.c.}\ ,\nonumber\\
&&\cA=P(\5W,\5\lambda,\5\varphi)+\cD^2 \Omega(T)
\label{act14}\eea
where $i_a$ are the abelian $i$ different from $\dR$
and the $\mu$'s are arbitrary constants.
$L_{FI}$ is the Fayet-Iliopoulos contribution
(redefining the abelian super-YM multiplets by
introducing appropriate linear  combinations of them, one
can achieve that at most one $\mu_{i_a}$ is different from zero).
Actually $L_{pure,new}$ is of the same type as
the contributions to $L_{FI}$: in fact it might be viewed
as the ``Fayet-Iliopoulos contribution'' of the
$R$-transformation because of (\ref{Id1}). 
$\Omega$ and $P$ are again subject to (\ref{rigid}).
The discussion of $L_2$ proceeds as the
discussion of (\ref{act2}) in old minimal SUGRA.

\appendix

\mysection{Lorentz algebra, spinors, Grassmann parity}
\label{`A'}\label{spin}

\subsection{Lorentz algebra}

$D$-dimensional Minkowski metric: 
\[
\eta_{ab}=diag(1,-1,\dots,-1),\quad a,b\in\{0,\dots,D-1\}.
\]
Lorentz algebra:
\[
[l_{ab},l_{cd}]=\eta_{ad}l_{bc}-\eta_{ac}l_{bd}-(a\lra b),\quad
l_{ab}=-l_{ba}.
\]
Vector representation of the Lorentz algebra:
\[
l_{ab}V_c=\eta_{cb}V_a-\eta_{ca}V_b,\quad
l_{ab}V^c=\delta_b^cV_a-\delta^c_aV_b.
\]

\subsection{Spinor representation in even dimensions} 

Dirac algebra ($\gamma_a$: complex $2^{D/2}\times 2^{D/2}$-matrices):
\[
\{\gamma_a,\gamma_b\}=2\eta_{ab}\umat.
\]
The Dirac algebra implies that the matrices
\[
\Sigma_{ab}=\sfrac 14 [\gamma_a,\gamma_b]
\]
form a matrix representation $R$ of the Lorentz algebra
(spinor representation):
\[
[\Sigma_{ab},\Sigma_{cd}]=\eta_{ad}\Sigma_{bc}
-\eta_{ac}\Sigma_{bd}-(a\lra b).
\]
Spinors $\Psi$ are complex ``column vectors'' on which
the $\gamma$-matrices act.\\
The Dirac algebra implies that the matrix
\[
\7\gamma=(-\Ii)^{1+D/2}\gamma_0\gamma_1\dots\gamma_{D-1}
\]
satisfies
\[
\7\gamma^2=\umat,\quad\{\7\gamma,\gamma_a\}=0,\quad
[\7\gamma,\Sigma_{ab}]=0.
\]
Owing to $\7\gamma\not\propto\umat$ and
$[\7\gamma,\Sigma_{ab}]=0$, $R$ is reducible
(Schur's lemma).
It decomposes into two
inequivalent irreducible representations  $R_+$ and $R_-$ of
the Lorentz algebra, $R=R_+\oplus R_-$. The corresponding
spinors $\Psi_+$, $\Psi_-$ are called Weyl spinors,
\[
\Psi=\Psi_++\Psi_-,\quad\7\gamma\Psi_\pm=\pm\Psi_\pm.
\]
Projectors $P_+$, $P_-$: owing to $\7\gamma^2=\umat$, one has
\[
P_\pm=\sfrac 12(\umat\pm\7\gamma),\quad P_\pm^2=P_\pm,\quad
P_+P_-=0=P_-P_+,\quad
P_++P_-=1,\quad
\Psi_\pm=P_\pm\Psi.
\]
Dirac conjugation, 
Majorana conjugation,
charge conjugation (the terminology used in the literature
varies a bit):
\beann
-\kappa\,\eta\, \gamma_a^\dagger= A\gamma_a A^{-1},&& \overline\Psi=\Psi^\dagger A
\quad\mbox{(Dirac conjugation)};
\\
\kappa\, \gamma_a^*= B^{-1}\gamma_a B,&& \Psi^c=B\Psi^*
\quad\mbox{(Majorana conjugation)};
\\
-\eta\, \gamma_a^T= C^{-1}\gamma_a C,&& \4\Psi^{cT}=\Psi^T C^{-1}
\quad\mbox{(charge conjugation)}
\eeann
where $\kappa,\eta\in\{1,-1\}$.
Majorana spinors: $\Psi=B\Psi^*$.

\subsection{Spinor representation in odd dimensions}

Can be obtained from a spinor representation in $D=2k$
by choosing $\gamma_0$,\dots,$\gamma_{2k-1}$ as
in $D=2k$ and $\gamma_{2k}=\pm\Ii\7\gamma$ with the
$\7\gamma$ of the representation in $D=2k$.
There are no Weyl spinors in $D=2k+1$ (in particular
one has $\gamma_0\gamma_1\dots\gamma_{2k}\propto
\gamma_{2k}^2=-\umat$).
\medskip

\noindent
For further details see, e.g., \cite{toine}.

\subsection{Spinors in 4 dimensions}\label{spin4D}

Weyl representation of $\gamma$-matrices:
\beann
&&\gamma^a=
\left(\ba{cc}0&\sigma^a \\ \5\sigma^a&0\ea\right),\quad
\gamma_a=\eta_{ab}\gamma^b,\quad a,b\in\{0,1,2,3\},
\\
&&\sigma^0=\left( \begin{array}{rr}1&0\\ 0&1\end{array}\right),\quad
\sigma^1=\left( \begin{array}{rr}0&1\\ 1&0\end{array}\right),\quad
\sigma^2=\left( \begin{array}{rr}0&-\Ii\\ \Ii&0\end{array}\right),\quad
\sigma^3=\left( \begin{array}{rr}1&0\\ 0&-1\end{array}\right),
\\[6pt]
&&\5\sigma^0=\sigma^0,\quad \5\sigma^1=-\sigma^1,\quad
\5\sigma^2=-\sigma^2,\quad\5\sigma^3=-\sigma^3
\eeann
Properties:
\beann
&1.&\mbox{$\7\gamma$ is diagonal:}\quad
\7\gamma=\left(\ba{rr}\umat&0 \\ 0&-\umat\ea\right)\quad
\then\quad
P_+=\left(\ba{cc}\umat&0 \\ 0&0\ea\right),\quad
P_-=\left(\ba{cc}0&0 \\ 0&\umat\ea\right)
\\
&&\then\ \mbox{Weyl spinors reduce to 2-component 
spinors:}\
\Psi_+={\varphi_+ \choose 0},\
\Psi_-={0\choose \chi_-}
\\
&2.& \Sigma_{ab}=\left(\ba{cc}\sigma_{ab}&0 \\ 0&\5\sigma_{ab}\ea\right),
\quad \sigma_{ab}=
\sfrac 14(\sigma_a\5\sigma_b-\sigma_b\5\sigma_a),\quad
\5\sigma_{ab}
=\sfrac 14(\5\sigma_a\sigma_b-\5\sigma_b\sigma_a)
\\[6pt]
&3.&\mbox{all $\gamma$-matrices are unitary:}\quad
\gamma_a^{-1}=\gamma_a^\dagger
\\[6pt]
&4.& A=\gamma^0,\quad 
B= \left(\ba{cc}0&-\epsilon \\ \epsilon&0\ea\right),\quad
C=\left(\ba{rr}-\epsilon &0 \\ 0&\epsilon\ea\right),\quad
\epsilon=\left( \begin{array}{rr}0&1\\ -1&0\end{array}\right)
\\
&5.& \mbox{Majorana spinors:}\quad 
\Psi={\varphi_+\choose \epsilon\varphi^*_+}
\eeann
Infinitesimal Lorentz transformations of $\Psi$:
\[
l_{ab}\Psi=-\Sigma_{ab}\Psi.
\]
Finite Lorentz transformations with
real parameters $\xi^{ab}=-\xi^{ba}$:
\beann
\Psi'&=&\exp(-\sfrac 12\xi^{ab}\Sigma_{ab})\Psi
={\Lambda_+\varphi_+ \choose \Lambda_-\chi_-},
\\
\Lambda_+&=&\exp(-\sfrac 12\xi^{ab}\sigma_{ab})
\in\mbox{SL}(2,\mathbb{C})
\quad[\mbox{SL}(2,\mathbb{C})\ \mathrm{because\ of}\ 
\sigma_{ab}\in\{\pm\sfrac 12\sigma^i,\pm\sfrac {\Ii}2\sigma^i\}],\\
\Lambda_-&=&
\exp(-\sfrac 12\xi^{ab}\5\sigma_{ab})
\stackrel{\5\sigma_{ab}=-\sigma_{ab}^\dagger}{=}
\exp(\sfrac 12\xi^{ab}\sigma_{ab})^\dagger
=(\Lambda_+)^{-1\dagger}\in\mbox{SL}(2,\mathbb{C}).
\eeann
In general: if $D(g)$ is a matrix representation of
a group $G$, i.e., $D(g_1)D(g_2)=D(g_1g_2)$ for all $g_1,g_2\in G$, then
$[D(g)]^*$, $[D(g)]^{-1T}$ and $[D(g)]^{-1\dagger}$ are
also matrix representations of $G$ (owing to
$M^*N^*=(MN)^*$ and $M^{-1T}N^{-1T}=(MN)^{-1T}$
for all matrices $M,N$).
Therefore: in addition to $\Lambda_+$ and $\Lambda_-=(\Lambda_+)^{-1\dagger}$
one automatically has two further representations of the Lorentz group
given by $(\Lambda_+)^{-1T}$ and $(\Lambda_+)^*
=(\Lambda_-)^{-1T}$. However, the
latter are equivalent to 
$\Lambda_+$ and $\Lambda_-=(\Lambda_+)^{-1\dagger}$, respectively:
\[
\forall M\in\mbox{SL}(2,\mathbb{C}):\quad
M^{-1T}=\ep\, M\, \ep^{-1}=-\ep\, M\, \ep.
\]
Hence, $\ep\varphi_+$ and $\ep\chi_-$ transform under the Lorentz
group according to $(\Lambda_+)^{-1T}$ and $(\Lambda_+)^*$,
respectively.\\
Remark: the last equation is equivalent to
$\ep=M\ep M^T$, i.e., $\ep$ is $\mbox{SL}(2,\mathbb{C})$-invariant
tensor.

\paragraph{Change of notation: undotted and dotted spinor indices:}
indices $\alpha\in\{1,2\}$, 
$\da\in\{\dot 1,\dot 2\}$ indicating
the transformation properties under the Lorentz group:
\beann
\ba{|c|c|l|l|}
\hline
\mbox{new notation} &\mbox{old notation} &
\mbox{representation}& \mbox{transformation}\\
\hline\rule{0em}{5ex}
\varphi_\alpha & \varphi_+ &
\Lambda_+ & l_{ab}\varphi_\alpha=-(\sigma_{ab}\varphi)_\alpha=
-\sigma_{ab\alpha}{}^\beta\varphi_\beta\\
\rule{0em}{3ex}
\varphi^\alpha & \epsilon\,\varphi_+ &
(\Lambda_+)^{-1T} & l_{ab}\varphi^\alpha
=(\varphi\sigma_{ab})^\alpha
=\varphi^\beta\sigma_{ab\beta}{}^\alpha\\
\rule{0em}{3ex}
\5\chi^\da & \chi_- & \Lambda_-= (\Lambda_+)^{-1\dagger} & 
l_{ab}\5\chi^\da=-(\5\sigma_{ab}\5\chi)^\da=
-\5\sigma_{ab}{}^\da{}_\dbeta \5\chi^\dbeta\\
\rule{0em}{3ex}
\5\chi_\da & -\epsilon\,\chi_-&(\Lambda_-)^{-1T}= (\Lambda_+)^* & 
l_{ab}\5\chi_\da=(\5\chi\5\sigma_{ab})_\da
=\5\chi_\dbeta\5\sigma_{ab}{}^\dbeta{}_\da \\
&&&\\
\hline
\ea
\eeann
Indices of $\sigma$-matrices:
\beann
\sigma_a\equiv \sigma_{a\alpha\da},\quad
\5\sigma_a\equiv \5\sigma_a{}^{\da\alpha},\quad
\sigma_{ab}\equiv \sigma_{ab\alpha}{}^\beta,\quad
\5\sigma_{ab}\equiv\5\sigma_{ab}{}^\da{}_\dbeta,\quad
\sigma^a=\eta^{ab}\sigma_b\equiv\sigma^a{}_{\alpha\da}\quad etc.
\eeann
Raising and lowering of spinor indices with
$\epsilon$ (``spinor metric''):
\beann
&
\varphi^\alpha=\epsilon^{\alpha\beta}\varphi_\beta,\quad
\varphi_\alpha=\epsilon_{\alpha\beta}\varphi^\beta,\quad
\5\chi_\da=\epsilon_{\da\dbeta}\5\chi^\dbeta,\quad
\5\chi^\da=\epsilon^{\da\dbeta}\5\chi_\dbeta,\quad
\sigma_a{}^\alpha{}_\da=\ep^{\alpha\beta}\sigma_{a\beta\da}\quad etc,
&
\\
&
\epsilon^{\alpha\beta}=-\epsilon^{\beta\alpha},\quad
\epsilon_{\alpha\beta}=-\epsilon_{\beta\alpha},\quad
\epsilon^{12}=\epsilon_{21}=1,
&
\\
&
\epsilon^{\da\dbeta}=-\epsilon^{\dbeta\da},\quad
\epsilon_{\da\dbeta}=-\epsilon_{\dbeta\da},\quad
\epsilon^{\dot 1\dot 2}=\epsilon_{\dot 2\dot 1}=1,
&
\\
&
\then\quad \epsilon^{\alpha\gamma}\epsilon_{\gamma\beta}
=\delta^\alpha_\beta,\quad
\epsilon^{\da\dgamma}\epsilon_{\dgamma\dbeta}
=\delta^\da_\dbeta.
&
\eeann
Complex conjugation:
\beann
(\psi_\alpha)^*=\5\psi_\da,\quad
(\psi^\alpha)^*=\5\psi^\da,\quad
(\5\psi_\da)^*=\psi_\alpha,\quad
(\5\psi^\da)^*=\psi^\alpha,\quad
(\psi_{\alpha\beta\dgamma})^*=\5\psi_{\da\dbeta\gamma}\quad etc.
\eeann
Dirac and Majorana spinors:
\beann
\mbox{Dirac spinor:}\ 
{\varphi_\alpha\choose \5\chi^\da}, \quad
\mbox{Majorana spinor:}\ 
{\psi_\alpha\choose\5\psi^\da}.
\eeann
Notation for
contraction of undotted and dotted spinor indices:
\[
\psi\chi\equiv\psi^\alpha\chi_\alpha,\quad
\5\psi\5\chi\equiv\5\psi_\da\5\chi^\da,\quad
\sigma_a\5\sigma_b\equiv(\sigma_a\5\sigma_b)_\alpha{}^\beta\equiv
\sigma_{a\alpha\da}\5\sigma_b{}^{\da\beta}\quad etc.
\]
Vector indices $\rightarrow$ spinor indices:
\[
V_{\alpha\da}=\sigma^a{}_{\alpha\da}V_a,\quad
V^{\da\alpha}=\5\sigma_a{}^{\da\alpha}V^a.
\]
\paragraph{Remark.} 
Since every vector index can be converted to
a pair of spinor indices, vector indices are actually superfluous
and so are $\gamma$-matrices and $\sigma$-matrices. In particular,
every Lagrangian, EOM, transformation etc can be written
in terms of objects carrying only spinor indices,
without $\gamma$-matrices or $\sigma$-matrices.
When this is done, an expression is only Lorentz invariant
if all undotted spinor indices are contracted
with $\epsilon_{\alpha\beta}$, $\epsilon^{\alpha\beta}$ or
$\delta_\alpha^\beta$, and
all dotted spinor indices are
contracted with $\epsilon_{\da\dbeta}$, $\epsilon^{\da\dbeta}$ or
$\delta_\da^\dbeta$.

Even though vector indices are superfluous,
they are nevertheless still useful, and
so are the $\sigma$-matrices (for instance, the use of
vector indices may reduce the total number of indices of an object,
because one vector index can substitute for two spinor indices).
For dealing with the $\sigma$-matrices, the following
identities are often useful:
\beann
&&
\5\sigma_a{}^{\da\alpha}=\sigma_a{}^{\alpha\da}=\epsilon^{\alpha\beta}
\epsilon^{\da\dbeta}\sigma_{a\beta\dbeta},\quad
\sigma_{a\alpha\da}=\5\sigma_{a\da\alpha}=
\epsilon_{\alpha\beta}\epsilon_{\da\dbeta}
\5\sigma_a{}^{\dbeta\beta},
\\
&&
\sigma_{ab}{}^{\alpha\beta}
=\sigma_{ab}{}^{\beta\alpha},\quad
\sigma_{ab\alpha\beta}=\sigma_{ab\beta\alpha},\quad
\5\sigma_{ab}{}^{\da\dbeta}
=\5\sigma_{ab}{}^{\dbeta\da},\quad
\5\sigma_{ab\da\dbeta}
=\5\sigma_{ab\dbeta\da},
\\
&&
(\sigma^a\5\sigma^b)_\alpha{}^\beta=\eta^{ab}\delta_\alpha^\beta
+2\sigma^{ab}{}_\alpha{}^\beta,\quad
(\5\sigma^a\sigma^b)^\da{}_\dbeta=
\eta^{ab}\delta^\da_\dbeta+2\5\sigma^{ab\da}{}_\dbeta,
\\
&&
\sigma_{a\alpha\da}\sigma^a{}_{\beta\dbeta}=
2\epsilon_{\alpha\beta}\epsilon_{\da\dbeta},\quad
\5\sigma^{a\da\alpha}\5\sigma_{a}^{\dbeta\beta}
=2\epsilon^{\da\dbeta}\epsilon^{\alpha\beta},\quad
\sigma^a{}_{\alpha\da}\5\sigma_a{}^{\beta\dbeta}
=2\delta^\beta_\alpha\delta^\dbeta_\da,
\\
&&
\ep^{abcd}\sigma_{cd}=2\Ii\sigma^{ab},\quad
\ep^{abcd}\5\sigma_{cd}=-2\Ii\5\sigma^{ab},\quad\ep^{0123}=1,
\\
&&
\sigma^{ab}\sigma^c=\sfrac 12(\eta^{bc}\sigma^a-\eta^{ac}\sigma^b
+\Ii\ep^{abcd}\sigma_d),
\\
&&
\sigma^c\5\sigma^{ab}=\sfrac 12(-\eta^{bc}\sigma^a+\eta^{ac}\sigma^b
+\Ii\ep^{abcd}\sigma_d),
\\
&&
\5\sigma^{ab}\5\sigma^c=\sfrac 12(\eta^{bc}\5\sigma^a-\eta^{ac}\5\sigma^b
-\Ii\ep^{abcd}\5\sigma_d),
\\
&&
\5\sigma^c\sigma^{ab}=\sfrac 12(-\eta^{bc}\5\sigma^a+\eta^{ac}\5\sigma^b
-\Ii\ep^{abcd}\5\sigma_d).
\eeann
\subsection{Grassmann parity}
  
Generalization of wedge product for differential forms:

\[
XY=(-)^{|X|\,|Y|}YX,\quad
|T_{\alpha_1\dots\alpha_n}^{\da_1\dots\da_m}|=(m+n+\mbox{form-degree})\ 
\mbox{mod\ 2},
\]
where $X$, $Y$, $T$ are fields or differential forms.
$|X|$ is called the Grassmann parity (or simply the parity)
of $X$.\footnote{In the BRST approach the definition of the
Grassmann parity involves the ghost number in addition to
the number of spinor indices and the form-degree.}
\\
Complex conjugation of products:
\[
(XY)^*=
(-)^{|X|\,|Y|}X^*Y^*.
\]
Simple consequences:
\beann
&&\psi\chi=\psi^\alpha\chi_\alpha=
\epsilon^{\alpha\beta}\psi_\beta\chi_\alpha
=-\epsilon^{\alpha\beta}\chi_\alpha\psi_\beta
=\epsilon^{\beta\alpha}\chi_\alpha\psi_\beta=\chi\psi,
\\
&&(\psi\chi)^*=(\psi^\alpha\chi_\alpha)^*=
-\5\psi^\da\5\chi_\da=+\5\chi_\da\5\psi^\da=\5\chi\5\psi.
\eeann

\mysection{Explicit verification of local SUSY}\label{`B'}\label{ver}

\subsection{Local SUSY of (\ref{action1})}
\label{ver1}

\paragraph{1.5 order formalism.}
This is a ``trick'' to
simplify the variation of a second order action
if it derives from a first order one. The argument is
simple and general: suppose a Lagrangian $L(\phi,H)$
involves fields $\phi^i$ and $H^A$ such that the EOM
for the $H^A$ have the algebraic solution $H^A=H^A(\phi)$.%
\footnote{To simplify formulae, we use here the notation
$L(\phi,H)$ and $H(\phi)$ in place of $L([\phi,H])$ and $H([\phi])$.}
Let us now
consider the second order Lagrangian $L(\phi,H(\phi))$
and vary the fields $\phi^i$. We obtain
\[
\delta L(\phi,H(\phi))\sim
\Big[
\delta\phi^i\,\frac{\7\6 L(\phi,H)}{\7\6\phi^i}
+\delta H^A(\phi)\,\frac{\7\6 L(\phi,H)}{\7\6 H^A}
\Big]_{H=H(\phi)}
=\Big[
\delta\phi^i\,\frac{\7\6 L(\phi,H)}{\7\6\phi^i}\Big]_{H=H(\phi)}\ .
\]
Here $L(\phi,H)$ is the first order Lagrangian,
$\sim$ denotes equality up to a total divergence,
and $\delta H^A(\phi)=H^A(\phi+\delta\phi)-H^A(\phi)$ is
the variation of  $H^A(\phi)$. The terms with
$\delta H^A(\phi)$ on the right hand side vanish 
(no matter what the $\delta H^A(\phi)$ are) 
because the
$H^A(\phi)$ algebraically solve the EOM of the $H$'s which means 
\[
\frac{\7\6 L(\phi,H)}{\7\6 H^A}\Big|_{H=H(\phi)}=0\quad
(\mbox{identically}).
\]
We observe that, up to a total derivative, the variation of the second order
Lagrangian $L(\phi,H(\phi))$ is obtained from
varying only the $\phi^i$
(but not the $H^A$) in the first order Lagrangian $L(\phi,H)$
and substituting $H^A(\phi)$ for $H^A$ afterwards. Hence,
one uses the first order action to compute
the variation of the second order one. This motivates
the term ``1.5 order formalism''.
Notice that the argument applies to all variations $\delta$.
In particular it shows that the EOM of the second order
formulation can be obtained from those of the
first order formulation according to
\bea
\frac{\7\6 L(\phi,H(\phi))}{\7\6\phi^i}=
\frac{\7\6 L(\phi,H)}{\7\6\phi^i}\Big|_{H=H(\phi)}\ .
\label{EOM1-2}
\eea
Furthermore it can be used to verify invariance
of the second order Lagrangian under symmetry transformations.

\paragraph{Verification of SUSY.}
Using the 1.5 order formalism, we shall
now demonstrate
the SUSY of the Lagrangian (\ref{action1})
in the second order formulation under the SUSY transformations
(\ref{susye}) through (\ref{susy5psi}). The advantage of
the 1.5 order formalism is that we do not need to transform
the spin connection $\omega$ but only the vierbein and gravitino, using the
first order Lagrangian. In fact,
we can further simplify the calculation by using only
the part $\delta_+$ of the SUSY transformations of the vierbein and
gravitino which involve the SUSY parameters
$\xi^\alpha$ but not their complex conjugates $\5\xi^\da$:
\[
\delta_+ e_\mu^a=2\Ii\xi\sigma^a\5\psi_\mu,\quad
\delta_+\psi_\mu^\alpha=\nabla_\mu\xi^\alpha,\quad
\delta_+\5\psi_\mu^\da=0,\quad
\delta_+\omega_\mu{}^{ab}=0.
\]
The reason is that the other part
$\delta_-$, involving the $\5\xi^\da$,
is the complex conjugate of
$\delta_+$, and thus, since the Lagrangian is real:
\[
\delta_-L=(\delta_+ L)^*.
\]
Hence $[\delta_+ L(e,\psi,\omega)]_{\omega
=\omega(e,\psi)}=\6_\mu K_+^\mu$ implies
$[\delta_- L(e,\psi,\omega)]_{\omega
=\omega(e,\psi)}=\6_\mu K_-^\mu$
with $K_-^\mu=(K_+^\mu)^*$. Conversely,
$[\delta_\mathrm{susy}L(e,\psi,\omega)]_{\omega
=\omega(e,\psi)}=\6_\mu K^\mu$
requires that $[\delta_+ L(e,\psi,\omega)]_{\omega
=\omega(e,\psi)}$ be a total divergence
[remember that local SUSY
requires invariance up to a total divergence for
{\em arbitrary} complex parameters, i.e., we
may consider $\xi$ and $\5\xi$ as
independent fields (instead of their real and imaginary
parts)]. Hence $[\delta_+ L(e,\psi,\omega)]_{\omega
=\omega(e,\psi)}\sim 0$ is necessary and sufficient
for $\delta_\mathrm{susy}L\sim 0$
(again, ``$\sim$'' denotes equality up to a total divergence).

Transformation of the ``Einstein-part'':
\bea
\sfrac 12\delta_+ [e E_b^\mu E_a^\nu R_{\mu\nu}{}^{ab}(\omega)]
=\sfrac 12\underbrace{(\delta_+ e)}_{eE^\rho_c\delta_+ e_\rho^c}
E_b^\mu E_a^\nu R_{\mu\nu}{}^{ab}(\omega)
+e\underbrace{(\delta_+ E_b^\mu)}_{-E_b^\rho E_c^\mu\delta_+ e^c_\rho}
E_a^\nu R_{\mu\nu}{}^{ab}(\omega)
\nonumber\\
=e(\delta_+ e_\rho^c)(\sfrac 12 E^\rho_c R-R_c{}^\rho)
=\underbrace{\Ii e(\xi\sigma^\mu\5\psi_\mu R
-2\xi\sigma^a\5\psi_\mu R_a{}^\mu)}_{\mbox{\fbox{1}}},
\label{term0}
\eea
where $R_a{}^\mu=R_{\rho\nu}{}^{bc}E^\rho_a E^\nu_b E^\mu_c$.
Transformation of the ``gravitino-part'':
\beann
2\ep^{\mu\nu\rho\sigma}
\delta_+(\nabla_\mu\psi_{\nu}\sigma_\rho\5\psi_\sigma+
\psi_\sigma\sigma_\rho\nabla_\mu\5\psi_{\nu})
=\underbrace{2\ep^{\mu\nu\rho\sigma}(\nabla_\mu
\delta_+\psi_{\nu})\sigma_\rho\5\psi_\sigma}_{\mbox{\fbox{2}}}
+\underbrace{2\ep^{\mu\nu\rho\sigma}\nabla_\mu\psi_{\nu}
(\delta_+\sigma_\rho)\5\psi_\sigma}_{\mbox{\fbox{3}}}
\\
+\underbrace{2\ep^{\mu\nu\rho\sigma}
(\delta_+\psi_\sigma)\sigma_\rho\nabla_\mu\5\psi_{\nu}}_{\mbox{\fbox{4}}}
+\underbrace{2\ep^{\mu\nu\rho\sigma}
\psi_\sigma(\delta_+\sigma_\rho)\nabla_\mu\5\psi_{\nu}}_{\mbox{\fbox{5}}}.
\eeann
Individual terms:
$\nabla_{[\mu}\delta_+\psi_{\nu]}=\nabla_{[\mu}\nabla_{\nu]}\xi
=\sfrac 12[\nabla_{\mu},\nabla_{\nu}]\xi=
-\sfrac 14R_{\mu\nu}{}^{ab}(\omega)l_{ab}\xi\then$
\[
\mbox{\fbox{2}}=-\sfrac 12\ep^{\mu\nu\rho\sigma}
R_{\mu\nu}{}^{ab}(\omega)\xi\sigma_{ab}\sigma_\rho\5\psi_\sigma
\]
$\delta_+\sigma_{\rho\alpha\da}
=\sigma_{a\alpha\da}\delta_+ e^a_\rho
=2\Ii\sigma_{a\alpha\da}\xi\sigma^a\5\psi_\rho
=4\Ii\xi_\alpha\5\psi_{\rho\da}\then$
\beann
\mbox{\fbox{3}}=8\Ii\ep^{\mu\nu\rho\sigma}\,\xi\nabla_\mu\psi_{\nu}
\underbrace{\5\psi_\rho\5\psi_\sigma}_{=\5\psi_{(\sigma}\5\psi_{\rho)}}=0
,\quad
\mbox{\fbox{5}}=
8\Ii\ep^{\mu\nu\rho\sigma}\,\xi\psi_\sigma\
\5\psi_\rho\nabla_\mu\5\psi_{\nu}\ .
\eeann
Fourth term:
``integration by parts'' to remove derivatives from $\xi$:
\beann
\mbox{\fbox{4}}&=&
\underbrace{\nabla_\sigma (2\ep^{\mu\nu\rho\sigma}
\xi\sigma_\rho\nabla_\mu\5\psi_{\nu})}_{\6_\sigma (2\ep^{\mu\nu\rho\sigma}
\xi\sigma_\rho\nabla_\mu\5\psi_{\nu})}
\underbrace{-2\ep^{\mu\nu\rho\sigma}
\xi(\nabla_\sigma\sigma_\rho)\nabla_\mu\5\psi_{\nu}}_{\mbox{\fbox{4a}}}
\underbrace{-2\ep^{\mu\nu\rho\sigma}\xi\sigma_\rho
\nabla_\sigma\nabla_\mu\5\psi_{\nu}}_{\mbox{\fbox{4b}}},
\\
\mbox{\fbox{4a}}|_{\omega=\omega(e,\psi)}&=&
-2\ep^{\mu\nu\rho\sigma}(\nabla|^{}_{[\sigma} e_{\rho]}^a)\,\xi\sigma_a
\nabla|_\mu\5\psi_{\nu}
\stackrel{(\ref{omega1})}{=}
-4\Ii\ep^{\mu\nu\rho\sigma}(\psi_\sigma\sigma^a\5\psi_\rho)\,\xi\sigma_a
\nabla|_\mu\5\psi_{\nu}
\\
&=&-8\Ii\ep^{\mu\nu\rho\sigma}\,\xi\psi_\sigma\
\5\psi_\rho\nabla|_\mu\5\psi_{\nu}\quad
\mbox{where\ }\nabla|_\mu=\6_\mu-\sfrac 12\omega_\mu{}^{ab}(e,\psi)l_{ab},
\\
\mbox{\fbox{4b}}&=&-2\ep^{\mu\nu\rho\sigma}\xi\sigma_\rho
\sfrac 12[\nabla_\sigma,\nabla_\mu]\5\psi_{\nu}
=-\sfrac 12\ep^{\mu\nu\rho\sigma}\xi\sigma_\rho R_{\sigma\mu}{}^{ab}(\omega)
\5\sigma_{ab}\5\psi_{\nu}\ .
\eeann
Terms 1, 2 and 4b cancel out [computation is similar to
a computation before (\ref{omega1})]:
\[
\mbox{\fbox{2}}+\mbox{\fbox{4b}}
=-\sfrac 12\ep^{\mu\nu\rho\sigma}
R_{\mu\nu}{}^{ab}(\omega)\xi
\underbrace{(\sigma_{ab}\sigma_\rho+\sigma_\rho\5\sigma_{ab})}_
{=\Ii\ep_{ab\rho c}\sigma^c
}\5\psi_\sigma=\dots
=-\mbox{\fbox{1}}
\]
\[
\then\ [\delta_+ L(e,\psi,\omega)]_{\omega=\omega(e,\psi)}
\sim
\Big[\underbrace{\mbox{\fbox{1}}+\mbox{\fbox{2}}+\mbox{\fbox{4b}}}_{=0}
+\underbrace{\mbox{\fbox{3}}}_{=0}
+\underbrace{
\mbox{\fbox{4a}}+\mbox{\fbox{5}}\ \Big]_{\omega
=\omega(e,\psi)}}_{=0}
=0,\quad \mbox{qed.}
\]

\subsection{Local SUSY of (\ref{act2})}\label{appB}

Let us verify explicitly the  
invariance of (\ref{act2}) up to a total divergence
under the local SUSY-transformations given in section \ref{oldtrafos}
(using unhatted parameters). Let us start
with the terms coming from the transformation
of $e$ which is given by
\[
\delta_\mathrm{susy}e=eE_a^\mu\delta_\mathrm{susy} e_\mu^a
=eE_a^\mu (2\Ii\xi\sigma^a\5\psi_\mu-2\Ii\psi_\mu\sigma^a\5\xi)
=2\Ii e(\xi\sigma^\mu\5\psi_\mu-\psi_\mu\sigma^\mu\5\xi).
\]
This gives:
\bea
\lefteqn{
(\delta_\mathrm{susy}e)(\5\cD^2-4\Ii\psi_\mu\sigma^\mu\5\cD-3M
+16\psi_\mu\sigma^{\mu\nu}\psi_\nu)\, \cA
}
\nonumber\\
&&=2\Ii e(\xi\sigma^\rho\5\psi_\rho-\psi_\rho\sigma^\rho\5\xi)
(\5\cD^2-4\Ii\psi_\mu\sigma^\mu\5\cD-3M
+16\psi_\mu\sigma^{\mu\nu}\psi_\nu)\cA.
\label{V0}
\eea
To evaluate the other contributions we shall use that
$\cA$ by construction is antichiral:
\bea
\cD_\alpha\cA=0.
\label{V1}
\eea
This holds because $P$ is antichiral, as it is a function
of antichiral tensor fields,
\[
\cD_\alpha \5W_{\da\dbe\dg}=0,\quad
\cD_\alpha \5\lambda^i_\da=0,\quad
\cD_\alpha \5\varphi=0,
\]
and because $(\cD^2-\5M)\Omega(T)$ is also antichiral,
since $(\cD^2-\5M)f(T)$ is antichiral
for every $l_{\alpha\beta}$-invariant
function $f(T)$:
\bea
l_{\alpha\beta}f(T)=0\quad\then\quad
\cD_\alpha(\cD^2-\5M)f(T)=0
\label{V2}
\eea
(\ref{V2}) can be deduced from the calculation of
$\cD_\alpha F$ in section \ref{chiral}, see
(\ref{DDchi}) and the equations subsequent to it: namely,
the result of that calculation
was $\cD_\alpha F=-\sfrac 12 \5M\chi_\alpha$
which can also be written as 
$-\sfrac 12\cD_\alpha\cD^2\varphi=-\sfrac 12 \5M\cD_\alpha\varphi$ or,
equivalently, as $\cD_\alpha(\cD^2-\5M)\varphi=0$.
As one can check, the derivation 
given in section \ref{chiral} made only use
of the (anti-)commutators 
$[\cD_\alpha,\cD_\beta]=\5M l_{\alpha\beta}$ and of
$l_{\alpha\beta}\varphi=0$. Hence, it actually goes also through
with $\varphi$ replaced by any $l_{\alpha\beta}$-invariant
function of tensor fields, which yields (\ref{V2}).

Let us now consider
$e\delta_\mathrm{susy}\5\cD^2\cA$. 
Since $\5\cD^2\cA$ is a (composite) tensor field, we have
\beann
e\,\delta_\mathrm{susy}\5\cD^2\cA
=e(\xi^\alpha\cD_\alpha+\5\xi_\da\5\cD^\da)\5\cD^2\cA.
\eeann
By the complex conjugate of (\ref{V2}) the
second term on the right hand side is
\beann
e\5\xi_\da\5\cD^\da\5\cD^2\cA=eM\5\xi_\da\5\cD^\da\cA.
\eeann
The evaluation  of $e\xi^\alpha\cD_\alpha\5\cD^2\cA$
requires more work. We treat it as follows: we use the
graded commutator algebra  (\ref{spinoralg}) to pass
$\cD_\alpha$ through $\5\cD^2$ until it hits $\cA$ where it
produces a 0 because of (\ref{V1}). Furthermore we
bring the covariant derivatives which arise
to the left of the spinor transformations,
using again the graded commutator algebra:
\beann
e\,\xi^\alpha\cD_\alpha\5\cD^2\cA&=&
e\,\xi^\alpha([\cD_\alpha,\5\cD_\da]\5\cD^\da
-\5\cD_\da [\cD_\alpha,\5\cD^\da])\cA
\\
&=&e\,\xi^\alpha(-2\Ii\cD_{\alpha\da}\5\cD^\da
-2B_{\alpha\dbe}\5l_\da{}^\dbe\5\cD^\da
-2\Ii\5\cD^\da\cD_{\alpha\da})\cA
\\
&=&e\,\xi^\alpha(-2\Ii\cD_{\alpha\da}\5\cD^\da+3B_{\alpha\da}\5\cD^\da
-2\Ii[\5\cD^\da,\cD_{\alpha\da}]-2\Ii\cD_{\alpha\da}\5\cD^\da)\cA
\\
&=&e\,\xi^\alpha(-2\Ii\cD_{\alpha\da}\5\cD^\da+3B_{\alpha\da}\5\cD^\da
+8\lambda^\dR_\alpha\delta_\dR-5B_{\alpha\da}\5\cD^\da
-2\Ii\cD_{\alpha\da}\5\cD^\da)\cA
\\
&=&e\,\xi^\alpha(-4\Ii\cD_{\alpha\da}\5\cD^\da-2B_{\alpha\da}\5\cD^\da
-16\Ii\lambda^\dR_\alpha)\cA,
\eeann
where we used (\ref{V1}) and (\ref{rigid}).
Finally we evaluate analogously the gravitino dependent terms
of the supercovariant derivative
in the last line:
\beann
\cD_{\alpha\da}\5\cD^\da\cA&=&
\sigma^\mu_{\alpha\da}(\nabla_\mu-\psi_\mu^\beta\cD_\beta
-\5\psi_{\mu\dbe}\5\cD^\dbe)\5\cD^\da\cA
\\
&=&\sigma^\mu_{\alpha\da}(\nabla_\mu\5\cD^\da
-\psi_\mu^\beta[\cD_\beta,\5\cD^\da]+\sfrac 12\5\psi_{\mu}^\da\5\cD^2)\cA
\\
&=&\sigma^\mu_{\alpha\da}(\nabla_\mu\5\cD^\da-2\Ii\psi_{\mu\beta}
\5\sigma^{\nu\da\beta}(\nabla_\nu-\5\psi_{\nu}\5\cD)
+\sfrac 12\5\psi_{\mu}^\da\5\cD^2)\cA
\\
&=&(\sigma^\mu_{\alpha\da}\nabla_{\mu}\5\cD^\da
-2\Ii(\sigma^\mu\5\sigma^\nu\psi_\mu)_\alpha
(\nabla_\nu-\5\psi_{\nu}\5\cD)
+\sfrac 12(\sigma^\mu\5\psi_{\mu})_\alpha\5\cD^2)\cA
\eeann
where $\nabla_\mu$ is covariant with regard to Lorentz and 
$R$-transformations.
Collecting all terms we obtain
\bea
e\,\delta_\mathrm{susy}\5\cD^2\cA=
e\,(-4\Ii\xi\sigma^\mu\nabla_{\mu}\5\cD
-8\xi\sigma^\mu\5\sigma^\nu\psi_\mu\nabla_\nu
+8\xi\sigma^\mu\5\sigma^\nu\psi_\mu\5\psi_{\nu}\5\cD
\nonumber\\
-2\Ii\xi\sigma^\mu\5\psi_{\mu}\5\cD^2
-2B_a\xi\sigma^a\5\cD
-16\Ii\xi\lambda^\dR+M\5\xi\5\cD)\cA.
\label{V5}
\eea

Next we compute the SUSY transformation of
$-4\Ii e\delta_\mathrm{susy}(\psi_\mu\sigma^\mu\5\cD\cA)$.
We obtain, using $\delta_\mathrm{susy}E^\mu_a
=-E^\mu_bE^\nu_a\delta_\mathrm{susy}e_\nu^b$
and manipulations as above:
\bea
\lefteqn{
-4\Ii e\delta_\mathrm{susy}(\psi_\mu\sigma^\mu\5\cD\cA)
}
\nonumber\\
&=&
  -4\Ii e[(\delta_\mathrm{susy}\psi_\mu)\sigma^\mu\5\cD
  +(\delta_\mathrm{susy}E^\mu_a)\psi_\mu\sigma^a\5\cD
  +\psi_\mu^\alpha\sigma^\mu_{\alpha\da}
  (\xi^\beta\cD_\beta+\5\xi_\dbe\5\cD^\dbe)\5\cD^\da]\cA
  \nonumber\\
&=&-4\Ii e
  (\nabla_\mu\xi
  -\Ii B_\mu\xi+\Ii B^\nu \xi\sigma_{\nu\mu}
  +\sfrac {\Ii}8 M\5\xi\5\sigma_\mu)
  \sigma^\mu\5\cD\cA
\nonumber\\
&&
  +4\Ii eE^\mu_bE^\nu_a(2\Ii\xi\sigma^b\5\psi_\nu-2\Ii\psi_\nu\sigma^b\5\xi)
  \psi_\mu\sigma^a\5\cD\cA
\nonumber\\
&&
  +8e\psi_\mu\sigma^\mu\5\sigma^\nu\xi
  (\nabla_\nu-\5\psi_\nu\5\cD)\cA
  +2\Ii e\psi_\mu\sigma^\mu\5\xi\5\cD^2\cA
\nonumber\\
&=&
  -4\Ii e \nabla_\mu\xi\sigma^\mu\5\cD\cA
  +2eB_\mu\xi\sigma^\mu\5\cD\cA
  +2eM\5\xi\5\cD\cA
\nonumber\\
&&
  +8 e(\psi_\nu\sigma^\mu\5\xi-\xi\sigma^\mu\5\psi_\nu)
  \psi_\mu\sigma^\nu\5\cD\cA
\nonumber\\
&&
  +8e\psi_\mu\sigma^\mu\5\sigma^\nu\xi(\nabla_\nu-\5\psi_\nu\5\cD)\cA
  +2\Ii e\psi_\mu\sigma^\mu\5\xi\5\cD^2\cA.
\label{V6}
\eea

To compute $-3e\,\delta_\mathrm{susy}(M\cA)$ we use
that (\ref{T12b}) gives explicitly:
\bea
S_\alpha=-(\sigma^{ab}T_{ab})_\alpha=
(-2\sigma^{\mu\nu}\nabla_\mu\psi_\nu+\sfrac{3\Ii}2B^\mu\psi_\mu
-\sfrac{3\Ii}8M\sigma^\mu\5\psi_\mu)_\alpha.
\label{V7a}
\eea
This yields:
\bea
\lefteqn{
-3e\,\delta_\mathrm{susy}(M\cA)=-16 e(\xi S-\Ii\xi\lambda^\dR)\cA
-3eM\5\xi\5\cD\cA}
\nonumber\\
&=&e(32\xi\sigma^{\mu\nu}\nabla_\mu\psi_\nu
-24\Ii B^\mu\xi\psi_\mu+6\Ii M\xi\sigma^\mu\5\psi_\mu
+16\Ii\xi\lambda^\dR
-3M\5\xi\5\cD)\cA.\quad
\label{V7}
\eea
Finally we compute $16e\delta_\mathrm{susy}
(\psi_\mu\sigma^{\mu\nu}\psi_\nu\cA)$:
\bea
\lefteqn{
16e\,\delta_\mathrm{susy}
(\psi_\mu\sigma^{\mu\nu}\psi_\nu\cA)}
\nonumber\\
&=&32e\,[(\delta_\mathrm{susy}\psi_\mu)\sigma^{\mu\nu}\psi_\nu
+(\delta_\mathrm{susy}E^\mu_a)\psi_\mu\sigma^{a\nu}\psi_\nu
+\sfrac 12\psi_\mu\sigma^{\mu\nu}\psi_\nu \5\xi\5\cD]\cA
\nonumber\\
&=&
e\,(32\nabla_\mu\xi\sigma^{\mu\nu}\psi_\nu+24\Ii B^\mu\xi\psi_\mu
-6\Ii M\psi_\mu\sigma^\mu\5\xi
\nonumber\\
&&
+64\Ii \psi_\mu\sigma^{\rho\nu}\psi_\nu
(\psi_\rho\sigma^\mu\5\xi-\xi\sigma^\mu\5\psi_\rho)
+16\psi_\mu\sigma^{\mu\nu}\psi_\nu \5\xi\5\cD)\cA.
\label{V8}\eea
Summing up (\ref{V0}), (\ref{V5}), (\ref{V6}), (\ref{V7}) and (\ref{V8})
one sees that indeed all terms cancel out except for
terms containing $\nabla_\mu$ and terms at least quadratic in the
gravitino. Playing a bit with spinor indices and using
(\ref{omegaold1}), one can check that these therms combine
to a total divergence:
\bea
\delta_\mathrm{susy}L_{old}=\6_\mu(
32e\,\xi\sigma^{\mu\nu}\psi_\nu \cA
-4\Ii e\,\xi\sigma^\mu\5\cD\cA)+\mathrm{c.c.}
\label{V9}
\eea

\providecommand{\href}[2]{#2}
\begingroup\raggedright\endgroup


\begin{thebibliography}{30}

\bibitem{LouisdeWit}
B.~de~Wit and J.~Louis, ``Supersymmetry and dualities in various dimensions,''
  in {\em Strings, Branes and Dualities}.
\newblock L. Baulieu et al (eds), Kluwer, Dordrecht, The Netherlands, 1999, pp.
  33-101.
\newblock
\href{http://arXiv.org/abs/hep-th/9801132}{{\tt hep-th/9801132}}.
\newblock

\bibitem{toine}
A.~Van~Proeyen, ``Tools for supersymmetry,''
\href{http://arXiv.org/abs/hep-th/9910030}{{\tt hep-th/9910030}}.

\bibitem{PvN}
P.~Van~Nieuwenhuizen, ``Supergravity,'' {\em Phys. Rept.} {\bf 68} (1981)
189--398.

\bibitem{WessBagger}
J.~Wess and J.~Bagger, {\em Supersymmetry and Supergravity}.
\newblock Princeton Univ. Press, Princeton, NJ, 1992.

\bibitem{1001}
S.~J. Gates, M.~T. Grisaru, M.~Rocek, and W.~Siegel, ``Superspace, or one
  thousand and one lessons in supersymmetry,'' {\em Front. Phys.} {\bf 58}
  (1983) 1--548,
\href{http://arXiv.org/abs/hep-th/0108200}{{\tt hep-th/0108200}}.

\bibitem{Nilles}
H.~P. Nilles, ``Supersymmetry, supergravity and particle physics,'' {\em Phys.
  Rept.} {\bf 110} (1984)
1.

\bibitem{West}
P.~C. West, {\em Introduction to Supersymmetry and Supergravity}.
\newblock World Scientific, Singapore, Singapore, 1990.

\bibitem{DragonES}
N.~Dragon, U.~Ellwanger, and M.~G. Schmidt, ``Supersymmetry and supergravity,''
  {\em Prog. Part. Nucl. Phys.} {\bf 18} (1987)
1--91.

\bibitem{Sergei}
I.~L. Buchbinder and S.~M. Kuzenko, {\em Ideas and Methods of Supersymmetry and
  Supergravity: Or a Walk Through Superspace}.
\newblock IOP, Bristol, UK, 1998.

\bibitem{goldschmidt}
H.~Goldschmidt, ``Integrability criteria for systems of non-linear partial
  differential equations,'' {\em J. Diff. Geom.} {\bf 1} (1967) 269--307.

\bibitem{pommaret}
F.~J. Pommaret, {\em Systems of Partial Differential Equations and Lie
  Pseudogroups}.
\newblock Gordan and Breach, New York, 1978.

\bibitem{olver}
P.~J. Olver, {\em Applications of Lie Groups to Differential Equations}.
\newblock Graduate Texts in Mathematics Vol. 107, Springer, Berlin, 1998.

\bibitem{saunders}
D.~J. Saunders, {\em The Geometry of Jet Bundles}.
\newblock Cambridge Univ. Press, Cambridge, 1989.

\bibitem{anderson}
I.~M. Anderson, ``The variational bicomplex,'' tech. rep., Utah State Univ.,
  1989,
  http://www.math.usu.edu/$\sim$fg$_{-}$mp/pages/publications/publications.htm%
l.

\bibitem{report}
G.~Barnich, F.~Brandt, and M.~Henneaux, ``Local {BRST} cohomology in gauge
  theories,'' {\em Phys. Rept.} {\bf 338} (2000) 439--569,
\href{http://arXiv.org/abs/hep-th/0002245}{{\tt hep-th/0002245}}.

\bibitem{noether}
E.~Noether, ``Invariante Variationsprobleme,'' {\em Nachrichten Kgl. Ges. d.
  Wiss. z. G{\"o}ttingen, Math.-phys. Kl.} {\bf 2} (1918) 235--257.

\bibitem{henteit}
M.~Henneaux and C.~Teitelboim, {\em Quantization of Gauge Systems}.
\newblock Princeton Univ. Press, Princeton, NJ, 1992.

\bibitem{FNF}
D.~Z. Freedman, P.~van Nieuwenhuizen, and S.~Ferrara, ``Progress toward a
  theory of supergravity,'' {\em Phys. Rev.} {\bf D13} (1976)
3214--3218.

\bibitem{DZ}
S.~Deser and B.~Zumino, ``Consistent supergravity,'' {\em Phys. Lett.} {\bf
  B62} (1976)
335.

\bibitem{offshell1}
K.~S. Stelle and P.~C. West, ``Minimal auxiliary fields for supergravity,''
  {\em Phys. Lett.} {\bf B74} (1978)
330.

\bibitem{offshell2}
S.~Ferrara and P.~van Nieuwenhuizen, ``The auxiliary fields of supergravity,''
  {\em Phys. Lett.} {\bf B74} (1978)
333.

\bibitem{ten}
F.~Brandt, ``Local {BRST} cohomology and covariance,'' {\em Commun. Math.
  Phys.} {\bf 190} (1997) 459--489,
\href{http://arXiv.org/abs/hep-th/9604025}{{\tt hep-th/9604025}}.

\bibitem{jetletter}
F.~Brandt, ``Jet coordinates for local {BRST} cohomology,'' {\em Lett. Math.
  Phys.} {\bf 55} (2001) 149--159,
\href{http://arXiv.org/abs/math-ph/0103006}{{\tt math-ph/0103006}}.

\bibitem{dragon}
N.~Dragon, ``Torsion and curvature in extended supergravity,'' {\em Z. Phys.}
  {\bf C2} (1979)
29--32.

\bibitem{diss}
F.~Brandt, ``Lagrangean densities and anomalies in four-dimensional
  su\-per\-symme\-tric theories.''
\newblock Ph.D. thesis (in German), Hannover, 1991,
  http://home.nexgo.de/fbrandt/pub.html.

\bibitem{grimmetal}
R.~Grimm, J.~Wess, and B.~Zumino, ``A complete solution of the {B}ianchi
  identities in superspace,'' {\em Nucl. Phys.} {\bf B152} (1979)
255--265.

\bibitem{sugra}
F.~Brandt, ``Local {BRST} cohomology in minimal {D}=4, {N}=1 supergravity,''
  {\em Annals Phys.} {\bf 259} (1997) 253--312,
\href{http://arXiv.org/abs/hep-th/9609192}{{\tt hep-th/9609192}}.

\bibitem{Freedman}
D.~Z. Freedman, ``Supergravity with axial gauge invariance,'' {\em Phys. Rev.}
  {\bf D15} (1977)
1173--1174.

\bibitem{Sohnius:1981tp}
M.~F. Sohnius and P.~C. West, ``An alternative minimal off-shell version of
  {N}=1 supergravity,'' {\em Phys. Lett.} {\bf B105} (1981)
353--357.

\end{thebibliography}
\end{document}